\documentclass[11pt]{article}
\pdfoutput=1

\usepackage{my-j}
\usepackage[T1]{fontenc}

\usepackage{amssymb}
\usepackage{amsfonts}
\usepackage{amsbsy}
\usepackage{amsmath}
\usepackage{amsthm}
\usepackage{graphicx}
\usepackage{epstopdf}
\usepackage[vcentermath]{youngtab}
\usepackage{multirow}
\usepackage{latexsym} 
\usepackage{array}
\usepackage{color}

\usepackage{tikz}
\usetikzlibrary{arrows}

\definecolor{Red}{rgb}{1.00, 0.00, 0.00}
\definecolor{Blue}{rgb}{0.00, 0.00, 1.00}
\definecolor{Purple}{cmyk}{0.45,0.86,0,0}%%%PANTONE PURPLE

\input cyracc.def %sha
\newfont{\twelvecyr}{wncyr10 at 12pt}
\renewcommand{\mod}[1]{\ (\mathrm{mod}\ #1)}

\def\F{\mathbb{F}}

\def\P{\mathbb{P}}

\def\n3a{t}

\def\O{\mathcal{O}}

\newcommand{\gsu}[0]{\mathrm{SU}}

\newcommand{\gsp}[0]{\mathrm{Sp}}

\newcommand{\bs}[1]{\ensuremath{\boldsymbol{#1}}}

\newcommand{\beq}{\begin{equation}}
\newcommand{\eeq}{\end{equation}}
\newcommand{\ba}{\begin{array}}
\newcommand{\ea}{\end{array}}
\newcommand{\bea}{\begin{eqnarray}}
\newcommand{\eea}{\end{eqnarray}}
\newcommand{\bean}{\begin{eqnarray*}}
\newcommand{\eean}{\end{eqnarray*}}

\newcommand{\nn}{\nonumber}

\newcommand{\comment}[1]{}

\newcommand{\musteq}{\overset{!}{=}}

\newcommand{\loc}[1]{\widetilde{#1}} % Formatting for parameters only in normalization ring
\newcommand{\quotring}[1]{R/\langle #1 \rangle}
\newcommand{\normring}[1]{\widetilde{R/\langle #1 \rangle}}

\newcommand{\stcurve}[0]{\sigma}
\newcommand{\sta}[0]{\xi}
\newcommand{\stb}[0]{\eta}
\newcommand{\stcoeff}[0]{b}
\newcommand{\stL}[0]{\loc{B}}
\newcommand{\stPhiL}[0]{\loc{\Phi}}
\newcommand{\stlambda}[0]{\lambda}

\newcommand{\sdcosqrt}[0]{\beta}

\newcommand{\sdcurve}[0]{\sigma}
\newcommand{\sda}[0]{\xi}
\newcommand{\sdb}[0]{\eta}
\newcommand{\sdcoeff}[0]{b}
\newcommand{\sdL}[0]{\loc{B}}
\newcommand{\sdphi}[0]{\phi}
\newcommand{\sdPhizeroL}[0]{\loc{\Phi}_0}
\newcommand{\sdpsi}[0]{\loc{\Psi}_1}

\newcommand{\unorm}[0]{\loc{U}}
\newcommand{\phinorm}[0]{\loc{\Phi}}

\newcommand{\dblcurve}[0]{h}
\newcommand{\dbla}[0]{\eta_{a}}
\newcommand{\dblb}[0]{\eta_b}
\newcommand{\dblL}[0]{\loc{H}}
\newcommand{\etaa}[0]{p_{(2)}}
\newcommand{\etab}[0]{p_{(1)}}
\newcommand{\etac}[0]{p_{(0)}}
\newcommand{\phidbl}[0]{\phi}
\newcommand{\PhizeroL}[0]{\loc{\Phi}_0}
\newcommand{\nua}[0]{\nu_{a}}
\newcommand{\nub}[0]{\nu_{b}}
\newcommand{\nubar}[0]{\overline{\nu}}
\newcommand{\PsiL}[0]{\loc{\Psi}}
\newcommand{\psia}[0]{\psi_{a}}
\newcommand{\psib}[0]{\psi_{b}}
\newcommand{\psibar}[0]{\overline{\psi}}
\newcommand{\gena}[0]{\alpha}
\newcommand{\genb}[0]{\beta}
\newcommand{\genc}[0]{\gamma}
\newcommand{\gend}[0]{\delta}
\newcommand{\genaone}[0]{\alpha_{a}}
\newcommand{\genatwo}[0]{\alpha_{b}}
\newcommand{\gencone}[0]{\gamma_{a}}
\newcommand{\genctwo}[0]{\gamma_{b}}
\newcommand{\genbalt}[0]{\check{\beta}}
\newcommand{\gendalt}[0]{\check{\delta}}

\newcommand{\trplcurve}[0]{t}
\newcommand{\trpla}[0]{\eta_{a}}
\newcommand{\trplb}[0]{\eta_{b}}
\newcommand{\trplL}[0]{\loc{T}}
\newcommand{\trplLone}[0]{\tau_{\eta b}}
\newcommand{\trplLzero}[0]{\tau_{\eta a}}
\newcommand{\trplLsq}[0]{\tau_{\text{sq}}}
\newcommand{\trplLcu}[0]{\tau_{\text{cu}}}
\newcommand{\ta}[0]{t_{(3)}}
\newcommand{\tb}[0]{t_{(2)}}
\newcommand{\tc}[0]{t_{(1)}}
\newcommand{\td}[0]{t_{(0)}}
\newcommand{\PhitrplL}[0]{\loc{\Phi}}
\newcommand{\phia}[0]{\phi_a}
\newcommand{\phib}[0]{\phi_b}
\newcommand{\ha}[0]{h_{a}}
\newcommand{\hb}[0]{h_{b}}
\newcommand{\hc}[0]{h_{c}}
\newcommand{\phione}[0]{\phi} % Initial component of Phi in R (not R~)
\newcommand{\phibar}[0]{\overline{\phi}}
\newcommand{\lambdaa}[0]{\lambda_{a}}
\newcommand{\lambdab}[0]{\lambda_{b}}
\newcommand{\trplLsqzero}{\tau_{\text{sq},a}}
\newcommand{\trplLsqone}{\tau_{\text{sq},b}}

\newcommand{\qcurve}[0]{q}
\newcommand{\qLab}[0]{\loc{Q}_{ab}}
\newcommand{\qLbc}[0]{\loc{Q}_{bc}}
\newcommand{\qLca}[0]{\loc{Q}_{ca}}

\newcommand{\quarta}[0]{\eta_{a}}
\newcommand{\quartb}[0]{\eta_{b}}
\newcommand{\quartc}[0]{\eta_{c}}

\newcommand{\qetaa}[0]{q_a}
\newcommand{\qetab}[0]{q_b}
\newcommand{\qetac}[0]{q_c}
\newcommand{\qetad}[0]{q_d}
\newcommand{\qetae}[0]{q_e}
\newcommand{\qetaf}[0]{q_f}

\newcommand{\qnua}[0]{\nu_{a}}
\newcommand{\qnub}[0]{\nu_{b}}
\newcommand{\qnuc}[0]{\nu_{c}}
\newcommand{\qnubara}[0]{\overline{\nu}_{a}}
\newcommand{\qnubarb}[0]{\overline{\nu}_{b}}
\newcommand{\qnubarc}[0]{\overline{\nu}_{c}}

\newcommand{\qpsia}[0]{\psi_{a}}
\newcommand{\qpsib}[0]{\psi_{b}}
\newcommand{\qpsic}[0]{\psi_{c}}
\newcommand{\qpsibara}[0]{\overline{\psi}_{a}}
\newcommand{\qpsibarb}[0]{\overline{\psi}_{b}}
\newcommand{\qpsibarc}[0]{\overline{\psi}_{c}}

\title{\boldmath Exotic matter on singular divisors in F-theory 
}

\author[1]{Denis Klevers,}
\author[2]{David R. Morrison,}
\author[3]{Nikhil Raghuram,}
\author[]{and}
\author[3]{Washington Taylor}

\affiliation[1]{Theoretical Physics Department\\ 
CERN\\
CH-1211 Geneva 23\\
 Switzerland}
\affiliation[2]{Departments of Mathematics and Physics\\
University of California, Santa Barbara\\
Santa Barbara, CA 93106, USA}
\affiliation[3]{Center for Theoretical Physics\\
Department of Physics\\
Massachusetts Institute of Technology\\
77 Massachusetts Avenue\\
Cambridge, MA 02139, USA}

\emailAdd{{\tt denis.klevers}
{\rm at} {\tt cern.ch}}
\emailAdd{\tt drm {\rm at} math.ucsb.edu}
\emailAdd{{\tt nikhilr} {\rm at} {\tt mit.edu}}
\emailAdd{{\tt wati} {\rm at} {\tt mit.edu}}

\preprint{\today \\
\hfill
CERN-TH-2016-252\\ \hfill
MIT-CTP-4878\\ \hfill
UCSB Math 2017-07.}

\abstract{
We analyze exotic matter representations that arise on singular
seven-brane configurations in F-theory.  We develop a general
framework for analyzing such representations, and work out explicit
descriptions for models with matter in the 2-index and 3-index symmetric representations
of SU($N$) and SU(2) respectively, associated with double and triple
point singularities in the seven-brane locus.  These matter
representations are associated with  Weierstrass models whose
discriminants vanish to high order thanks to
nontrivial cancellations possible only in the presence of
a non-UFD algebraic structure.
This structure can be described using the normalization of the ring of intrinsic
local functions on a singular divisor.  We consider the connection
between geometric constraints on singular curves and corresponding
constraints on the low-energy spectrum of 6D theories, identifying
some new examples of apparent ``swampland'' theories that cannot be
realized in F-theory but have no apparent low-energy inconsistency.
 }
  
\begin{document}
\maketitle

\flushbottom

%--------------------------------
\section{Introduction}

The relationship between geometric structure and the physical content
of quantum field theories and gravity theories has been a theme in
string theory and related research for several decades.  The
formulation of F-theory \cite{Vafa-F-theory, Morrison-Vafa-I,
  Morrison-Vafa-II} has given perhaps the most general geometric
approach yet to the construction of physical theories with varied
gauge groups and matter content.  While the F-theory ``dictionary''
that relates geometry and gauge symmetry is well understood both
mathematically and physically, the corresponding connection between
geometric structure and the representation theory content of matter
fields is still under development.  In this paper we analyze some new
aspects of the geometry-matter F-theory correspondence, associated
with nonperturbative features of singular seven-brane configurations
that carry exotic matter representations in the associated physical
picture.

In standard perturbative type II string theory, a stack of D-branes
carries a U($N$) gauge symmetry, and only certain relatively simple
matter representations can arise.  In particular, on supersymmetric
branes in flat space, intersecting branes carrying U($N$) and U($M$)
gauge groups give rise to bifundamental $(N, \bar{M})$ and $(\bar{N}, M)$ matter fields.  The two-index nature of the matter fields in
perturbative type II constructions comes from the realization of these
matter fields through strings, where the Chan-Paton factors on the two
ends of the string correspond to the two indices on the matter fields.
In the nonperturbative framework of
F-theory, the range of matter fields that can be realized is much
broader.  In  F-theory compactifications where an SU($N$) gauge
group is realized (e.g.\ via a type $I_N$ Kodaira singular fiber) over a
smooth 7-brane locus, the generic types of matter that  arise are adjoint
($\bs{N^2 -1}$), fundamental   ($\bs{N}$), and two-index
antisymmetric 
($\bs{N \times (N -1)/2}$)
matter fields.  These correspond again to 
two-index representations with origins common to those in the
perturbative formulation of the theory.  Another set of matter
fields that can arise in F-theory are the 3-index antisymmetric
representations ({\bf 20}, {\bf 35}, {\bf 56}) of SU(6), SU(7), and
SU(8), which can arise through nonperturbative F-theory constructions
over a smooth seven-brane locus \cite{Bershadsky-all,
mt-singularities,  Grassi-Morrison-2,  transitions}.
These antisymmetric representations can be realized explicitly through
relatively standard
Weierstrass models in F-theory.  

A more exotic set of SU($N$) representations
in F-theory
are those for which the
Young diagram has more than one column, corresponding to some indices
over which the representation is symmetric.  Such representations can
only arise over seven-brane configurations that are singular
\cite{KumarParkTaylor}.  The possibility of a two-index symmetric
representation arising at a double point singularity was suggested by
Sadov \cite{Sadov}, and considered further in \cite{mt-singularities},
but can only be distinguished from an adjoint through global
geometric considerations.  Explicit examples of such two-index
symmetric representations of SU(3) were found and explored in
\cite{ckpt, transitions}.  These explicit models exhibit rather
subtle structure in the Weierstrass model involving a nontrivial
cancellation in the ring of functions on the divisor carrying the
gauge group, which depends crucially on the structure of the
singularity.  Similar explicit representations of 3-index symmetric
representations of SU(2) were found in \cite{KleversTaylor} to have a
related structure.  In this paper we develop a systematic approach to
understanding these kinds of representations, using the non-UFD (UFD =
unique factorization domain) nature of the ring of functions on
singular seven-brane loci.

The structure of this paper is as follows: In \S\ref{sec:background}
we review some basic relevant background on F-theory constructions and
low-energy 6D supergravity theories.  Most of the explicit examples in
the paper are given in the context of 6D models, where the
understanding is most complete, though the same principles will apply
for 4D F-theory models.  In \S\ref{sec:simple-examples} we give two
very simple examples of the kinds of construction needed to realize
exotic non-UFD matter realizations, to illustrate the general structure
of these models.  In \S\ref{sec:math-description} we give a concise
description of the mathematical framework needed to describe the
Weierstrass models for these kinds of constructions.
In \S\ref{sec:double-points} we go into detail in analyzing the general
construction of models with two-index symmetric matter at double
points, and in \S\ref{sec:details-3} we describe the construction
of models with three-index symmetric matter at triple points.  In
\S\ref{sec:mattertransitions} we show how these geometric constructions are
connected to more standard matter constructions through
``matter transitions''
analogous to those studied in \cite{transitions}.
We then in \S\ref{sec:allowed}
consider how the configurations that contain these exotic
matter fields are constrained both in F-theory and from low-energy
considerations, and identify cases where the F-theory constraints are
stronger than those that are known in the low-energy theory, giving
some new examples of theories in the 6D supergravity ``swampland''.
In \S\ref{sec:allowed-representations}
we consider the more general question of what exotic matter
representations are allowed in any F-theory models, and conclude that
those studied here
seem to essentially exhaust the interesting possibilities
for matter charged under nonabelian gauge groups,
though some  more complicated representations are not ruled
out from low-energy considerations and currently lie in the swampland.
\S\ref{sec:conclusions} contains some concluding remarks.

%------------------------------------------------------------

\section{Background on F-theory and 6D supergravity}
\label{sec:background}

We review here very briefly some basics of F-theory and summarize
the important features of the
6D supergravity theories that are the focus of the explicit examples
in this paper.
Further background on F-theory can be found in \cite{Vafa-F-theory,
  Morrison-Vafa-I, Morrison-Vafa-II} or in the review notes
 \cite{Taylor:TASINotes, Morrison-TASI}.

\subsection{SU($N$) gauge factors in F-theory}
\label{sec:sun-factors}

We will consider F-theory models on a base $B$, defined by a
Weierstrass model
\begin{equation}
y^2=x^3+fx+g \,.
\label{eq:Weierstrass}
\end{equation}
Here $f, g$ are functions depending on  local coordinates in
$B$ that define an elliptic curve at each point in $B$.  More
formally, these are sections of line bundles
$f\in \Gamma (\O(-4K))$, $g\in \Gamma   (\O(-6K))$, where $K$ is the canonical class of
the base; this fixes the total space of the elliptic fibration over
$B$ to be an elliptic Calabi-Yau manifold.  The elliptic fibration is
singular along the
seven-brane locus defined by the discriminant
\begin{equation}
\Delta:= 4f^3+27g^2 = 0 \,.
\end{equation}
We will focus here primarily on type $I_n$ Kodaira singularities, which
locally are like perturbative stacks of $n$ D7-branes.  Such a
singularity occurs when the discriminant vanishes to order $n$ in a
local coordinate $z$.  In a local expansion in $z$,
\begin{eqnarray}
f & = &  f_0 + f_1 z + f_2z^2 + \cdots \label{eq:f-expansion}\\
g & = &  g_0 + g_1 z + g_2z^2 + \cdots\label{eq:g-expansion}
\end{eqnarray}
To realize an SU(2) gauge symmetry along $z = 0$, we must then have
$\Delta =  \Delta_2z^2 + \cdots$.  For vanishing at order 0, we have
$4f_0^3 + 27g_0^2 = 0$, which can be satisfied if $f_0 = -\phi^2/48, g_0 = \phi^3/864$ for some $\phi$.  For vanishing at order 1 we then
have $12f_0^2 f_1 + 54g_0g_1 = 0$, which can be solved by $g_1 = -2f_0^2 f_1/9g_0= -\phi f_1/12$.  This gives a local
construction of the Weierstrass model with an SU(2) gauge symmetry
over the locus $z = 0$.

This  analysis is extended to higher order in $z$ in
\cite{mt-singularities}.  To get an SU(3) gauge group, there are several
conditions.  First, the ``split'' condition states that $\phi$ must be
a perfect square $\phi = \phi_0^2$.  Second, the vanishing of $\Delta$
at order 2 gives the further conditions that $f_1 = \phi_0 \psi_1/2$ for some
function $\psi_1$ and that $g_2 = \psi_1^2/4-\phi_0^2 f_2/12$.

One of the principal goals of this paper is to generalize this kind of
analysis to situations where the SU($N$) gauge group is realized on a
general divisor $D$ that can have singularities.  In such a situation
the local coordinate $z$ is replaced by the section $\sigma$, where
the equation $\sigma = 0$ defines
the divisor $D$.\footnote{Thus, $D$ is a Cartier divisor.  We assume in
this paper that the base $B$ is nonsingular, which implies that all divisors
are Cartier divisors.}

\subsection{Anomaly cancellation conditions and SU($N$) spectra}
\label{sec:anomalies}

In a 6D supergravity theory there are strong consistency conditions on
the massless spectrum from anomaly constraints \cite{gs-west,
  Sagnotti}.  Using the notation and formalism of \cite{KMT-II}, the
gauge and gauge-gravitational anomaly cancellation conditions can in
general be written as
\begin{align}
-a\cdot b &= -\frac{1}{6}\left(A_{\text{Adj}} - \sum_{R} n_R A_R\right),\label{eq:aanom}\\
0&= B_{\text{adj}} - \sum_{R} n_R B_R, \label{eq:banom}\\
b\cdot b &=  -\frac{1}{3}\left(C_{\text{Adj}} - \sum_{R} n_R C_R\right). \label{eq:canom}
\end{align}
Here $a, b$ are Green-Schwarz coefficients that live in a lattice of
signature $(1, T)$ and $A_R, B_R, C_R$ are group theory coefficients
defined in e.g.\ \cite{Erler}, while $n_R$ is the number of matter
(hypermultiplet) fields in the representation $R$.
There is also the gravitational anomaly constraint
\begin{equation}
H - V = 273 - 29 T,\label{eq:gravanom}
\end{equation}
where $T$ is the number of tensor multiplets, $V$ is the number of
vector multiplets, and $H$ is the total number of hypermultiplets.
In a model that comes from F-theory, $b$ represents the divisor class
of the seven-brane curve $D$ carrying the gauge group and $a = K$
is the canonical class of $B$.  In this case, the genus of the curve
$D$ satisfies $ 2g-2 = b \cdot b + a \cdot b$. We can take this more
generally as the definition of a quantity $g$
in the low-energy theory for any choice of $a, b$, and an associated
simple gauge factor $g$
satisfying the anomaly conditions.

For the explicit models in this paper we focus primarily on theories with gauge
group SU(2) and SU(3).  For each of these gauge groups there is no
quartic invariant, so $B = 0$ and (\ref{eq:banom}) is satisfied
automatically.  Furthermore, for each of these groups global anomaly
conditions constrain $b \cdot b$ and $a \cdot b$ to be integers.    We
discuss models with each of these gauge groups in turn, and then
briefly describe the story for SU($N$) for general $N$.

\begin{table}
\centering
\begin{tabular}{|c|c|c|c|c|c|}\hline
Representation & Dimension & $A_R$ & $B_R$ & $C_R$ & $g$\\\hline
${\tiny \yng(1)}$ & $\mathbf{2}$ & 1 [$\frac12$] & 0 & $\frac{1}{2}$ [$\frac14$]& 0\\\hline
$\mathbf{Adj}$ & $\mathbf{3}$ & 4 & 0 & 8 & 1\\\hline
${\tiny \yng(3)}$ & $\mathbf{4}$ & 10 [5] & 0 & 41
$\left[\frac{41}{2}\right]$ & 6 [3]\\\hline
${\tiny \yng(4)}$ & $\mathbf{5}$ &  20  & 0 &  136   &  21
\\\hline
\end{tabular}
\caption{Anomaly coefficients for $\gsu(2)$ representations. Numbers in square brackets refer to half-hypermultiplets for self-conjugate representations. Values calculated using formulae in \cite{KumarParkTaylor}}
\label{tab:su2anomcoeff}
\end{table}

The anomaly coefficients for $\gsu(2)$ are given in Table
\ref{tab:su2anomcoeff}. 
If we assume that the only $\gsu(2)$ representations that arise are
the fundamental, adjoint, and 3-index symmetric, then
Equations \eqref{eq:aanom} and \eqref{eq:canom}
can be solved to find:
\begin{align}
n_{\mathbf{4}} &= \frac{r}{2} & n_{\mathbf{Adj}} &= g - 3 r &
n_{\mathbf{2}} &= 16+6(b\cdot b)-16 g + 7r  \,.
\label{eq:spectrum-2}
\end{align}
The gravitational anomaly constraint then gives
\begin{equation}
n_{\mathbf{1}} = 244 -29 T + 29 g - 12(b\cdot b) - 7r\,.
\end{equation}
These are the spectra for the models we wish to describe explicitly
here through F-theory by explicit Weierstrass constructions.
The multiplicities for such $\gsu(2)$ tunings on the simplest base surfaces
$\mathbb{P}^2$ and $\F^n$ are given in Table
\ref{tab:mattermult}.

\begin{table}
\centering
\begin{tabular}{|c||c|c|c|c|c|c|c|c|}\hline
 Base & $\mathbb{P}^2$ & $\F^n$ \\ \hline
 $-K_B$ & $3H$ & $2 S + (n+2) F$ \\\hline
 Number of Tensors & 0 & 1\\\hline
 Divisor Class of Curve & $d H$ & $\frac{\alpha}{2}\left(S+\frac{n}{2}F\right) +\frac{\tilde{\alpha}}{2}F$\\\hline
 $-a\cdot b$ & $3d$ & $\alpha + \tilde{\alpha}$\\\hline 
 $b\cdot b$ & $d^2$ & $\frac{1}{2}\alpha \tilde{\alpha}$\\\hline 
 Genus $g$ & $\frac{1}{2}\left(d^2 - 3 d + 2\right) $ & $\frac{1}{2}\left(\frac{1}{2}\alpha\tilde{\alpha}-\alpha -\tilde{\alpha}+2\right) $\\\hline
 ${\tiny \yng(3)}$ Multiplicity & $\frac{1}{2}r$ & $\frac{1}{2}r$\\\hline
 Adjoint Multiplicity & $\frac{1}{2}\left(d^2 - 3 d + 2 - 6r\right)$ & $\frac{1}{4}(\alpha -2)(\tilde{\alpha}-2) - 3r$ \\\hline  
 Fundamental Multiplicity & $-2d^2 + 24 d + 7r$ & $-\alpha\tilde{\alpha} +8\left(\alpha+\tilde{\alpha}\right)+7r$\\\hline
 Singlet Multiplicity & $273+\frac{5}{2}d^2 -\frac{87}{2}d-7r$ & $244+ \frac{5}{4}\alpha\tilde{\alpha}-\frac{29}{2}(\alpha+\tilde{\alpha}) - 7 r$\\\hline
%Base & Divisor& $-a\cdot b$ & $b\cdot b$ & Genus & ${\tiny \yng(3)}$'s & Adjoints & Fundamentals & Singlets\\ \hline
%$\mathbb{P}^2$ & $\beta H$ & $3 \beta$ & $\beta^2$ & $\frac{\beta^2-3\beta + 2}{2}$ & 
\end{tabular}
\caption{Multiplicities for $\gsu(2)$ models on compactification bases $\mathbb{P}^2$ and $\mathbb{\F}^n$.}
\label{tab:mattermult}
\end{table}

One way of understanding the spectrum (\ref{eq:spectrum-2}) is to note
that the most generic model
(having the largest number $n_{\mathbf{1}}$
of uncharged scalar fields) with  given $a, b$ in most cases
corresponds to the $r = 0$ model, with $g$ adjoint
representations and $16 (1-g)+6 (b\cdot b)$ fundamental
representations.  Because there are only two independent anomaly
coefficients $A, C$, the contribution of any other representation can
be described in terms of the fundamental and adjoint, giving an {\it
  anomaly equivalence} \cite{mt-singularities, Grassi-Morrison-2} such as
\begin{equation}
3 \times\bs{3} + 7 \times\bs{1} \leftrightarrow
\frac{1}{2}\times\bs{4} + 7 \times\bs{2} \,.
\label{eq:3-transition}
\end{equation}
This means that, at least as far as anomalies are concerned, 3
adjoints and 7 uncharged scalars can be exchanged for a half
hypermultiplet in the 3-index symmetric (\bs{4}) representation and 7
fundamental fields.  In \cite{transitions}, it was shown that
$3$-index antisymmetric matter representations of SU($N$) that are
anomaly equivalent to simpler matter fields can be connected
explicitly to more generic fields through unusual 
``matter transitions'' in which
the gauge group and tensor content stay unchanged but the matter
representations change.  In \S\ref{sec:mattertransitions} we show that
in a similar fashion the transition (\ref{eq:3-transition}) can be
realized explicitly as a continuous phase transition between distinct
Weierstrass models.  Note that for some choices of $a, b$ there are no
allowed models with $r = 0$.  For example, if $a = -3H, b = dH =13H$ in a
model with $T = 0$ tensor multiplets, then the number of fundamentals
$7r + 2 d(12-d) =7r-26$ being nonnegative implies that there are at
least $r \geq 4$ 3-index symmetric representations in any valid
model.  Such examples have been encountered in \cite{KleversTaylor,
  Turner-WT} and are discussed further in \S\ref{sec:triple-p2}.

Note that there is also an anomaly equivalence in the low-energy
theory
\begin{equation}
{\bf 5} +64 \times {\bf 2} \leftrightarrow 21 \times {\bf 3} + 70
\times {\bf 1} \,.
% \label{eq:}
\end{equation}
From this we can see that there are low-energy 6D supergravity models
that contain 4-index symmetric representations of SU(2) that satisfy
all the anomaly constraints including the gravitational anomaly
\cite{KumarParkTaylor}.  For example, the generic $T = 0$ model with $d = 8$
has 21 adjoints, 64 fundamental representations, and 82 uncharged
scalars.  This is anomaly-equivalent to a model with 128 fundamentals,
a single {\bf 5} and 12 uncharged scalars.  As discussed further in
\S\ref{sec:allowed-representations}, we do not believe however that
this model has an F-theory realization.

\begin{table}
\centering
\begin{tabular}{|c|c|c|c|c|c|}\hline
Representation & Dimension & $A_R$ & $B_R$ & $C_R$ & $g$\\\hline
$\mathbf{Adj}$ & $\mathbf{8}$ & 6 & 0 & 9 & 1\\\hline
${\tiny \yng(1)}$ & $\mathbf{3}$ & 1 & 0 & $\frac{1}{2}$ & 0\\\hline
${\tiny \yng(2)}$ & $\mathbf{6}$ &  5 & 0 & $\frac{17}{2}$  & 1\\\hline
${\tiny \yng(3)}$ & $\mathbf{10} $ &  15 & 0 &  $\frac{99}{2}$  & 7\\\hline
\end{tabular}
\caption{Anomaly coefficients for $\gsu(3)$ representations.}
\label{tab:table-3}
\end{table}

A similar story holds for  SU(3) models.  The anomaly coefficients
of the simplest representations are given in Table~\ref{tab:table-3}.
A generic model has $g$ adjoints and $18 (1-g)+ 6(b \cdot b)$ fundamental
representations.  There is an anomaly equivalence for every SU($N$),
$N > 2$ that relates an adjoint (plus  an uncharged scalar) to a
combination of symmetric and antisymmetric two-index tensors
\begin{equation}
\bs{1} + {\rm Adj} (\bs{N^2 -1}) \leftrightarrow
\bs{N (N -1)/2}+\bs{N (N +1)/2}\,.
\label{eq:n-equivalence}
\end{equation}
This enables the exchange of adjoints and symmetric matter while
maintaining the total value of $g$, to which each contributes one.
For  SU(3), the two-index antisymmetric representation is equivalent
to the antifundamental, so this simply gives a fundamental hypermultiplet,
and the anomaly equivalence is $\bs{1} +\bs{8} \leftrightarrow\bs{3} +\bs{6}$.  
Note that there are anomaly-consistent
SU(3) spectra with choices of $a, b$ that must have
two-index symmetric representations.  For example, for $T = 0$ at $d = 9$ the generic model has 28 adjoint fields and 0 fundamentals.  At $d = 10$, there is an anomaly-allowed model with 6 adjoints and 30
\bs{6}'s, along with 45 uncharged scalar fields.  There are no
fundamentals, however, so despite the anomaly equivalence the \bs{6}'s
cannot be exchanged for adjoints. We return to these models in
\S\ref{subsec:exp-weier-dbl}.

The story is similar for SU($N$), $N > 3$ except that there
  are three independent representations since generically $B_R\neq 0$.
  Generic models will have $g$ adjoints, $16(1-g) + (8-N) (b \cdot b)$
  fundamental $\bs{N}$ matter fields, and $2(1-g) + b \cdot b$
  two-index antisymmetric matter fields.  Adjoints can then be
  exchanged for symmetric plus antisymmetric fields through
  (\ref{eq:n-equivalence}).
Finally, note that there is an anomaly equivalence for SU(3) representations in the low-energy
theory $ 27 \times\bs{3} +\bs{10} \leftrightarrow 7 \times\bs{8} + 25 \times\bs{1}$, so there are anomaly-consistent low-energy models with
a three-index symmetric tensor (\bs{10}) representation, such as the
$T = 0, d = 6$ model with 
3 adjoints, 81 fundamentals, one 3-index symmetric, and 56 uncharged
scalar fields.  Again, we argue in 
\S\ref{sec:allowed-representations}
that such models cannot be realized in F-theory.

This completes the overview of the low-energy theories that we
encounter in the various constructions later in this paper.  
Before moving on, we note that the anomaly equations suggest that, at least for 6D theories, gravity cannot be decoupled if certain representations are present. If a representation $R$ has a $C_R$ larger than $C_{\text{Adj}}$, Equation \eqref{eq:canom} implies that $b\cdot b$ must be positive if there are any hypermultiplets in the representation $R$. (If half-hypermultiplets are possible, this scenario occurs when $\frac{1}{2}C_R > C_{\text{Adj}}$.) Recall that the Green-Schwarz coefficients live in a lattice of signature $(1,T)$. The negative part of the signature corresponds to tensors living in tensor multiplets, whereas the positive part corresponds to the tensor living in the graviton multiplet. A positive $b\cdot b$ indicates that the tensor field in the graviton multiplet participates non-trivially in the Green-Schwarz mechanism. Thus, if gravity is decoupled, one cannot cancel anomalies if there are any representations with $C_R > C_{\text{Adj}}$ (or $\frac{1}{2} C_R > C_{\text{Adj}}$ for representations with half-hypermultiplets). The $\mathbf{4}$ representation of $\gsu(2)$ has a $C_R$ that leads to positive $b\cdot b$, as do the $\mathbf{35}$ of $\gsu(7)$ and the $\mathbf{56}$ of $\gsu(8)$. While these representations occur in known 6D supergravity theories coming from F-theory, they cannot be part of a 6D theory without gravity, explaining their absence from the classification in \cite{Bhardwaj}. The $\mathbf{5}$ representation of $\gsu(2)$ and the $\mathbf{10}$ representation of $\gsu(3)$, both of which we believe cannot be realized in F-theory, also have $C_R > C_{\text{Adj}}$. It may be interesting to further explore whether this fact gives new physical insights into these representations.

%------------------------------------------------------------

\section{Tuning with  a non-UFD ring: examples}
\label{sec:simple-examples}

Before getting into technical details, to give a sense of the spirit
of the constructions needed we give a pair of simple examples of how
nontrivial cancellations can arise in the Weierstrass models realizing
$I_2$ and $I_3$ singularities when the 
divisor $D$ supporting the gauge group is itself singular.  We take
$\stcurve$ to be a section of the line bundle associated with $D$, so
that in local coordinates $\stcurve = 0$ denotes the locus of points in $D$.

\subsection{Triple points and SU($2$) 3-symmetric matter}
\label{sec:example-triple}

As a simplest example,
we want to tune on $\stcurve =  \sta^3-\stcoeff \stb^3 = 0$.  
Here $\sta$, $\stb$, and $\stcoeff$ are some  functions (sections) that do not admit
any factorization.
In general, $\stcurve$
cannot be factorized and defines a divisor
that is singular at the locus of points
$\sta = \stb = 0$.  For example, if $\sta$, $\stb$  and $\stcoeff$
are respectively irreducible quadratic, linear, and cubic
functions in some local coordinates, then $\sta = \stb = 0$ gives a pair of triple point singularities.
The general idea is that we want to  expand the ring of functions on
$D$ to allow  $\stcurve$ to be factorized.  Formally this is done using
the mathematical notion of the {\it normalized intrinsic ring}, which is
developed in detail in the following section.  More informally, the idea is that
to generalize
the expansions (\ref{eq:f-expansion}, \ref{eq:g-expansion}) for
$\stcurve$ instead of $z$, the coefficients $f_0, \ldots$ must be in the natural
ring of
functions on $D$.
The auxiliary function $\phi$ from \S\ref{sec:sun-factors}, however,
can be in a larger ring
that is
given by adjoining an element $\stL$ such that
$\sta = \stL \stb$; note that  $\stL = \stcoeff^{1/3}$ solves the
cubic equation $\stL^3 = \stcoeff$.  
This gives the normalized intrinsic ring for $D$, which has somewhat the
flavor of a Galois field extension. 
For the cubic
$\stcurve =  \sta^3-\stcoeff\stb^3 = 0$
we choose an element $\stPhiL$, the analogue of $\phi$,\footnote{We have changed the symbol to agree with the notation used later: parameters that are well-defined only in the normalized intrinsic ring are capitalized and have a tilde.} to be the following element
of the normalized intrinsic ring
\begin{equation}
\stPhiL = \stL^2 \stb \,.
% \label{eq:}
\end{equation}
We can then define the leading terms of $f$ and $g$ in terms of $\stPhiL$
\begin{eqnarray}
f_0 & = &  -\stPhiL^2/48 = -\stL^4 \stb^2/48 \Rightarrow  -\stcoeff \sta \stb/48\\
g_0 & = &  \stPhiL^3/864 = \stL^6 \stb^3/864 \Rightarrow \stcoeff^2 \stb^3/864 \,.
\end{eqnarray}
and note that they are restrictions of functions on the F-theory base, as indicated by the righthand side of the above expressions.
We then have
\begin{equation}
\Delta_0  \rightarrow
4f_0^3 + 27g_0^2 =  (- \stcoeff^3 \sta^3 \stb^3 + \stcoeff^4 \stb^6)/27648
= -\stcoeff^3 \stb^3 \stcurve/27648 \,.
% \label{eq:}
\end{equation}
 We thus have a nontrivial cancellation in the discriminant made
possible by the form of $\stcurve$.  At the next order we have
\begin{equation}
\Delta_1 \rightarrow  12f_0^2 f_1 + 54g_0g_1   -\stcoeff^3 \stb^3/27648 
= g_1 (\stcoeff^2 \stb^3)/16 + (\stcoeff^2 \stb^2 \sta^2) f_1/192 
-\stcoeff^3 \stb^3/27648 \,.
% \label{eq:}
\end{equation}
This can be made to vanish by taking, for example, $f_1 = \stb\stlambda$ for
some $\stlambda$.  Then $ g_1 =-\sta^2\stlambda/12+ \stcoeff/1728$.  We then have the
expansion
\begin{eqnarray}
f & = & -\stcoeff \sta \stb/48+\stlambda\stb\stcurve +{\cal O} (\stcurve^2)
\\
g & = &  \stcoeff^2 \stb^3/864 +(-\sta^2\stlambda/12+ \stcoeff/1728) \stcurve +{\cal O} (\stcurve^2)\\
\Delta & = &{\cal O} (\stcurve^2)
\end{eqnarray}
This gives an SU(2) on the divisor $\stcurve$, which has triple points
at the loci $\sta = \stb = 0$ in a nonstandard Weierstrass form.

\subsection{Double points and SU($3$) symmetric matter}

Now consider SU(3) with a double point associated with
\begin{equation}
\sdcurve = \sda^2 -\sdcoeff \sdb^2 \,.
% \label{eq:}
\end{equation}
Again, the normalized intrinsic ring is given by adjoining $\sdL$ such
that $\sda = \sdL \sdb$; this time, we have $\sdL = \sqrt{\sdcoeff}$,
which solves the quadratic equation $\sdL^2=\sdcoeff$.
Working in the normalized intrinsic ring, we have $f_0$ proportional
to $\sdphi^2$ and $g_0$ proportional to $\sdphi^3$, but because the
split condition must be enforced to obtain $SU(3)$, we must take
$\sdphi = \sdPhizeroL^2$.  As a possible solution not in standard form,
we choose
$\sdPhizeroL= \sdL = \sqrt{\sdcoeff}$ in the normalized intrinsic ring, so that
\begin{equation}
\sdphi = \sdcoeff 
% \label{eq:}
\end{equation}
is well-defined in the ring of functions on $D$.

At leading order, $f_0 = -\sdphi^2/48 = -\sdcoeff^2/48, g_0 = \sdphi^3/864 = \sdcoeff^3/864$.
Cancelling $\Delta_1$ we have
\begin{equation}
g_1 = -\sdphi f_1/12 = -\sdcoeff f_1/12 \,.
% \label{eq:}
\end{equation}

At the next order, we have
\begin{equation}
\Delta_2 = -\sdcoeff^2f_1^2/16+\sdcoeff^4f_2/192+\sdcoeff^3g_2/16
\end{equation}
so we wish to solve
\begin{equation}
f_1^2-\sdcoeff^2f_2/12-\sdcoeff g_2=0 
\end{equation}
in the normalized intrinsic ring.  Since $\sdcoeff=\sdL^2$, for any solution
we must be able to write
\begin{equation}
f_1 = \sdL \sdpsi \,,
% \label{eq:}
\end{equation}
where $\sdpsi$  is in the normalized intrinsic ring.  
We then take
\begin{equation}
g_2=\sdpsi^2-\sdcoeff f_2/12 \,.
\end{equation}
The challenge is to ensure that $\sdL\sdpsi$ and $\sdpsi^2$ lie
in the appropriate ring of functions on $D$.

If we choose $f_1 =  \sdL \sdb = \sda$ then $g_2= \sdb^2 - \sdcoeff f_2/12$
and we have
\begin{equation}
\Delta_2 = -\sdcoeff^2\sda^2/16+\sdcoeff^4f_2/192+\sdcoeff^3(\sdb^2 - \frac1{12}\sdcoeff f_2)/16
=-\sdcoeff^2\sdcurve/16\,,
% \label{eq:}
\end{equation}
ensuring $SU(3)$ gauge symmetry.

%--------------------------------------------------

\section{Mathematical description of the normalized intrinsic ring}
\label{sec:math-description}

Let us review the history of how our understanding of 
the singular fibers in F-theory fibrations has evolved over time.  The
first step was Kodaira's classification \cite{Kodaira}, which related specific
geometric singular fibers to specific choices of monodromy
on the homology of elliptic curves along loops in the base surrounding
the singular fiber.  (In F-theory terms, this
classifies singular fibers according to the ways in which they
source the scalar field in type IIB supergravity \cite{whatF}.)
The total space of the corresponding Weierstrass model has an ADE
singularity, and this -- together with the known gauge theory behavior
for perturbative IIB 7-branes -- allowed the association of a gauge
algebra to each codimension one singularity (for eight-dimensional
theories).  A straightforward
method to ``read off'' the type of Kodaira singular fiber from a Weierstrass
equation is also known, in terms of the orders of vanishing of the
Weierstrass coefficients $f$ and $g$ as well as the discriminant
$4f^3+27g^2$.

The second step was the realization that in lower dimensional 
compactifications, another kind of monodromy comes into play:  monodromy
could act as  automorphisms of the Kodaira singular fibers themselves
\cite{Aspinwall:1996nk}.
We refer to this as ``Tate monodromy'' to distinguish it from the
original ``Kodaira monodromy'' because Tate's algorithm \cite{Tate}
(a refinement
of the Kodaira classification) allows one to fully classify gauge
algebras in lower dimension, including monodromy 
considerations.\footnote{From Tate's point of view, this arises because
the function field of the F-theory base  is not algebraically closed.}  
This was spelled out in \cite{Bershadsky-all},
with some clarifications in \cite{Grassi-Morrison-2}.

Tate's algorithm also allows one to ``read off'' the matter content
from certain codimension two singular loci, but it was realized
in \cite{mt-singularities} and \cite{Katz-etal-Tate} that the analysis from 
\cite{Bershadsky-all}  was not complete,
and that analysis was reexamined in those two 
papers.\footnote{Ref.~\cite{mt-singularities}
had the goal of describing as many matter configurations as possible,
whereas Ref.~\cite{Katz-etal-Tate} was devoted to exploring to what extent  the
original Tate algorithm was predictive in codimension two.}
The term ``Tate form'' has come to mean a model whose gauge algebra
is determined by an equation in
one of the forms studied in \cite{Bershadsky-all,Katz-etal-Tate}; the
goal of this paper is to begin a systematic study 
of models that are not in Tate form.

The key technique in both \cite{mt-singularities} and 
\cite{Katz-etal-Tate} was to find expansions
for the Weierstrass coefficients $f$ and $g$ as finite power series in
$\sigma$, when $\{\sigma=0\}$ defines a component $\Sigma$
of the discriminant
locus of the fibration.  More precisely,  sequences of functions
$f_0$, $f_1$, \dots, $f_N$ and $g_0$, $g_1$, \dots, $g_N$ were
found such that
\begin{equation}
\label{eq:expansions}
\begin{aligned}
f &\equiv f_0 + f_1 \sigma + \cdots + f_N\sigma^N \mod{\sigma^{N+1}} \\
g &\equiv g_0 + g_1 \sigma + \cdots + g_N\sigma^N \mod{\sigma^{N+1}} ,
\end{aligned}
\end{equation}
and satisfying other properties that clarify the structure of the 
corresponding singularities.  Each function $f_j$ or $g_j$ is
chosen for its properties as an intrinsic function on $\Sigma$.
That is, if we introduce the algebraic coordinate 
ring\footnote{This algebraic coordinate ring need only contain functions
defined in a neighborhood
of the point being studied, 
and for example might take the form $R=\mathbb{C}[s,t]$
for appropriate local coordinates $s$ and $t$.} $R$
of (an open
subset of) the F-theory
base $B$  with $f_j, g_j\in R$, then
the key properties of these
functions are determined by their images in $R/\langle\sigma\rangle$.
We can think of $R/\langle\sigma\rangle$ as the ring of
{\em intrinsic local functions on $\Sigma$}.\footnote{A note about
terminology:  every affine algebraic variety has an associated
 ``coordinate ring;'' this applies equally well to  open subsets of
the F-theory base $B$ and as well as to open subsets
of the divisor $\Sigma$ on $B$.  This terminology can be confusing when
more than one algebraic variety is under discussion, so we shall use
the word ``intrinsic'' to emphasize that the functions in question
need only be defined on the divisor $\Sigma$.}

In both \cite{mt-singularities} and \cite{Katz-etal-Tate}, 
a condition was imposed that 
this ring of intrinsic local functions on $\Sigma$
should be a unique factorization domain (UFD),
and that property was used extensively in
analyzing the expansion.  In this paper, we will go beyond that assumption,
and consider divisors $\Sigma$ whose ring of intrinsic local
functions is not a UFD.

A fundamental result in algebraic geometry says that any algebraic variety
$\Sigma$ has a ``normalization'' $\widetilde{\Sigma}$ that is nonsingular
in codimension one.  (If $\Sigma$ has dimension one, then $\widetilde{\Sigma}$
is in fact nonsingular.)  The functions on
$\widetilde{\Sigma}$ are described by the ``normalization'' of the
ring  $R/\langle\sigma\rangle$.
% of intrinsic local functions.
We shall refer to this normalization $\widetilde{R/\langle\sigma\rangle}$
 as the {\em normalized intrinsic ring}.

A key property that holds when $\Sigma$ has dimension one is that
the normalized intrinsic ring is a UFD.  This means that, at least for
6D theories, we will be
able to use aspects of the UFD analysis from \cite{mt-singularities}
but applied to elements of the normalized intrinsic ring rather than
elements of the intrinsic ring itself.  For all divisors studied in
this paper (of whatever dimension), we will assume that the normalized 
intrinsic ring is a UFD.

Algebraically, the normalized intrinsic ring is 
what is known as the ``integral closure
of $R/\langle\sigma\rangle$ in its field of fractions.''  Algorithms are
known for computing this normalization in very general settings: 
we refer the reader
to Chapter 1 of \cite{Cutkosky} for a very readable account of this.  
In this paper, we will focus on
examples that are closely connected to interesting matter representations
in F-theory.

We begin with a  simple example
of a normalized intrinsic ring: a cusp singularity on $\Sigma$.
That is, we assume that $\Sigma$ has a local equation of the form
$\sigma=t^3-s^2$.  The corresponding intrinsic ring $R/\langle \sigma\rangle$
takes the form
\begin{equation}
\mathbb{C}[s,t]/\langle t^3-s^2\rangle
\end{equation}
and is visibly not a UFD, since $s\cdot s = t \cdot t \cdot t$ in that ring.

The algebraic prescription for finding the normalized intrinsic ring
is to add elements in the field of fractions of $R/\langle\sigma\rangle$
that satisfy a monic polynomial with coefficients in $R/\langle\sigma\rangle$.
(In general it may be necessary to shrink the open set in order to find
such elements, and the systematic algorithm can be complicated: see
\cite{Cutkosky}.)  In this particular case, we only need to adjoin the element 
$\unorm=s/t$,
which satisfies two equations:
\begin{equation}
\begin{aligned}
0&=\unorm t-s\\
0&=t-\unorm^2.
\end{aligned}
\end{equation}
That is, we have
\begin{equation}
\widetilde{R/\langle\sigma\rangle} = \mathbb{C}[s,t,\unorm]/\langle
t^3-s^2, \unorm t-s, t-\unorm^2 \rangle \,,
\end{equation}
which can be rewritten in the form $\widetilde{R/\langle\sigma\rangle} =\mathbb{C}[\unorm]$ since
%$s-\unorm^3 = \unorm(t-\unorm^2)-(\unorm t-s)$ 
%is also contained in the  ideal
%$\langle t^3-s^2, \unorm t-s, t-\unorm^2 \rangle$,
%allowing both $s$ and $t$ to be eliminated.
$s$ can be eliminated using $s- \unorm t$ and then $t$ can be eliminated using $t - \unorm^2$.

The geometric interpretation is this:  the function $s/t$ is well-defined
away from the cusp and has a well-defined limit on the smooth divisor
$\widetilde{\Sigma}$, so it should be added to the ring of functions.
Note that adding this function resolves the UFD issue, 
since $s^2=\unorm^6=t^3$ in the larger ring.

This structure now gives us some additional flexibility in building
F-theory models.  For $\Sigma$ to be contained in the discriminant locus,
we need $4f_0^3+27g_0^2$ to be identically zero.  If the ring of intrinsic
functions is itself a UFD, this implies that there is a function $\phi$
such that\footnote{We are following 
the normalization used in \cite{mt-singularities}.}
$f_0=-\phi^2/48$ and $g_0=\phi^3/864$.  In the case of a cusp
singularity, although the ring of intrinsic functions is not a UFD,
the normalized intrinsic ring {\em is}\/ a UFD.  We will get a solution
to the problem of putting $\Sigma$ into the discriminant locus if
we can find a function 
$\phinorm\in \widetilde{R/\langle\sigma\rangle}$ 
with the
property that $f_0:=-\phinorm^2/48$ and $g_0:=\phinorm^3/864$ both lie in the 
subring $R/\langle\sigma\rangle$ of functions coming by restriction from 
the F-theory base.
Choosing $\phinorm=\unorm$ satisfies this property without $\phinorm$ itself
being the restriction of a function from $B$.  Thus, we can take
$f_0=-t/48$ and $g_0=s/864$ to obtain a solution.

This is a gratifying result, since one of the first observations one
makes about F-theory in dimension six or lower
is that the multiplicity one part of the discriminant
almost always contains cusp singularities, at points where $f$ and $g$
both vanish.  Here we see this arising from a local analysis in a
non-UFD case.
While in this situation the discriminant generically does not support
a gauge group and there is no charged matter, the non-UFD structure
here is a simple example of the kind of thing that we encounter in the
cases here with matter at double point and triple point singularities.

The examples in \S\ref{sec:simple-examples} were also phrased in terms of the
normalized intrinsic ring.  We will be more systematic about the
structure of that ring in subsequent sections.  We will also use a
notation aimed at distinguishing between elements of the various
rings. Variables that are well-defined only in the normalized
intrinsic ring are capitalized and marked with a tilde. For the most
part, variables that are in either the coordinate ring or the ring of
intrinsic local functions are lowercase; the main exceptions are the
discriminant $\Delta$ and variables related to it (such as terms in a
power series expansion of $\Delta$).

%---------------------------------------------------
\section{Detailed analyses of constructions: double points}
\label{sec:double-points}

In this section, we describe how to derive more general $\gsu(N)$
tunings using the normalized intrinsic ring techniques. Specifically, we focus on tuning $\gsu(N)$ on curves of the form
\begin{equation}
\dblcurve \equiv \etaa \dbla^2 + 2 \etab \dbla \dblb + \etac \dblb^2 = 0, \label{eq:Hcurve}
\end{equation}
with symmetric matter localized at the $\dbla=\dblb=0$ double points. The previously derived $\gsu(3)$ models with symmetric matter use curves that can be written in this form, making this case an important one to consider. Before performing the tuning, we describe some of the physical and conceptual ideas behind the tuning. These conceptual insights in fact foreshadow some of the features of the Weierstrass model. We then give the algebraic derivation of the $\gsu(N)$ tuning and discuss the resulting matter spectrum. The final tunings are also given in Appendix \ref{app:summaryOfModels}. 

The quantities $\dbla$, $\dblb$, $\etac$, $\etab$, and $\etaa$ are all
elements of the coordinate ring of the F-theory base.  However, for some
purposes it is convenient to work with \eqref{eq:Hcurve} more abstractly,
and to do computations in an auxiliary ring 
$\mathbb{C}[\dbla,\dblb,\etac,\etab,\etaa]$ and to regard $h$ as an
element of that ring.

\subsection{Geometry, monodromy and symmetric matter}

\label{sec:symmonodromy}

 In field theory, one can Higgs an $\gsu(N\geq4)$ gauge group to
$\gsp(\lfloor\frac{N}{2}\rfloor)$ by giving a VEV to an antisymmetric
hypermultiplet. The corresponding branching rules
for the $\gsu(N)$ representations are
\begin{align}
{\tiny \yng(1)} &\rightarrow {\tiny \yng(1)} & {\tiny \yng(1,1)}
&\rightarrow {\tiny \yng(1,1)} + \mathbf{1} & {\tiny\yng(2)}
&\rightarrow {\tiny \yng(2)} &\textbf{Adj} &\rightarrow {\tiny
  \yng(2)}+{\tiny \yng(1,1)}
\label{eq:branching-ss}
\end{align}
From \S\ref{sec:anomalies}, there are anomaly-equivalent $\gsu(N)$ matter spectra related by the exchange
\begin{equation}
\textbf{Adj} + \textbf{1} \leftrightarrow {\tiny\yng(2)} + {\tiny\yng(1,1)} \label{eq:adjsymex}
\end{equation}
The branching rules in (\ref{eq:branching-ss}) imply that both sides of \eqref{eq:adjsymex} branch to the same $\gsp(\lfloor\frac{N}{2}\rfloor)$ representations. In other words, two anomaly-equivalent $\gsu(N)$ models Higgs down to the same $\gsp(\lfloor\frac{N}{2}\rfloor)$ model, even though the two models initially have different matter spectra. A similar story holds for $\gsu(3)$.
Giving a VEV to two fundamental hypermultiplets Higgses $\gsu(3)$ down to $\gsp(1)$, with the branching rules given by
\begin{align}
\mathbf{3} &\rightarrow \mathbf{2} + \mathbf{1} & \mathbf{6} &\rightarrow \mathbf{3} + \mathbf{2} + \mathbf{1} & \mathbf{8} &\rightarrow \mathbf{3}+2\times\mathbf{2}+\mathbf{1}
\end{align}
There are anomaly equivalent $\gsu(3)$ spectra related by the exchange
\begin{equation}
\mathbf{8}+\mathbf{1} \leftrightarrow \mathbf{6} + \mathbf{3}.
\end{equation}
Again, both sides of the exchange branch to the same $\gsp(1)$ representations, implying that the anomaly-equivalent $\gsu(3)$ models Higgs down to the same $\gsp(1)$ model.

An F-theory $\gsu(N)$ model with symmetric matter should have a
non-UFD Weierstrass tuning. 
This follows for $N < 6$ from the fact that a UFD Weierstrass tuning
  always has a Tate description \cite{Katz-etal-Tate} and has only
  the generic fundamental, adjoint, and two-index antisymmetric matter representations.
After Higgsing, the model contains no
exotic matter and would presumably not require non-UFD structure. Therefore,
the Weierstrass model deformation corresponding to the
$\gsu(N)\rightarrow\gsp(\lfloor \frac{N}{2}\rfloor)$ Higgsing process
should remove non-UFD structure. If we know the specifics of the
deformation, we may be able to guess where non-UFD structure appears
in the $\gsu(N)$ tuning.

Fortunately, the $\gsu(N)\rightarrow\gsp(\lfloor \frac{N}{2}\rfloor)$
Higgsing process is part of the well-known story of the split
condition. In six and fewer dimensions, the singularity type of a
codimension-one singularity may not fully specify the gauge
group. Suppose the discriminant vanishes to order $N$ along some
codimension-one locus $\sdcurve=0$, while $f$ and $g$ do not vanish
along the locus. The resulting gauge group can be either $\gsu(N)$ or
$\gsp(\lfloor \frac{N}{2}\rfloor)$. To distinguish between the two
possibilities, one must consider Tate monodromy. When one 
goes around a closed loop in
the gauge divisor, exceptional curves in the resolved fiber may or
may not be interchanged. If no interchange
occurs, the gauge group is $\gsu(N)$; otherwise, the gauge group is
$\gsp(\lfloor \frac{N}{2}\rfloor)$. At the level of the Weierstrass
model, information about the monodromy is encoded in the split
condition, namely, whether there exists $\psi$ defined on $\sdcurve=0$
such that 
\begin{equation}
\psi^2 + \frac{9g}{2f}\Bigg|_{\sdcurve=0} = 0
\end{equation}
Essentially, the condition asks whether $9g/2f$ is a perfect square along $\sdcurve=0$. If the condition is satisfied, the gauge group is $\gsu(N)$; otherwise, monodromy effects are present, and the gauge group is $\gsp(\lfloor \frac{N}{2}\rfloor)$.

When $N$ is even, the standard UFD tunings for $\gsu(N)$ and $\gsp(\frac{N}{2})$ are identical except for the split condition. From the arguments above, any non-UFD structure in the $\gsu(N)$ model with symmetrics must disappear after Higgsing. Therefore, non-UFD structure can only appear at the level of the split condition when $N$ is even. The split condition is evaluated only on $\sdcurve=0$, so one needs to consider only the leading terms $f_0$ and $g_0$ in $f$ and $g$. In both the UFD and non-UFD tunings, $f_0$ and $g_0$ will respectively be proportional to $\sdphi^2$ and $\sdphi^3$, and $9g_0/2f_0$ will be proportional to $\sdphi$. For the UFD case, the only way to satisfy the split condition is for $\sdphi$ to be a perfect square. The non-UFD case allows for more possibilities. If $\sdcurve=\sda^2-\sdcoeff\sdb^2$, the choice $\sdphi=\sdcoeff$ satisfies the split condition on $\sdcurve=0$, as
\begin{equation}
\sdcoeff = \left(\frac{\sda}{\sdb}\right)^2 - \frac{1}{\sdb^2}\sdcurve.
\end{equation}

These observations suggest the form that the non-UFD $\gsu(N)$ tunings
should take. For even $N$, one starts with the non-split UFD tuning
and implements the split condition in a non-UFD fashion. Importantly,
all of the discriminant cancellations occur exactly, and all of the
non-UFD structure is contained in the split condition. For odd $N$,
there are minor differences between the split and non-split UFD
tunings, so the prescription for the non-UFD tunings is more
complicated. Nevertheless, the odd $N$ tunings implement that split
condition in a non-UFD fashion, and there are only relatively minor
changes from the UFD tuning. In the remainder of this section, we show
via direct calculation that these insights hold in the $\gsu(3)$
tunings and $\gsu(4)$ tunings. The explicit formulas for higher
$\gsu(N)$ are described in \S\ref{sec:highsun-double}.

This picture also explains  from a geometric perspective
why the non-UFD tunings give the
symmetric matter representation.  As
described in \cite{mt-singularities}, the difference between the
adjoint and the symmetric + antisymmetric matter representations at a
double point of a divisor supporting an $A_{N -1}$ singularity comes
from the two distinct ways in which the two copies of $A_{N -1}$
associated with the gauge factors on the two branches of the divisor
are embedded into the $A_{2 N -1}$ Dynkin diagram associated with
$\P^1$'s in the resolution of the singularity in the total space of
the fibration over the double point.
When $\phi$ is a perfect square, so $\phi_0$ lives in the ring of intrinsic local functions as in the UFD case, this embedding gives the adjoint representation of SU(N).  When, on the other hand, $\phi_0$ lives only in the normalized intrinsic ring
$\widetilde{R/\langle\sigma\rangle}$, which is a quadratic extension
of $R/\langle \sigma\rangle$, there is a change of sign between the
two branches of the divisor that intersect at the double point, which flips the orientation of one of the $A_{N -1}$ Dynkin diagrams relative to the other, giving the symmetric + antisymmetric representations of SU(N).  An example of how this works is given explicitly in Appendix~\ref{sec:resolution}.  

\subsection{Generators of the normalized intrinsic ring}
\label{sec:local-double}
To find the generators that must be added to the ring of intrinsic
local functions in order to obtain the normalized intrinsic ring,
it is helpful to rewrite the expression (\ref{eq:Hcurve}) in the more suggestive form
\begin{equation}
\left(\etaa \dbla + \etab\dblb\right)\dbla = -\left(\etab\dbla + \etac \dblb\right)\dblb. \label{eq:Hrewrite}
\end{equation}
Thus, in the field of fractions we have two expressions for a single element
$\dblL$:
\begin{equation}
\dblL = \frac{\etaa\dbla+\etab\dblb}{\dblb}
= - \frac{\etab\dbla+\etac\dblb}{\dbla}
\end{equation}
Moreover, we can see that 
\begin{equation}
\dblL^2 = \frac{\etaa^2\dbla^2+2\etaa\etab\dbla\dblb+\etab^2\dblb^2}{\dblb^2} = \frac{\etaa \dblcurve}{\dblb^2} - \etaa\etac+\etab^2
\end{equation}
so that $\dblL$ satisfies the monic polynomial
in $R/\langle \dblcurve\rangle$
\begin{equation}
\dblL^2 = \etab^2-\etaa\etac \,.
\end{equation}
If $R/\langle \dblcurve\rangle$ denotes the ring of intrinsic local functions, then
the normalized intrinsic ring is\footnote{We have actually only established
that $\dblL$ is an element of this ring, not that it is the only element
that needs to be added.  However, that will be true if everything else
about $R$ and the elements $\eta_j$, $\sigma_j$ is sufficiently 
general.\label{fn:generic}}
\begin{equation}
\widetilde{R/\langle \dblcurve\rangle} = R[\dblL]/\langle 
\dblb \dblL-\etaa\dbla-\etab\dblb,
\dbla \dblL+ \etab\dbla+\etac\dblb,
\dblL^2-\etab^2+\etaa\etac \rangle
\end{equation}

Note that
\begin{equation}
\dblcurve = \dblb (\dbla \dblL+ \etab\dbla+\etac\dblb) 
- \dbla (\dblb \dblL-\etaa\dbla-\etab\dblb)
\end{equation}
so that $\dblcurve$ vanishes in $\widetilde{R/\langle \dblcurve\rangle}$, as
expected.
  Note also that $4\dblL^2$ is  the
discriminant of the quadratic (\ref{eq:Hcurve}) considered as a
function of $\dbla/\dblb$.  Thus, extending the ring by
$\dblL$ is closely related to the natural extension by the root
$\alpha$ of the quadratic $\etaa (\dbla \alpha)^2 + 2 \etab (\dbla \alpha) + \etac = 0$. Using $\dblL$, however, 
gives a particularly simple and clear way to understand the algebraic structure of
the models.  
Our discussion of triple points in \S\ref{sec:details-3} takes a similar form.

\subsection{Monomials and polynomials in the normalized intrinsic ring}
\label{sec:dblptmonos}
To perform the Weierstrass tunings, we need to determine when a product of polynomials in $\widetilde{R/\langle \dblcurve \rangle}$ lies in $R/\langle \dblcurve \rangle$. It is helpful to first focus on individual monomials before turning to polynomials. Consider a monomial in $\widetilde{R/\langle \dblcurve \rangle}$ of the form $\dbla^i \dblb^j \dblL^k$. Monomials for which $k$ is even are automatically in $R/\langle \dblcurve \rangle$, as are monomials with $i+j \geq k$. Thus, the only monomials potentially not in $R/\langle \dblcurve \rangle$ are those with $i+j < k$, where $k$ is odd. Given a monomial with odd $k$, we can repeatedly convert factors $\dblL^2$ to $\etab^2-\etaa\etac$ until we are left with a single factor of $\dblL$. Therefore, all monomials in $\widetilde{R/\langle \dblcurve \rangle}$ that do not lie in $R/\langle \dblcurve \rangle$ can be written as $\dblL$ times an expression in $\quotring{\dblcurve}$. 

A generic polynomial in $\widetilde{R/\langle \dblcurve \rangle}$ thus takes the form
\begin{equation}
\gena + \genb \dblL,
\end{equation}
where $\gena$ and $\genb$ are polynomials in $R/\langle \dblcurve \rangle$. We will be interested in situations where $\genb$ has at least one term that is not proportional to either $\dbla$ or $\dblb$. This condition in turn implies that $\genb\dblL$ is not in $R/\langle \dblcurve \rangle$, as will be necessary for a non-Tate Weierstrass tuning.

We now consider the product of two polynomials
\begin{equation}
\left(\gena+\genb\dblL\right)\left(\genc+\gend\dblL\right) = \gena \genc + \left(\genb \genc + \gena \gend\right)\dblL + \genb \gend \dblL^2.
\end{equation}
To ensure that this product lies in $R/\langle \dblcurve \rangle$, we 
need $\genb\genc+\gena \gend$ to be a linear combination of $\dbla$ and $\dblb$.
The general solution to this\footnote{Assuming that all polynomials
are sufficiently general: see footnote~\ref{fn:generic}.} takes the form
\begin{align}
\gena &= \genaone \dbla + \genatwo \dblb + \lambda\genbalt  & \genc &= \gencone \dbla + \genctwo \dblb-\lambda\gendalt,
\end{align}
where $\genbalt$ and $\gendalt$ are  parts of $\genb$ and $\gend$ which are
not divisible by $\dbla$ or $\dblb$ (which implies that
$(\genb-\genbalt)\dblL$ and $(\gend-\gendalt)\dblL$ 
both lie in $R/\langle \dblcurve\rangle$).
We would then have that
\begin{multline}
(\genb \genc+\gena\gend)\dblL = (\genb-\genbalt)\genc\dblL + \genbalt(\gencone \dbla + \genctwo \dblb)\dblL
-\genbalt\lambda\gendalt\dblL \\
+ \gena(\gend-\gendalt)\dblL + (\genaone \dbla + \genatwo \dblb)\gendalt\dblL + \lambda\genbalt\gendalt\dblL
\end{multline}
which we  see lies in $R/\langle \dblcurve\rangle$ after canceling the
$\lambda\genbalt\gendalt\dblL$ terms.  Note that if $(\gena+\genb\dblL)^2\in R/\langle \dblcurve\rangle$ then $\lambda$ must be $0$.

%\noindent
%Thus, for example, the product of  $\left(a_0 \dbla +a_1 \dblb+b\dblL\right)$ and $\left(c_0 \dbla +c_1 \dblb+d\dblL\right)$ lies in $R/\langle \dblcurve \rangle$, even though individually the two polynomials are only elements of  $\widetilde{R/\langle \dblcurve \rangle}$.

\subsection{Tuning process}
\label{sec:tuning-double}
We start by expanding $f$ and $g$ as
\begin{align}
f &= f_0 + f_1 \dblcurve + f_2 \dblcurve^2 + \ldots & g &= g_0 + g_1 \dblcurve + g_2 \dblcurve^2+\ldots \,.
\end{align}
In other words, we find an algebraic function\footnote{Ideally, this would be
a polynomial in some projective or affine coordinate ring, in practice
it may be easier to treat it as a rational function in some situations.} $f_0$ such that $f-f_0$ is divisible
by $\dblcurve$, and then an algebraic function $f_1$ such that $f-f_0-f_1\dblcurve$ is divisible
by $\dblcurve^2$, and so on.  The functions $f_i$ and $g_i$ are not unique,
and in fact may not exist on the entire base: they might only exist
in open subsets \cite{Katz-etal-Tate}.

For any choice of such an expansion, the discriminant can be expanded as
\begin{multline}
\Delta = 4 f^3 + 27 g^2 = \left(4 f_0^3 + 27 g_0^2\right) + \left(12 f_0^2 f_1 + 54 g_0 g_1\right)\dblcurve \\
+ \left(12 f_0 f_1^2 + 12 f_0^2 f_2 + 27 g_1^2 + 54 g_0 g_2\right)\dblcurve^2\\
\left(4 f_1^3 + 24 f_0 f_1 f_2+12 f_0^2 f_3 + 54 g_1 g_2 + 54 g_0 g_3\right)\dblcurve^3+ \ldots \, .
\end{multline}
Although the $f_i$ and $g_i$ are not unique, their images in the
quotient ring $R/\langle \dblcurve\rangle$ have important properties which are
independent of choices.

\subsubsection{Tuning $I_1$}
For the $I_1$ singularity, we require that 
\begin{equation}
4 f_0^3 + 27 g_0^2 \propto \dblcurve.
\end{equation}
Thanks to unique factorization in the normalized intrinsic ring,
there must exist an element $\loc{\Phi}$ in that ring such that
\begin{align}
f_0 &\equiv -\frac{1}{48}\loc{\Phi}^2 \mod \dblcurve& g_0 &\equiv\frac{1}{864}\loc{\Phi}^3 \mod \dblcurve. \label{eq:F0G0equiv}
\end{align}
In principle, $\loc{\Phi}$ might not be well-defined as an element of $R/\langle \dblcurve \rangle$.
However, if we write $\loc{\Phi}=\phidbl_1+\phidbl_2\dblL$ then having both
$\loc{\Phi}^2$ and $\loc{\Phi}^3$ in $\quotring{\dblcurve}$ implies that $\phidbl_2^3\dblL^3$ is
in $\quotring{\dblcurve}$ (since by the argument above $\phi_1$
only contains terms proportional to $\sigma_a, \sigma_b$), so that $\phidbl_2^3$ (and hence $\phidbl_2$) is a combination
of $\dbla$ and $\dblb$.  That in turn implies that $\loc{\Phi}$ itself
lies in $\quotring{\dblcurve}$.
We can therefore solve \eqref{eq:F0G0equiv} with $\phidbl \in R/\langle \dblcurve \rangle$, i.e., we can choose an algebraic function $\phidbl\in R$
that solves \eqref{eq:F0G0equiv} (mod $\dblcurve$) and then define
\begin{align}
f_0 &:= -\frac{1}{48}\phidbl^2 & g_0 &:=\frac{1}{864}\phidbl^3 . \label{eq:F0G0def}
\end{align}

With such a choice, the zeroth order term of the discriminant vanishes
exactly:
\begin{equation}
4 f_0^3 + 27 g_0^2 = 0.
\end{equation} 

This may naively seem to imply that $f_0$ and $g_0$ lack any non-Tate structure. There is a remaining condition yet to be implemented, however: the split condition. Since our focus is on tuning $\gsu(N)$ gauge groups with $N\geq 3$, we must satisfy the split condition by letting
\begin{equation}
\phidbl \equiv \PhizeroL^2 \mod \dblcurve.
\end{equation}
Here, $\PhizeroL$ is an element of $\normring{\dblcurve}$, while $\phidbl$ must be an element of $\quotring{\dblcurve}$.  From the discussion in \S\ref{sec:dblptmonos}, $\PhizeroL$ can therefore be written as
\begin{equation}
 \nua \dbla + \nub \dblb + \nubar \dblL \label{eq:phi0exp}
\end{equation}
with $\nua$, $\nub$, and $\nubar$ all algebraic functions in $R$.
$\phidbl$ is now given by
\begin{multline}
\phidbl = \left(\nua \dbla + \nub \dblb\right)^2 -2 \nubar \nua \left(\etac\dblb + \etab \dbla\right)\\+ 2 \nubar \nub \left(\etaa \dbla + \etab \dblb\right) + \nubar^2 \left(\etab^2-\etaa\etac\right) .\label{eq:phiglobaldef}
\end{multline}

\subsubsection{Tuning $I_2$}
The discriminant now reads
\begin{equation}
\Delta = \left(12 f_0^2 f_1 + 54 g_0 g_1\right)\dblcurve + \mathcal{O}(\dblcurve^2) = \frac{1}{192}\phidbl^3\left(12 g_1 + f_1 \phidbl\right)\dblcurve + \mathcal{O}(\dblcurve^2).
\end{equation}
To remove the order one term, we simply let
\begin{equation}
g_1 = -\frac1{12}f_1\PhizeroL^2=-\frac{1}{12}f_1 \phidbl. \label{eq:G1def}
\end{equation}
The order one term of $\Delta$ now vanishes exactly, leaving an $I_2$ singularity on the locus $\dblcurve=0$.

\subsubsection{Tuning $I_3^s$ to obtain $\gsu(3)$}

\label{sec:su3doubletuning}
$\Delta$ is now given by
\begin{multline} 
\label{eq:DeltaSU(3)}
\Delta = \left(12 f_0 f_1^2 + 12 f_0^2 f_2 + 27 g_1^2 + 54 g_0 g_2\right)\dblcurve^2 + \mathcal{O}(\dblcurve^3) \\
= \frac{1}{192}\phidbl^2\left(12 \phidbl g_2 + f_2 \phidbl^2 -12 f_1^2\right)\dblcurve^2 + \mathcal{O}(\dblcurve^3).
\end{multline}
To tune an $I_3^s$ singularity and obtain an $\gsu(3)$ model, we must have that
\begin{equation}
12 \phidbl g_2 + f_2 \phidbl^2 -12 f_1^2 \propto \dblcurve.
\end{equation}

Working in $\normring{\dblcurve}$, we have the condition
\begin{equation}
12 \PhizeroL^2\left(g_2 + \frac{1}{12}f_2 \PhizeroL^2\right) - 12 f_1^2 \equiv 0 \mod \dblcurve.
\end{equation}
The UFD nature of $\normring{\dblcurve}$ implies that we should tune $f_1$ and $g_2$ as
\begin{align}
f_1 &\equiv \PsiL \PhizeroL \mod \dblcurve & g_2 &\equiv \PsiL^2 - \frac{1}{12}f_2 \PhizeroL^2 \mod \dblcurve, \label{eq:su3loctuning}
\end{align}
where $\PsiL \in \normring{\dblcurve}$. 

Of course, the expressions for $f_1$ and $g_2$ should be well-defined in $\quotring{\dblcurve}$. To ensure Equation \eqref{eq:su3loctuning} is consistent with this requirement, $\PsiL$ should be expanded as
\begin{equation}
\PsiL = \psia \dbla + \psib \dblb + \psibar \dblL. \label{eq:lambdaexp}
\end{equation}
From the analysis of \S\ref{sec:dblptmonos}, $f_1$ and $g_2$ can now be written as
\begin{multline}
f_1 = \left(\psia \dbla + \psib \dblb\right)\left(\nua \dbla + \nub \dblb\right)  - \left(\psibar \nua + \nubar \psia\right)\left( \etab\dbla+\etac \dblb \right) \\
+ \left(\psibar \nub + \nubar \psib\right) \left(\etaa \dbla + \etab \dblb\right)  + \psibar \nubar \left(\etab^2 -\etaa\etac\right) \label{eq:F1globaldef}
\end{multline}
and
\begin{multline}
g_2 = \left(\psia \dbla + \psib \dblb\right)^2 -2 \psibar \psia \left( \etab \dbla+\etac\dblb\right)\\
+ 2 \psibar \psib \left(\etaa \dbla + \etab \dblb\right) + \psibar^2 \left(\etab^2-\etaa\etac\right) - \frac{1}{12}f_2 \phidbl. \label{eq:G2globaldef}
\end{multline}
After these expressions are plugged in, Equation \eqref{eq:DeltaSU(3)} takes the form
\begin{multline} 
\label{eq:defDelta2SU(3)}
\Delta = \frac{1}{16} \Big[2\left(\psia \nub-\psib \nua\right)\left(\nubar \left[\psia \dbla + \psib \dblb\right]-\psibar\left[\nua \dbla + \nub \dblb\right]\right) \\
-\nubar^2\left(\etaa \psib^2 - 2\etab \psia \psib + \etac \psia^2\right)-\psibar^2\left(\etaa \nub^2 - 2 \etab \nua \nub + \etac \nua^2\right)\\
+2 \nubar \psibar \left(\etac \psia \nua - \etab\left(\psib \nua + \psia \nub\right)+\etaa \psib \nub\right)\Big]\phidbl^2 \dblcurve^3 + \mathcal{O}(h^3).
\end{multline}
We will refer to one-sixteenth of the quantity in square brackets as $\Delta_2^\prime$. $\Delta$ is proportional to $\dblcurve^3$, and we have an $I^s_3$ singularity and an $\gsu(3)$ gauge group. Importantly, this is the first step with non-trivial cancellations in the discriminant.

Before proceeding to higher orders, let us summarize the $\gsu(3)$ model. The Weierstrass model is described by
\begin{align}
\label{eq:SU(3)WS}
f &= -\frac{1}{48}\phidbl^2 + f_1 \dblcurve + f_2 \dblcurve^2 & 
g &= \frac{1}{864}\phidbl^3 -\frac{1}{12} \phidbl f_1 \dblcurve + g_2 \dblcurve^2 + g_3 \dblcurve^3,
\end{align}
with $\phidbl$, $f_1$, and $g_2$ given respectively by Equations \eqref{eq:phiglobaldef}, \eqref{eq:F1globaldef}, and \eqref{eq:G2globaldef}. Full, expanded expressions for $f$ and $g$ are given in Appendix \ref{app:summaryOfModels}. The homology classes of the parameters are given in Table \ref{tab:su3parameters}. For some choices of $[\dblcurve]$, $[\dbla]$, $[\dblb]$, and $-K_B$, certain parameters may have ineffective homology classes. It might be possible to obtain a valid model in such situations by setting the ineffective parameters to zero. In many cases, setting a parameter to zero has only benign effects, giving a valid model. For example, if $f_2$ is set to zero, the model does not change significantly and is free of problems. Other cases may lead to an invalid model, however. There are situations in which $\nubar$ and $\psibar$ are ineffective and are forced to be zero; $(f,g,\Delta)$ then vanish to orders $(4,6,12)$ at the $\dbla=\dblb=0$ loci. Meanwhile, if $\psia$, $\psib$, $\psibar$, $f_2$, and $g_2$ are all set to zero, the discriminant vanishes exactly. The effectiveness of the various parameters therefore constrains the set of possible models. In particular, this Weierstrass tuning cannot realize certain matter spectra, as discussed further in \S\ref{sec:double-p2}. 

\begin{table}
\begin{center}
\begin{tabular}{|c|c|c|c|}\hline
Parameter & Homology Class & Equivalent in \cite{ckpt} & Equivalent in \cite{transitions}\\\hline
$\dbla$ & $[\dbla]$ & $a_1$& $\sigma$\\
$\dblb$ & $[\dblb]$ & $-b_1$& $\epsilon_1$\\
$\etaa$ & $[\dblcurve]-2[\dbla]$ & $2 s_8$& $1$\\
$\etab$ & $[\dblcurve]-[\dbla]-[\dblb]$& $s_6$& $0$\\
$\etac$ & $[\dblcurve]-2[\dblb]$& $2 s_3$& $-\frac{h}{4}$\\
$\nua$ & $-K_B - [\dbla]$& $0$& $0$\\
$\nub$ & $-K_B - [\dblb]$& $0$ & $\nu$\\
$\nubar$ & $-K_B -[\dblcurve] + [\dbla]+ [\dblb]$& $1$ & $2\beta$\\
$\psia$ & $-3K_B - [\dblcurve]-[\dbla]$ & $\frac{1}{4}s_5$& $0$ \\
$\psib$ & $-3K_B -[\dblcurve] - [\dblb]$ & $\frac{1}{4}s_2$ & $-\frac{3}{2}\lambda$\\
$\psibar$ & $-3K_B - 2[\dblcurve] + [\dbla]+[\dblb]$& $0$ & $-\frac{1}{3}\phi_2$\\
$f_2$ & $-4K_B - 2[\dblcurve]$ & $0$ & $f_4+f_5 \sigma$\\
$g_3$ & $-6 K_B - 3[\dblcurve]$ & $-\frac{1}{8}s_1$ & $g_6$ \\\hline
\end{tabular}
\end{center}
\caption{Homology classes for the $\gsu(3)$ model tuned on the generic quadratic $\dblcurve\equiv \etaa \dbla^2 + 2 \etab \dbla \dblb + \etac \dblb^2$. The homology classes are given in terms of the canonical class $K_B$ for the base and the homology classes for $\dbla$, $\dblb$ and $\dblcurve$. The third and fourth columns give the map between the parameters used here and the $\gsu(3)$ models in \cite{ckpt,transitions}.}
\label{tab:su3parameters}
\end{table}

Table \ref{tab:su3parameters} also gives the map from the $\gsu(3)$
model considered here to the two previous $\gsu(3)$ models with
symmetric matter \cite{ckpt,transitions}. In order to obtain either of
the two previous models, one of the parameters, either $\nubar$ or
$\etaa$, must be set to a constant. This restriction suggests that
both of the previous models are in fact specializations of the model
derived here. In particular, setting a parameter to a constant forces
a relationship between the three unspecified homology classes in Table
\ref{tab:su3parameters}. If $\nubar$ is set to $1$ as in \cite{ckpt},
the homology class of the curve $\dblcurve$ is fixed:
\begin{equation}
[\dblcurve] = -K_B + [\dbla] + [\dblb].
\end{equation}
Likewise, forcing $\etaa$ to be a constant, as in \cite{transitions} leads to the constraint that
\begin{equation}
[\dblcurve] = 2[\dbla].
\end{equation}
As a result, the two previous $\gsu(3)$ models have only two unspecified homology classes, and the model given here has an extra degree of freedom. The extra unspecified homology class is important physically, as it allows for matter spectra not possible with the previous two $\gsu(3)$ models. Making a parameter constant also affects matter transitions that exchange adjoints for symmetric matter, as discussed in \S\ref{sec:mattertransitions}. 

\subsubsection{Tuning $I^s_4$ to obtain $\gsu(4)$}
For the discriminant to be proportional to $\dblcurve^4$, we require that
\begin{equation}
\label{eq:Delta3}
\Delta_2^\prime \phidbl^2 +4 f_1^3 + 24 f_0 f_1 f_2+12 f_0^2 f_3 + 54 g_1 g_2 + 54 g_0 g_3 \propto \dblcurve 
\end{equation}
Working in $\normring{\dblcurve}$ and using \eqref{eq:F0G0equiv}, \eqref{eq:G1def}, and \eqref{eq:su3loctuning}, this condition can be written as
\begin{equation}
\frac{1}{192}\PhizeroL^3\left(-96 \PsiL^3 + 192\Delta_2^\prime \PhizeroL - 24 f_2 \PsiL \PhizeroL^2 +12 g_3 \PhizeroL^3 + f_3 \PhizeroL^5\right) \equiv 0 \mod \dblcurve. \label{eq:delta3loc}
\end{equation}
We therefore need $\PsiL$ to be proportional to $\PhizeroL$, which can be accomplished by letting
\begin{align}
\psia & = -\frac{1}{6}\phi_1 \nua & \psib &= -\frac{1}{6}\phi_1 \nub & \psibar &= -\frac{1}{6}\phi_1 \nubar \label{LpropF}
\end{align}
for some $\phi_1 \in R$ (i.e., we are solving \eqref{LpropF} in $\quotring{\dblcurve}$,
not in $\normring{\dblcurve}$). With these redefinitions, $\Delta_2^\prime$ is now zero, and \eqref{eq:delta3loc} is now 
\begin{equation}
\frac{1}{192}\PhizeroL^6\left(\frac{4}{9}\phi_1^3+ 4 f_2 \phi_1 +12 g_3  + f_3 \PhizeroL^2\right)= 0
\end{equation}
We thus redefine $g_3$ as
\begin{equation}
g_3 = -\frac{1}{27}\phi_1^3-\frac{1}{3}\phi_1 f_2  -\frac{1}{12}\phidbl f_3.
\end{equation}
$\Delta$ is now proportional to $\dblcurve^4$, and we have an $I^s_4$
singularity.
Note that in this discussion
since $f_2$ is untuned, we could in principle set $f_3$ to vanish by
setting $f_2 \rightarrow f_2 - f_3 h$, but we have left the
appearances of $f_3$ explicit for clarity.

To summarize the $\gsu(4)$ model, the $f$ and $g$ for the Weierstrass model are given by 
\begin{align} 
\label{eq:SU(4)WS}
f&= -\frac{1}{48}\phidbl^2 - \frac{1}{6}\phidbl \phi_1 \dblcurve + f_2 \dblcurve^2 + f_3 \dblcurve^3, \\ 
g&= \frac{1}{864}\phidbl^3 +\frac{1}{72} \phi_1 \phidbl^2 \dblcurve + \frac{1}{36}\phidbl\left(\phi_1^2 - 3 f_2\right)\dblcurve^2 + \left(-\frac{1}{12}\phidbl f_3-\frac{1}{3}\phi_1 f_2 -\frac{1}{27}\phi_1^3 \right)\dblcurve^3 + g_4 \dblcurve^4,
\end{align}
with $\phidbl$ given by Equation \eqref{eq:phiglobaldef}. The homology classes for the parameters are given in Table \ref{tab:su4parameters}. As before, ineffective parameters should be set to zero, which may lead to an invalid model.

\begin{table}
\begin{center}
\begin{tabular}{|c|c|}\hline
Parameter & Homology Class \\\hline
$\dbla$ & $[\dbla]$ \\
$\dblb$ & $[\dblb]$ \\
$\etaa$ & $[\dblcurve]-2[\dbla]$ \\
$\etab$ & $[\dblcurve]-[\dbla]-[\dblb]$\\
$\etac$ & $[\dblcurve]-2[\dblb]$\\
$\nua$ & $-K_B - [\dbla]$\\
$\nub$ & $-K_B - [\dblb]$\\
$\nubar$ & $-K_B -[\dblcurve] + [\dbla]+ [\dblb]$\\
$\phi_1$ & $-2 K_B -[\dblcurve]$ \\
$f_2$ & $-4K_B - 2[\dblcurve]$\\ 
$f_3$ & $-4K_B - 3[\dblcurve]$\\
$g_4$ & $-6 K_B - 4[\dblcurve]$\\\hline
\end{tabular}
\end{center}
\caption{Homology classes for the $\gsu(4)$ model tuned on the generic quadratic $\dblcurve\equiv \etaa \dbla^2 + 2 \etab \dbla \dblb + \etac \dblb^2$. The homology classes are given in terms of the canonical class $K_B$ for the base and the homology classes for $\dbla$, $\dblb$ and $\dblcurve$.}
\label{tab:su4parameters}
\end{table}

The $\gsu(4)$ tuning is essentially a UFD non-split $I_4$ tuning with
a specialized
non-UFD tuning for $\phidbl$. As mentioned in \S\ref{sec:symmonodromy}, this result matches the expectation that, for $\gsu(2N)$ symmetric representations, the only non-UFD structure should appear when implementing the split condition. 

\subsubsection{Tuning higher $\gsu(N)$}
\label{sec:highsun-double}
The tunings for larger $\gsu(N)$ symmetries with symmetric matter
follow from the general principles described in
\S\ref{sec:symmonodromy}. In fact, the procedure requires only small
modifications of the known UFD tunings. Note that we only discuss
models with fundamental, two-index antisymmetric, adjoint, and
two-index symmetric matter; the strategies we discuss may not apply to situations with three-index antisymmetric matter, for example. The $\gsu(N)$ tuning for even $N$ is less complicated than the odd $N$ tuning, so we first focus on the even $N$ case.

An $\gsu(2k)$ gauge group with symmetric matter can be Higgsed down to an $\gsp(k)$ model without any singular higher-genus matter. For the $\gsu(2k)$, we therefore start with the UFD tuning for a non-split $I^{ns}_{2k}$ singularity tuned on the curve $\dblcurve=0$. As discussed in \cite{mt-singularities}, this tuning takes the form\footnote{Note that, for clarity, we have used different variables and notations than \cite{mt-singularities}.}
\begin{align}
f&= -\frac{1}{3}\upsilon^2 + \mathcal{O}(\dblcurve^k) & g&= -\frac{1}{27}\upsilon^3 -\frac{1}{3} \upsilon f + \mathcal{O}(\dblcurve^{2k}),
\end{align}
with
\begin{equation}
\upsilon = \frac{1}{4}\phidbl + \phi_1 \dblcurve + \phi_2 \dblcurve^2 + \ldots + \phi_{k-1}\dblcurve^{k-1}. 
\end{equation}
We now specify that $\dblcurve$ has the quadratic form given in \eqref{eq:Hcurve}. To enhance the gauge symmetry to $\gsu(2k)$, we must perform further tunings to satisfy the split condition.  In the UFD case, this is accomplished by letting $\phidbl=\phi_0^2$. But for the non-UFD situation we are interested in here, we use $\PhizeroL$ instead of $\phi_0$, where $\PhizeroL$ is an element of $\normring{\dblcurve}$ as in Equation \eqref{eq:phi0exp}. $\phidbl$ therefore takes the form of Equation \eqref{eq:phiglobaldef}:
\begin{equation}
\phidbl = \left(\nua\dbla+\nub\dblb\right)^2-2\nua\nubar \left(\etab\dbla+\etac\dblb\right)+2\nub\nubar\left(\etaa\dbla+\etab\dblb\right)+\nubar^2\left(\etab^2-\etaa\etac\right)
\end{equation}
The split condition is satisfied and the gauge symmetry is enhanced to $\gsu(2k)$. The double points at $\dbla=\dblb=0$ now contribute symmetric matter. 

For the $\gsu(2k+1)$ tunings, we start with the UFD tuning of a split $I^s_{2k+1}$ singularity, which is also given in \cite{mt-singularities}:
\begin{align}
f&=-\frac{1}{3}\upsilon + \phi_0 \psi_k\dblcurve^k + \mathcal{O}(\dblcurve^{k+1}) & g&= -\frac{1}{27}\upsilon^3-\frac{1}{3}\upsilon f +\psi_k^2\dblcurve^{2k}+\mathcal{O}(\dblcurve^{2k+1}),
\end{align}
with
\begin{equation}
\upsilon = \frac{1}{4}\phi_0^2 + \phi_1 \dblcurve + \phi_2 \dblcurve^2 +\ldots + \phi_{k-1}\dblcurve^{k-1}.
\end{equation}
We assume that the singularity is tuned on a curve $\dblcurve$ as in Equation \eqref{eq:Hcurve}. To convert the UFD model to one with symmetric matter, we consider $\PhizeroL$ and $\Psi_k$, elements of $\normring{\dblcurve}$ that are expanded as
\begin{align}
\PhizeroL &= \nua \dbla + \nub \dblb + \nubar \dblL & \PsiL_k &= \psia \dbla + \psib \dblb + \psibar \dblL.
\end{align}
We then replace $\phi_0^2$, $\phi_0\psi_k$, and $\psi_k^2$ with the well-defined expressions for $\PhizeroL^2$, $\PhizeroL\PsiL$, and $\PsiL^2$ in $\quotring{h}$:
\begin{align}
\phi_0^2 \rightarrow  \PhizeroL^2=&\left(\nua \dbla + \nub \dblb\right)^2 -2 \nubar \nua \left(\etac\dblb + \etab \dbla\right)\notag\\&+ 2 \nubar \nub \left(\etaa \dbla + \etab \dblb\right) + \nubar^2 \left(\etab^2-\etaa\etac\right) \\
\psi_k^2 \rightarrow  \PsiL^2 =& \left(\psia \dbla + \psib \dblb\right)^2 -2 \psibar \psia \left(\etac\dblb + \etab \dbla\right)\notag\\&+ 2 \psibar \psib \left(\etaa \dbla + \etab \dblb\right) + \psibar^2 \left(\etab^2-\etaa\etac\right)\\
\phi_0\psi_k \rightarrow  \PhizeroL\PsiL =& \left(\psia \dbla + \psib \dblb\right)\left(\nua \dbla + \nub \dbla\right)  - \left(\psibar \nua + \nubar \psia\right)\left(\etac \dblb + \etab\dbla\right) \notag \\
& + \left(\psibar \nub + \nubar \psib\right) \left(\etaa \dbla + \etab \dblb\right)  + \psibar \nubar \left(\etab^2 -\etaa\etac\right) 
\end{align}
The replacements give an $\gsu(2k+1)$ model in which both the discriminant cancellations and the split condition involve non-Tate structure.

\subsection{The matter spectrum}
\label{eq:MatterDoublePoints}

Finally, we determine the structure of codimension two singularities
and the associated matter spectrum of F-theory models with gauge
symmetry from Kodaira singularities over divisors with double point
singularities. We explicitly analyze the resulting matter spectrum for
SU(3) and SU(4), while our general techniques are readily applicable
to SU(N) for general $N$.

We will focus in the following on two-dimensional base manifolds $B$
of the elliptic fibration corresponding to 6D F-theory models.  While
in principle the matter spectrum for 6D models is essentially
determined by anomaly constraints once the symmetric matter content is
known, for the non-Tate models constructed here explicitly identifying
the  structure of the
algebra-geometric loci and multiplicity of the matter fields is much more subtle
than in the simpler case of the UFD-based constructions.  We note that the 
results obtained for the
matter representations and the homology class of their corresponding
codimension two loci in $B$  are the same for F-theory compactifications
to 4D with three-dimensional base manifolds $B$; in fact, following
the analysis of \cite{CveticGrassiKleversPiragua}, in much of the
discussion here we use a language more appropriate for four dimensions,
where each matter type is associated with an irreducible codimension
two locus.

\subsubsection{General comments}

We begin with some general facts and observations on the determination
of the matter loci and spectrum in F-theory models that is common to
both examples discussed in the following. We recall that in general
the matter content of F-theory (except for  adjoint matter) is
encoded in the singularities of the elliptic fibration at codimension
two in the base $B$. 
The variety in $B$ defined in this way is in
general reducible with each of its irreducible components yielding a
particular  enhancement of the singularity type of the elliptic
fibration, which  corresponds to a particular matter representation.
Each matter representation can, in principle, occur on multiple
irreducible components, and all of those representations must be collected
together to describe the entire matter content of the model.
Note that
for compactifications to 6D there are typically many such components for
each representation (since irreducible components are points), while for
compactifications to 4D each representation might be associated to only
one irreducible component.\footnote{But note that even in this case, it is
possible for there to be more than one component associated to a given
matter representation.}

More explicitly, in the examples at hand, we have two types 
of codimension two singularities. We have a factorization of the discriminant as 
\beq
\label{eq:DeltaFactorizedGeneral}
	\Delta=\dblcurve^N \Delta_N\,
\eeq
with $N\geq 2$ and $\dblcurve$ singular at $\dbla=\dblb=0$.\footnote{The
  case $N=1$ is special as there is no codimension one singularity
  giving rise to gauge symmetry, despite the appearance of
 a codimension two symmetry of type $I_2$.}
\begin{itemize}
\item The first type of codimension two singularities are the common zeros at codimension two in 
$B$ of $\Delta_N=\dblcurve=0$.
These contain the conventional matter representations of SU(N), 
\textit{i.e.}~the fundamental representation $ {\tiny\yng(1)}$ and, for $N>3$, also the two-index anti-symmetric tensor 
representation $ {\tiny\yng(1,1)}$ of SU(N). 
\item  Second, there are codimension two 
singularities from the singularities at $\dbla=\dblb=0$ of
$\dblcurve=0$. The discriminant vanishes to order $2N$ at these loci.
They support the two-index symmetric tensor representations $
{\tiny\yng(2)}$ of SU(N).
In general, singularities of this type could also support localized
adjoint matter, as discussed in more detail in
\S\ref{sec:mattertransitions}; we assume in the analysis here that we
have a non-UFD Weierstrass model where all the double point
singularities in the discriminant locus support symmetric matter
through the kind of mechanism analyzed explicitly  in Appendix
\ref{sec:resolution}.
\end{itemize}

From a technical perspective, the determination of the irreducible components
of the codimension two loci described by the ideal $\Delta_N=\dblcurve=0$ is
the most challenging. In simple situations, such as local analyses
with $\dblcurve$ being a normal coordinate to a smooth divisor, which is
assumed for example in the standard analysis of Tate forms, $\dblcurve=0$ can simply
be inserted into $\Delta_N=0$.  
In particular, in the UFD-based analysis of \cite{mt-singularities},
the structure of $\Delta_N$ clearly decomposes into a contribution
from $\phi_0$ corresponding to antisymmetric matter fields, and a
residual discriminant component capturing the fundamental matter fields.
However, in the situation at hand,
with $\dblcurve$ given by \eqref{eq:Hcurve}, we can not solve $\dblcurve=0$
globally. This poses a problem if we want to compute the homology
classes of the codimension two matter loci. One way to circumvent this
problem is introduced in \cite{CveticGrassiKleversPiragua} to which we
refer for further details.  There, a general primary decomposition of
the locus $\Delta_N=\dblcurve=0$ is performed, yielding its associated prime
ideals, each of which corresponds to an irreducible component of the
codimension two singularities of the elliptic fibration.  Then, one
determines the homology class of each of these irreducible components
using their respective prime ideals.
(If we can, it is desirable to only partially decompose the prime ideal,
grouping various irreducible components together when they correspond
to the same matter representation.)

Note that part of the challenge in identifying the irreducible
components of the codimension two locus arises from the non-UFD
structure of the ring of intrinsic local functions.  We could in
principle analyze the codimension two structure in the normalized
intrinsic ring, in which we could write e.g.  $\Delta_3 = \PhizeroL^3
\tilde{\Delta}_{\rm  fund}$.   Since the vanishing locus of   $\PhizeroL$ is the
same as that of  $\PhizeroL^2$,  and the latter  is in  the  ring of
intrinsic local functions, we can identify in a fairly direct fashion the geometric locus of
vanishing $h = \tilde{\Delta}_{\rm  fund}= 0$, which will support fundamental
matter.  In the analysis of this section, however, we work more
generally in the geometry of the ring of intrinsic local functions,
which will in principle automatically handle issues such as
multiplicity.
These analyses should agree; the results of this section should in
part be interpreted as a confirmation of the structure of the matter
spectrum that can be derived more directly through anomaly analysis or
other approaches.

We now outline the procedure described above in the context of the two F-theory models with SU(3) and SU(4) gauge symmetry.

\subsubsection{Matter spectrum of SU(3) models}

We consider the F-theory model with $I_3^s$ singularity over the
divisor $\dblcurve=0$ defined in \eqref{eq:Hcurve}. It is specified by the
non-Tate Weierstrass form in \eqref{eq:SU(3)WS} with the tunings \eqref{eq:F1globaldef} and
\eqref{eq:G2globaldef}. 
We refer to Appendix
\ref{app:summaryOfModels} for a concise summary of the Weierstrass
model. The discriminant, given by Equation \eqref{eq:SU3discapp}, is proportional to $\dblcurve^3$, so that we have in our notation from
\eqref{eq:DeltaFactorizedGeneral} \beq \Delta=\dblcurve^3 \Delta_3\,. \eeq

For the convenience of the reader, we begin by summarizing the findings of the determination of matter content based on the analysis of codimension two singularities of this model along with the corresponding 6D matter content of 
F-theory in Table \ref{tab:SU3matter}. 
Note that in the absence of exotic (symmetric) matter, 
these multiplicities can be understood directly from the UFD
Weierstrass expansion in e.g.\ \cite{mt-singularities}, where the
discriminant takes the form $\Delta_3 = h^3 \phi_0^3 \Delta_{\text{fund}}
+{\cal O} (h^4),$ and fundamental matter arises at the zeros of
$\Delta_{\text{fund}}$,
which is in the class $-12K_B + 3K_B-3[h]$ since
$[\phi_0] = -K_B$.  This direct interpretation is more difficult to
make, however, in the more intricate non-UFD models we have
constructed here, as discussed above, although from the point of view of the ``matter
transitions'' described in \S\ref{sec:mattertransitions} the
multiplicities in this table can also be reproduced by starting with a
UFD construction and trading adjoint matter for symmetric  matter
through matter transitions.
\begin{table}[ht!]
\begin{center}
\footnotesize
\renewcommand{\arraystretch}{1.2}
\begin{tabular}{|c|c|c|@{}c@{}|}
\hline
 SU(3)-rep & Multiplicity & Fiber & Locus  \\ \hline
${\tiny\yng(2)}=\mathbf{6}$ & $x_{\mathbf{6}}=[\dbla]\cdot [\dblb]$ & $I_6$  & $V(I_{\text{Sing}})=\{\dbla = \dblb = 0\}$ \rule{0cm}{.4cm}  \\[.1cm] \hline

$\mathbf{8}$ &  $x_{\mathbf{8}}=\frac12[\dblcurve]\cdot([\dblcurve]+K_B) + 1-  x_{\bf 6}$ & $I_3$
 &  $V_{\text{SU}(2)}=\{\dblcurve=0\}$ \rule{0cm}{.4cm}
 \\[0.1cm] \hline
$\mathbf{3}$  &\rule{0cm}{0.2cm}
$  x_{\mathbf{3}}=3[\dblcurve]\cdot(-3K_B-[\dblcurve])+ x_{\mathbf{6}} $ & $I_4$ & $ V(\mathfrak{p}_1)\cup V(I_{\text{Sing}})$\\[0.1cm] \hline
\end{tabular}
\caption{\label{tab:SU3matter} Matter spectrum of the elliptic fibration \eqref{eq:SU(3)WS} with a singularity of type $I_3$ over a divisor $\dblcurve=0$ with ordinary double point singularities. Shown are the SU(3) representations, the multiplicity of full hypermultiplets in 6D, corresponding fiber types and loci in the base. We denote the variety described by the vanishing set of an ideal $I$ by $V(I)$.}
\end{center}
\end{table}

We begin with the discussion of the adjoint as well as the matter localized at the singularities of $\dblcurve=0$. As pointed out in \cite{Sadov,mt-singularities}, both representations arise from the arithmetic genus $g$ of the curve $\dblcurve=0$. Decomposing the genus $g$ of $\dblcurve=0$ into its geometric genus $p_g$ and contributions from the  $[\dbla]\cdot[\dblb]$ double point singularities, we obtain   
\beq \label{eq:p_g}
	g=p_g+ [\dbla]\cdot[\dblb]=1+\frac12 [\dblcurve]\cdot([\dblcurve]+K_B) \,.
\eeq
Here we employ that a double point contributes $1$ to the arithmetic
genus $g$, which we compute via adjunction in the second equality.  As
has been shown in \cite{Witten-phases}, there are $x_{\mathbf{8}}=p_g$
hypermultiplets in the adjoint where $p_g$ is the geometric genus
of the curve, i.e., the genus of its normalization.
As discussed above, we assume that we have a construction where all
double points correspond to the two-index symmetric tensor
$\mathbf{6}$ of SU(3).
For the constructions of this paper,
this follows from the geometric logic
described earlier, is confirmed in 
\S\ref{sec:mattertransitions} via a matter transition argument, and is
shown explicitly through resolution in an example in Appendix \ref{sec:resolution}.  Thus,
we identify the intersection number $[\dbla]\cdot[\dblb]$ as the
multiplicity $x_{\mathbf{6}}$ of matter fields in the representation
$\mathbf{6}$, and we arrive at the matter multiplicities in the  first
and second lines in Table \ref{tab:SU3matter}.

We note that each double point contributes also one hypermultiplet in the $\mathbf{3}$ of SU(3) as noticed in
\cite{ckpt}. There this was shown by Higgsing the model to an Abelian model with two U(1)'s. The presence of an additional $\mathbf{3}$ can be motivated by viewing the double points locally as the collision
of two different 7-branes carrying an SU(3) gauge group; the intersection points support matter in the bi-fundamental representation $(\mathbf{3},\mathbf{3})$.\footnote{It is the non-trivial point of our analysis that the relevant representation is $(\mathbf{3},\mathbf{3})$ and \textit{not} $(\mathbf{3},\bar{\mathbf{3}})$.} As the two 7-branes are really part of one single brane in the global geometry, 
we have to identify them and view the bi-fundamental as  the reducible
representation $\mathbf{3}\otimes \mathbf{3}$, which exhibits the
group theory decomposition $\mathbf{3}\otimes \mathbf{3}=\mathbf{6}\oplus \bar{\mathbf{3}}$
in the non-Tate situation where a symmetric matter representation is present.

Next we turn to the conventional matter localized at the intersection
loci $\Delta_3=\dblcurve=0$.  We first gain some intuition about the possible
matter loci by solving $\dblcurve=0$ locally and away from its double point
singularities and inserting the solution into $\Delta_3=0$.  We
immediately observe a factorization of $\Delta_3$ into two components,
which indicates the existence of two irreducible varieties inside
$\dblcurve=\Delta_3=0$.  

In order to find the varieties inside
$\dblcurve=\Delta_3=0$ which correspond to different types of matter,
we make a computation in the auxiliary ring 
$\mathbb{C}[\dbla,\dblb,\etac,\etab,\etaa]$.
We have performed
a rigorous primary
decomposition of $\dblcurve=\Delta_3=0$ in that ring
using Singular \cite{Singular},
obtaining two prime ideals denoted by $\mathfrak{p}_1$ and
$\mathfrak{p}_2$. 
(Note that these ideals are prime in $\mathbb{C}[\dbla,\dblb,\etac,\etab,\etaa]$ although they may factor further once specific elements of the ring $R$
are chosen to represent the variables in the auxiliary ring.  For
compactifications to 6D, they will almost certainly factor further since
each codimension two prime ideal is supported at a single point of the 
F-theory base.)

Explicitly, we find
\beq \label{eq:SU3p2}
	\mathfrak{p}_2=\left\{\dbla \etaa\nubar+\dblb(\dbla\nua +\dblb\nub+\etab\nubar) ,\dblb \etac\nubar-\dbla(\dbla\nua +\dblb\nub-\etab\nubar)\right \} / (I_{\text{sing}}\cup I')\,,
\eeq
where we have to quotient by the ideals 
\beq \label{eq:SU3redundant}
	I_{\text{sing}}=\{\dbla,\dblb\}\,,\qquad I'=\{ \dbla \nua+ \dblb \nub,\nubar\}
\eeq	
in order to obtain a prime ideal. The prime ideal $\mathfrak{p}_1$ is too lengthy to be reproduced here; it is generated by several large polynomials
in the parameters in $f$ and $g$. It can be obtained by computing  the saturation of the ideal $\dblcurve=\Delta_3=0$
w.r.t.~the ideal $\mathfrak{p}_3$, \textit{i.e.}~the repeated quotient ideal
\beq
	\mathfrak{p}_1=\lim_{n\rightarrow \infty}\{\Delta_3,\dblcurve\} / (\mathfrak{p}_2)^n\,.
\eeq

Next we analyze the singularity type of the elliptic fibration along the  varieties $V(\mathfrak{p}_1)$, $V(\mathfrak{p}_2)$ defined by the vanishing loci 
of $\mathfrak{p}_1$, $\mathfrak{p}_2$ inside the variety $\Delta_3=\dblcurve=0$. By investigation of the orders of vanishing of $(f,g,\Delta)$, we find that
\beq
		\left.(f,g,\Delta)\right\vert_{V(\mathfrak{p}_1)}\sim (0,0,4)\,, \qquad  \left.(f,g,\Delta)\right\vert_{V(\mathfrak{p}_2)}\sim (2,2,4)\,,
\eeq
which indicates Kodaira singularities of type $I_4$ and $IV$ respectively \cite{Kodaira,Tate}. This means that the variety $V(\mathfrak{p}_1)$
supports matter in the fundamental $\mathbf{3}$ of SU(3) \cite{Bershadsky-all}, whereas $V(\mathfrak{p}_2)$ is the locus of a degeneration of 
an $I_3$ singularity to type $IV$. This does not correspond to the emergence of additional physical degrees of freedom due to the lack of new 
holomorphic curves to be wrapped by M2-branes. 

Finally, for the computation of the matter multiplicity $x_{\mathbf{3}}$ of fundamentals, we need to know the homology class of  $V(\mathfrak{p}_1)$. As 
we are on a two-dimensional base $B$, the variety  $V(\mathfrak{p}_1)$ is just a collection of points and its homology class is simply the number of such points. 
We start by computing the multiplicities of   $V(\mathfrak{p}_1)$, $V(\mathfrak{p}_2)$ inside $\Delta_3=\dblcurve=0$. Using the resultant technique discussed in \cite{CveticGrassiKleversPiragua}, 
we find the multiplicities to be $1$ and $3$, respectively, i.e.~we obtain the following relation in homology:
\beq 
\label{eq:Delta3HomRelation}
	[\Delta_3]\cdot[\dblcurve]=[V(\mathfrak{p}_1)]+3[V(\mathfrak{p}_2)]\,.
\eeq
The homology class of the left hand side of this equation is readily computed using the explicit expression for $\dblcurve$ in \eqref{eq:Hcurve} and $\Delta_3$ 
in \eqref{eq:Delta3}. We then compute $[V(\mathfrak{p}_2)]$ using its definition \eqref{eq:SU3p2} as being contained in 
a complete intersection among the additional components specified by the complete intersection ideals in \eqref{eq:SU3redundant}. The homology classes of the latter are 
easily computed noting their definition as irreducible complete intersections. Their multiplicities inside the complete intersection in \eqref{eq:SU3p2} are
computed using the resultant as $1$ and $1$, respectively. Thus, we obtain
\bea \label{eq:SU3p2Homology}
	[V(\mathfrak{p}_2)]&=&(-K_B+[\dblb])\cdot (-K_B+[\dbla])-[\dbla]\cdot[\dblb]-(-K_B-[\dblcurve]+[\dbla]+[\dblb])\cdot (-K_B)\nn\\
	&=& -K_B\cdot[\dblcurve]\,,
\eea
where we used the homology classes of all relevant sections given in Table \ref{tab:su3parameters}. The first term in the first equality is the homology class of the
complete intersection in \eqref{eq:SU3p2} and the second and third terms are the homology classes of the varieties corresponding to the ideals in \eqref{eq:SU3redundant}.
Putting everything together, we obtain the homology class of $[V(\mathfrak{p}_1)]$ using the homology relation \eqref{eq:Delta3HomRelation} as
\beq 
\label{eq:Vp1SU3}
	[V(\mathfrak{p}_1)]=[\dblcurve]\cdot(-12K_B-3[\dblcurve])-3(-K_B\cdot[\dblcurve])=3[\dblcurve]\cdot(-3K_B-[\dblcurve])\,.
\eeq
The first term in the first equality is the homology class of $\Delta_3=\dblcurve=0$ and the second term is \eqref{eq:SU3p2Homology}. 
We also double check the result  for $[V(\mathfrak{p}_1)]$ by directly working with the lengthy ideal  $\mathfrak{p}_1$, \textit{i.e.}~by finding a suitable complete intersection containing 
$V(\mathfrak{p}_1)$ among with other ``auxiliary'' varieties given as complete intersections. We then just have to compute the homology class of the complete intersection and 
subtract the homology classes of the auxiliary varieties with their appropriate multiplicities inside $V(\mathfrak{p}_1)$, which we compute using the resultant.

In summary, we obtain the contribution \eqref{eq:Vp1SU3} from Kodaira
singularities of type $I_4$ over the component $V(\mathfrak{p}_1)$
inside $\Delta_3=\dblcurve=0$ to the multiplicity $x_{\mathbf{3}}$ of
$\mathbf{3}$ matter fields.  As noted earlier, there are additional
matter fields in the $\mathbf{3}$ representation for each ordinary
double point singularity \cite{ckpt}.  The combined results leads to
the full matter multiplicity in the last line of Table
\ref{tab:SU3matter}.

We conclude by checking the consistency of the derived SU(3) matter
spectrum by testing anomaly freedom of the 6D theory.  Following the
discussion of \S\ref{sec:anomalies}, we identify $b=[\dblcurve]$ and
$a=K_B$. We then see that anomaly cancellation follows immediately for
the spectrum in Table \ref{tab:SU3matter} upon the identification
$r=[\dbla]\cdot[\dblb]$, $g=1+[\dblcurve]\cdot([\dblcurve]+K_B)$.

\subsubsection{Matter spectrum of SU(4) models}

Next, we consider an F-theory model with SU(4) gauge algebra arising from a Kodaira singularity of type $I_4^s$  over the divisor $\dblcurve=0$ defined in \eqref{eq:Hcurve}. The non-Tate Weierstrass form is given in \eqref{eq:SU(4)WS} with a discriminant as in \eqref{eq:DeltaFactorizedGeneral} of the form
\beq
	\Delta=\dblcurve^4 \Delta_4\,,
\eeq
with $\Delta_4$ given in \eqref{eq:Delta4} in Appendix \ref{app:summaryOfModels}.
We again first summarize the matter content of the 6D F-theory in Table \ref{tab:SU4matter}. 
\begin{table}[ht!]
\begin{center}
\footnotesize
\renewcommand{\arraystretch}{1.2}
\begin{tabular}{|c|c|c|@{}c@{}|}
\hline
 SU(4)-rep & Multiplicity & Fiber & Locus  \\ \hline
${\tiny\yng(2)}=\mathbf{10}$ & $x_{\mathbf{10}}=[\dbla]\cdot [\dblb]$ & $I_8$  & $V(I_{\text{Sing}})=\{\dbla = \dblb = 0\}$ \rule{0cm}{.4cm}  \\[.1cm] \hline
 ${\tiny\yng(1,1)}=\mathbf{6}$  &\rule{0cm}{0.2cm}
$  x_{\mathbf{6}}=-[\dblcurve]\cdot K_B+ x_{\mathbf{10}} $ & $I_0^*$ & $ V(\mathfrak{p}_2)\cup V(I_{\text{Sing}})$\rule{0cm}{0.4cm}\\[0.1cm] \hline
$\mathbf{15}$ &  $x_{\mathbf{15}}=\frac12[\dblcurve]\cdot([\dblcurve]+K_B) + 1-  x_{\bf 10}$ & $I_4$
 &  $V_{\text{SU}(2)}=\{\dblcurve=0\}$ \rule{0cm}{.4cm}
 \\[0.1cm] \hline
$\mathbf{4}$  &\rule{0cm}{0.2cm}
$  x_{\mathbf{4}}=4[\dblcurve]\cdot(-2K_B-[\dblcurve]) $ & $I_5$ & $ V(\mathfrak{p}_1)$\\[0.1cm] \hline
\end{tabular}
\caption{\label{tab:SU4matter} Matter spectrum of the elliptic fibration \eqref{eq:SU(4)WS} with an $I_4^s$-singularity over the singular divisor $\dblcurve=0$. Shown are the SU(4) representations, the multiplicity of full hypermultiplets in 6D, corresponding fiber types and loci $V(I)$ in the base.}
\end{center}
\end{table}

As in the previous discussion of SU(3),  the adjoint matter arises from the geometric genus $p_g$ given by the general formula \eqref{eq:p_g} while the symmetric matter arises from the $[\dbla]\cdot[\dblb]$ double point singularities on $\dblcurve=0$.   
Thus, we arrive at the matter multiplicities in the  first and third lines in Table 
\ref{tab:SU4matter}. 

We note that, as in the SU(3) case, each double point contributes also one hypermultiplet in the $\mathbf{6}$ of SU(4), which we can understand by decomposing the bi-fundamental $\mathbf{4}\otimes \mathbf{4}$ as $\mathbf{10}\oplus \mathbf{6}$ at the double points.

Next, we discuss the emergence of conventional matter localized at the intersection loci 
$\Delta_4=\dblcurve=0$.
Performing a primary decomposition in the auxiliary ring
$\mathbb{C}[\dbla,\dblb,\etac,\etab,\etaa]$,
 we immediately obtain two prime ideals denoted by 
$\mathfrak{p}_1$ and $\mathfrak{p}_2$ corresponding to two irreducible varieties inside 
$\dblcurve=\Delta_4=0$. Explicitly, we find
\beq \label{eq:SU4p2}
	\mathfrak{p}_2=\left\{\dbla \etaa\nubar+\dblb(\dbla\nua +\dblb\nub+\etab\nubar) ,\dblb \etac\nubar-\dbla(\dbla\nua +\dblb\nub-\etab\nubar)\right \} / (I_{\text{sing}}\cup I')\,,
\eeq
where we have to quotient by the ideals 
\beq \label{eq:SU4redundant}
	I_{\text{sing}}=\{\dbla,\dblb\}\,,\qquad I'=\{ \dbla \nua+ \dblb \nub,\nubar\}
\eeq	
in order to obtain a prime ideal, as in \eqref{eq:SU3p2}. The prime ideal $\mathfrak{p}_1$ is once again too lengthy to be reproduced here.
It can be obtained by computing  the saturation of the ideal $\dblcurve=\Delta_4=0$ w.r.t.~the ideal $\mathfrak{p}_2$, or as the quotient ideal 
\bea \label{eq:SU4p1}
	\mathfrak{p}_1=&\{&\dbla^2 \left(f_2+\tfrac13\phi _1^2\right)^2-g_4 \left(\dbla (\nua \dbla+\nub \dblb)-\bar{\nu } (\etab\dbla+\etac\dblb)\right)^2,\nn\\&\phantom{\{}&\etaa\dbla^2+\dblb (2 \etab\dbla+\etac\dblb)\}/(I_{\text{sing}}\cup \{\etac,\dbla\})
\eea

The singularity type of the elliptic fibration along the  two varieties $V(\mathfrak{p}_1)$, $V(\mathfrak{p}_2)$ is readily analyzed by investigation of the orders of vanishing of $(f,g,\Delta)$. They are given by
\beq
		\left.(f,g,\Delta)\right\vert_{V(\mathfrak{p}_1)}\sim (0,0,5)\,, \qquad  \left.(f,g,\Delta)\right\vert_{V(\mathfrak{p}_2)}\sim (2,3,6)\,,
\eeq
respectively,  indicating Kodaira singularities of type $I_5$ and $I_0^{*}$ respectively \cite{Kodaira,Tate}. This means that the variety $V(\mathfrak{p}_1)$
supports matter in the fundamental $\mathbf{4}$ of SU(4), indicated by a local enhancement to SU(5), while the variety $V(\mathfrak{p}_2)$, in contrast to the SU($N$) case with $N\leq 3$, supports matter in the anti-symmetric representation $\mathbf{6}$, indicated by the  enhancement to SO$(8)$. We note that this is completely analogous to the SU(5) case discussed, for example, in great detail in \cite{Esole-Yau}.

Finally, the matter multiplicities $x_{\mathbf{4}}$ of fundamentals and $x_{\mathbf{6}}$ 
require the knowledge of the homology classes of  $V(\mathfrak{p}_1)$ and 
$V(\mathfrak{p}_2)$.
The multiplicities of   $V(\mathfrak{p}_1)$, $V(\mathfrak{p}_2)$ inside $\Delta_4=\dblcurve=0$ are computed with the resultant technique of \cite{CveticGrassiKleversPiragua} to be $1$ and $4$, respectively, resulting in the homology relation
\beq 
\label{eq:Delta4HomRelation}
	[\Delta_4]\cdot[\dblcurve]=[V(\mathfrak{p}_1)]+4[V(\mathfrak{p}_2)]\,.
\eeq
The homology class of the complete intersection on the left hand side is readily computed using the explicit expression for $\dblcurve$ in \eqref{eq:Hcurve} and $\Delta_4$ 
in \eqref{eq:Delta4}. We then cross-check our computations for $[V(\mathfrak{p}_1)]$ and $[V(\mathfrak{p}_2)]$ using their respective definitions \eqref{eq:SU4p1} and \eqref{eq:SU4p2}. Indeed, we find
\bea \label{eq:SU4p1Homology}
	[V(\mathfrak{p}_1)]&=&2[\dblcurve]\cdot ([\dbla]-4K_B-2[\dblcurve])-4[\dbla]\cdot[\dblb]-2[\dbla]\cdot([\dblcurve]-2[\dblb])\nn\\
	&=& 4[\dblcurve]\cdot(-2K_B-[\dblcurve])\,,
\eea
and
\bea \label{eq:SU4p2Homology}
	[V(\mathfrak{p}_2)]&=&([\dblb]-K_B)\cdot ([\dbla]-K_B)-[\dbla]\cdot[\dblb]-(-K_B-[\dblcurve]+[\dbla]+[\dblb])\cdot (-K_B)\nn\\
	&=& -K_B\cdot[\dblcurve]\,,
\eea
which obey this consistency check. Here we used the homology classes of all relevant sections 
given in Table \ref{tab:su4parameters}. The numerical prefactors in front of the terms that are 
subtracted are the multiplicities of the redundant components \eqref{eq:SU4redundant}  and 
$\{\etac,\dbla\}$ computed using the resultant. 

In summary, we obtain that the number of fundamental hypermultiplets is given by \eqref{eq:SU4p1Homology} and  the number of hypermultiplets in the $\mathbf{6}$ contributed
from $I_0^{*}$ fibers is given by \eqref{eq:SU4p2Homology}.
Together with the additional matter fields in the $\mathbf{6}$  representation at each each ordinary double point singularity we obtain the third and  last lines of Table \ref{tab:SU4matter}.

We conclude with the consistency check on the derived SU(4) matter
spectrum via 6D anomalies. Following the discussion of 
\S\ref{sec:anomalies}, we again set $b=[\dblcurve]$ and $a=K_B$. We then see
that anomaly cancellation follows immediately for the spectrum in
Table \ref{tab:SU4matter} upon the identification
$r=[\dbla]\cdot[\dblb]$ and $g=1+[\dblcurve]\cdot([\dblcurve]+K_B)$.

The upshot of the analysis in this subsection is that we can
explicitly determine the loci where the distinct matter representation
types are localized, even in the more subtle non-Tate non-UFD cases
studied earlier in this section.

%----------------------------------------------------------------------

\section{Detailed analyses of constructions: triple points}
\label{sec:details-3}

In this section, we describe how to derive $\gsu(2)$ models with
three-index symmetric matter using the normalized intrinsic ring
techniques. We focus on tuning the $\gsu(2)$ singularity on curves of
the form 
\begin{equation} \label{eq:Tgen}
\trplcurve \equiv \ta \trpla^3 + 3 \tb \trpla^2 \trplb + 3 \tc \trpla \trplb^2 + \td \trplb^3 = 0,
\end{equation}
with the ${\tiny \yng(3)}$ matter localized at the $\trpla=\trplb=0$
triple points. This form of the gauge curve agrees with that used in
\cite{KleversTaylor}. However, the tuning derived here is more general
than the one in \cite{KleversTaylor}, even though both models use the
same form of the gauge curve. We first describe the normalized
intrinsic ring and give an algebraic derivation of the $\gsu(2)$
tuning. We then discuss the resulting matter spectrum. The final model
is summarized in Appendix \ref{app:su2summary}. 
Note that this construction can also be used to describe
tri-fundamental matter fields, by choosing a cubic form that
explicitly factorizes.
There is a corresponding auxiliary ring
$\mathbb{C}[\trpla,\trplb,\td,\tc,\tb,\ta]$ in which some of our
computations are done.

\subsection{Description of the normalized intrinsic ring}
Notice that because $\trplcurve$ is a homogeneous polynomial of degree $3$,
the equation $\trplcurve=0$ can be written in the form
\begin{equation}
\trpla \left( \frac13\,\frac{\partial\trplcurve}{\partial\trpla}\right) =
-\trplb \left( \frac13\, \frac{\partial\trplcurve}{\partial\trplb}\right).
\end{equation}
Thus, in the field of fractions of $R/\langle\trplcurve\rangle$
we have two expressions for
a single element $\trplL$:
\begin{equation}
\trplL = \frac{\frac13 (\partial\trplcurve/\partial\trpla)}\trplb
= \frac{-\frac13 (\partial\trplcurve/\partial\trplb)}\trpla,
\end{equation}
which leads us to relations 
$\trplL \trplb = \frac13 (\partial\trplcurve/\partial\trpla)$ and
$\trplL \trpla = -\frac13 (\partial\trplcurve/\partial\trplb)$ to
be used in the normalized intrinsic ring.  For ease of notation, we
introduce
\begin{align}
\trplLone &:= \frac13\,\frac{\partial\trplcurve}{\partial\trpla} 
= \ta \trpla^2 + 2 \tb \trpla \trplb + \tc \trplb^2 \\
\trplLzero &:= -\frac13\,\frac{\partial\trplcurve}{\partial\trplb}
= -\tb \trpla^2 - 2 \tc \trpla \trplb - \td \trplb^2 
\end{align}
so that the relations can be written $\trplL \trplb=\trplLone$
and $\trplL \trpla=\trplLzero$.

Furthermore,
\begin{equation}
(\trplL \trplb)^2 - (\ta \trpla + \tb \trplb)\,\trplcurve = 
\trplb^2 \trplLsq
\end{equation}
where
\begin{equation}
\trplLsq := \left(\tb^2-\ta \tc\right)\trpla^2 + \left(\tb \tc - \ta \td\right)\trpla \trplb+\left(\tc^2 - \tb \td\right)\trplb^2. \label{eq:globTsquared}
\end{equation}
so that $\trplL^2=\trplLsq$ is also a relation in the normalized intrinsic
ring.  In fact, if all parameters are generic, we can define the normalized
intrinsic ring as
\begin{equation}
\widetilde{R/\langle\trplcurve\rangle} = R[\trplL]/\langle \trplL \trpla-\trplLzero, \trplL \trplb-\trplLone, \trplL^2-\trplLsq\rangle.
\end{equation}

We will later need an expression for $\trplL^3$ in this ring, which we
derive by writing
\begin{equation}
\trplLsq = \trpla \trplLsqzero + \trplb \trplLsqone,
\end{equation}
where
\begin{align}
\trplLsqzero &= 
\left(\tb^2-\ta \tc\right)\trpla + \left(\tb \tc - \ta \td\right) \trplb
\\
\trplLsqone &= 
\left(\tc^2 - \tb \td\right)\trplb . 
\end{align}
Then
\begin{equation}
\begin{aligned} \trplL^3 &= \trplL \trplLsq = \trplLzero \trplLsqzero +
\trplLone \trplLsqone \\
&=
\left( \ta \tb \tc - \tb^3 \right)\trpla^3
+3\left(\ta \tc^2 -\tc \tb^2\right)\trpla^2 \trplb\\
&+3\left(\ta \tc \td-\tb^2\td\right)\trpla \trplb^2 
+ \left( \tc^3 - 2 \tb \tc \td + \ta \td^2\right)\trplb^3.
\end{aligned}
\end{equation}
We denote the right hand side of the previous equation by $\trplLcu$.

Note that $\trplL^2$ is in the intrinsic local ring
$R/\langle\trplcurve\rangle$, so that we are really only carrying out an
extension by a quadratic element of the ring.
This construction is parallel to the ring we found in the double point
case in \S\ref{sec:double-points}.

\subsection{Tuning process}
We start by expanding $f$ and $g$ as
\begin{align}
f &= f_0 + f_1 \trplcurve + f_2 \trplcurve^2 + \ldots & g&= g_0 + g_1 \trplcurve + g_2 \trplcurve^2 + \ldots,
\end{align}
just as done in \S\ref{sec:tuning-double}. The discriminant is then given by
\begin{equation}
\Delta = 4 f^3 + 27 g^2 = \left(4 f_0^3 + 27 g_0^2\right) + \left(12 f_0^2 f_1 + 54 g_0 g_1\right)\trplcurve + \mathcal{O}(\trplcurve^2).
\end{equation}

\subsubsection{Tuning $I_1$}
For an $I_1$ singularity, the zeroth order term of the discriminant must be proportional to $\trplcurve$: 
\begin{equation}
 4 f_0^3 + 27 g_0^2 \propto \trplcurve
\end{equation}
Because the normalized intrinsic ring is a UFD, there must be some element $\PhitrplL$ in $\normring{\trplcurve}$  such that
\begin{align}
f_0 &\equiv -\frac{1}{48}  \PhitrplL^2 \mod \trplcurve & g_0 &\equiv \frac{1}{864}\PhitrplL^3 \mod \trplcurve.
\end{align}
Of course, $ \PhitrplL^2$ and $\PhitrplL^3$ must have well defined expressions in $\quotring{\trplcurve}$, which places restrictions on the form of $\PhitrplL$. We start by expanding $\PhitrplL$ as
\begin{equation}
\PhitrplL = \phione + \phibar \trplL,
\end{equation}
where $\phione$ and $\phibar$ have well-defined expressions in $\quotring{\trplcurve}$. 
Focusing first on $\PhitrplL^2$, the only potentially problematic term in
\begin{equation}
\PhitrplL^2 = \phione^2 + 2  \phione \phibar \trplL + \phibar^2 \trplL^2
\end{equation}
is the $\phione \phibar \trplL$ term. To ensure that this term lies in $\quotring{\trplcurve}$, $\phione$ should take the form 
\begin{equation}
\phione := \phia \trpla + \phib \trplb.
\end{equation}
(Note that $\phibar$ cannot take this form if we want a non-UFD Weierstrass tuning.)

$f_0$ can then be defined as
\begin{multline}
f_0 := -\frac{1}{48}\Big[\left(\phia \trpla + \phib \trplb\right)^2 - 2  \phibar  \phia\left(\tb \trpla^2 + 2 \tc \trpla \trplb + \td \trplb^2\right)\\ 
+ 2  \phibar  \phib\left(\ta \trpla^2 + 2 \tb \trpla \trplb + \tc \trplb^2\right) + \phibar^2 \trplLsq\Big].
\end{multline}
Essentially, we have replaced the terms in $\PhitrplL^2$ involving $\trplL$ with the corresponding expressions in $\quotring{\trplcurve}$. 
$\PhitrplL^3$ also lies in $\quotring{\trplcurve}$ after the redefinition of $\phione$, and $g_0$ can now be defined as
\begin{multline}
g_0 := \frac{1}{864}\Big[\left(\phia \trpla + \phib \trplb\right)^3 - 3 \phibar \phia\left(\phia \trpla + \phib \trplb\right)\left(\tb \trpla^2 + 2 \tc \trpla \trplb + \td \trplb^2\right) \\ 
+ 3 \phibar \phib\left(\phia \trpla + \phib \trplb\right)\left(\ta \trpla^2 + 2 \tb \trpla \trplb + \tc \trplb^2\right)  + 3 \phibar^2 \left(\phia \trpla + \phib \trplb\right) \trplLsq + \phibar^3 \trplLcu\Big]
\end{multline}

We now have
\begin{equation}
4 f_0^3 + 27 g_0^2 = \Delta_0^\prime \trplcurve,
\end{equation}
and the discriminant is proportional to $\trplcurve$. 
$\Delta_0^\prime$ has a lengthy expression that we do not give here. 

\subsubsection{Tuning $I_2$ to obtain $\gsu(2)$}
For an $I_2$ singularity, the discriminant must be proportional to $\trplcurve^2$, and 
\begin{equation}
\Delta_0^{\prime} +  12 f_0^2 f_1 + 54 g_0 g_1 
\end{equation}
must be proportional to  $\trplcurve$.
In other words
\begin{equation}
\Delta_0^{\prime} +  12 f_0^2 f_1 + 54 g_0 g_1 \equiv 0 \mod\trplcurve.
\end{equation}

Let us first focus on the $\Delta_0^{\prime}$ term. It is easiest to work with the  field of fractions of $\quotring{\trplcurve}$, in which $\Delta_0^\prime$ is equivalent to
\begin{multline}
\frac{\phibar^2}{32}\frac{g_0}{\trpla \trplb}\Bigg[-3 \phia \trplb \left(\td \trplb + \tc \trpla\right)-3 \phib \trpla \left(\tb \trplb + \ta \trpla\right)\\
+\phibar\left(\ta \td- \tb\tc\right)\trpla\trplb\Bigg] \label{eq:delta0primefrac}
\end{multline}
If $g_1$ were allowed to be an element
in the field of fractions, we could immediately determine how $g_1$ should be defined to cancel the $\Delta_0^\prime$ contributions. But $g_1$ is an element of the coordinate ring, and the terms in Equation \eqref{eq:delta0primefrac} cannot be fully canceled using $g_1$. In particular, the $\phia \td \trplb^2$ and $\phib \ta \trpla^2$ terms in the square brackets are not proportional to $\trpla \trplb$ and cannot be canceled. To proceed further, $\phia$ and $\phib$ must be tuned so that
\begin{equation}
\phia \td \trplb^2 + \phib \ta \trpla^2 = r \trpla \trplb + s \trplcurve.
\end{equation}
where $r$ and $s$ are some expressions in the coordinate ring. Considering the situations in which either $\trpla=0$ or $\trplb=0$ leads to the conditions that
\begin{align}
(\phia - s \trplb) \td \trplb^2\Big|_{\trpla=0} &= 0 & (\phib - s \trpla) \ta \trpla^2\Bigg|_{\trplb=0} &= 0. 
\end{align}
These conditions suggest that $\phia$ and $\phib$ should be defined as
\begin{align}
\phia :=& \ha \trpla + \hb \trplb & \phib := & \hb \trpla + \hc \trplb.
\end{align}
With these definitions, $\Delta_0^\prime$ takes the form
\begin{multline}
\Delta_0^\prime = -g_0\frac{\phibar^2}{32}\Bigg(3 \ha \left(\td \dblb + \tc \dbla\right) - 6 \hb \left(\tc \dblb + \tb \dbla\right) + 3 \hc \left(\tb \dblb + \ta \dbla\right)\\
-\phibar\left(\ta \td-\tb\tc\right) + \mathcal{O}(\trplcurve)\Bigg),
\end{multline}
where all terms above are well defined in $R$. We now define $g_1$ to be
\begin{multline}
g_1 := \frac{\phibar^2}{576}\Bigg(\ha \left(\td \dblb + \tc \dbla\right) - 2 \hb \left(\tc \dblb + \tb \dbla\right) +  \hc \left(\tb \dblb + \ta \dbla\right)\\
-\frac{1}{3}\phibar\left(\ta \td-\tb\tc\right) \Bigg)+ \gamma_1,
\end{multline}
leaving
\begin{equation}
\Delta = \left(12 f_0^2 f_1 +54 g_0 \gamma_1\right)\trplcurve + \mathcal{O}(\trplcurve^2).
\end{equation}

We now turn to the $f_0^2 f_1$ term. Working in $\normring{\trplcurve}$, the condition for the $I_2$ singularity is now
\begin{equation}
\PhitrplL^4 f_1 + 12 \PhitrplL^3 \gamma_1 \equiv 0 \mod\trplcurve.
\end{equation}
$\gamma_1$ should therefore be identified with 
\begin{equation}
-\frac{1}{12}\PhitrplL f_1 = -\frac{1}{12}\left(\ha \trpla^2 + 2\hb \trpla \trplb + \hc \trplb^2 + \phibar \trplL\right)f_1.
\end{equation}
$\PhitrplL f_1$ must lie in $\quotring{\trplcurve}$, which implies that $f_1$ must take the form
\begin{equation}
f_1 := \lambdaa \trpla + \lambdab \trplb.
\end{equation}
$\gamma_1$ should in turn be defined as
\begin{equation}
\gamma_1 := -\frac{1}{12} \left(\ha \trpla^2 + 2 \hb \trpla \trplb + \hc \trplb^2\right)\left(\lambdaa \trpla + \lambdab \trplb\right) -\frac{1}{12} \phibar\left(\lambdaa \trplLzero + \lambdab \trplLone\right)
\end{equation}

With these redefinitions, the discriminant is now proportional to $\trplcurve^2$, indicating we have successfully tuned an $\gsu(2)$ model. To summarize, $f$ and $g$ are now given by
\begin{align} \label{eq:WSSU2}
f &= f_0 + f_1 \trplcurve + f_2 \trplcurve^2 & g= g_0 + g_1 \trplcurve + g_2 \trplcurve^2,
\end{align}
with
\begin{align} \label{eq:SU2ExpF0F1G0G1}
f_0 =& -\frac{1}{48}\left(\ha \trpla^2 + 2 \hb \trpla \trplb + \hc \trplb^2\right)^2 - 
\frac{1}{24}\phibar  \phia\trplLzero -\frac{1}{24}  \phibar  \phib\trplLone -\frac{1}{48} \phibar^2 \trplLsq\\
g_0 =& \frac{1}{864}\left(\ha \trpla^2 + 2 \hb \trpla \trplb + \hc \trplb^2\right)^3 \notag\\
&+ \frac{3}{864}\phibar\left(\ha \trpla^2 + 2 \hb \trpla \trplb + \hc \trplb^2\right)\left[\left(\ha \trpla+\hb \trplb\right) \trplLzero + \left(\hb \trpla + \hc\trplb\right) \trplLone\right]\notag\\
&+ \frac{3}{864}\phibar^2\left(\ha \trpla^2 + 2 \hb \trpla \trplb + \hc \trplb^2\right) \trplLsq +\frac{1}{864}\phibar^3 \trplLcu\\
f_1 =& \lambdaa \trpla + \lambdab \trplb\\
g_1 =&\frac{\phibar^2}{576}\left[\trpla\left(\hc \ta - 2 \hb \tb + \ha \tc\right) + \trplb\left(\hc \tb -2 \hb \tc + \ha \td\right)-\frac{\phibar}{3}\left(\ta \td-\tb\tc\right)\right]\notag\\
& -\frac{1}{12} \left(\ha \trpla^2 + 2 \hb \trpla \trplb + \hc \trplb^2\right)\left(\lambdaa \trpla + \lambdab \trplb\right) -\frac{1}{12} \phibar\left(\lambdaa \trplLzero + \lambdab \trplLone\right).\label{eq:SU2ExpF0F1G0G1end}
\end{align}
The homology classes of the various parameters are given in Table \ref{tab:su2parameters}.

Table \ref{tab:su2parameters} also gives the dictionary between this $\gsu(2)$ model and the previous $\gsu(2)$ model given in \cite{KleversTaylor}. The key difference between these models is that $\phibar$ is forced to be a constant in \cite{KleversTaylor}. This restricts the homology classes: for $\phibar$ to be a constant, 
\begin{equation}
[\trplcurve] \musteq -2K_B + [\trplb] + [\trpla].
\end{equation}
The $\gsu(2)$ model of \cite{KleversTaylor} thus has only two unspecified homology classes, whereas the model derived using the normalized intrinsic ring techniques has three unspecified homology classes. This extra freedom has physical consequences. In particular, the model derived here can support a wider array of matter spectra than the model in \cite{KleversTaylor}. Otherwise, the two models are fairly similar. In fact, if $f_2$ and $g_2$ are set to zero and $\phibar$ is set to a constant, the two models are equivalent. 

\begin{table}[ht!]
\begin{center}
\begin{tabular}{|c|c|c|}\hline
Parameter & Homology Class & Equivalent Symbol in \cite{KleversTaylor}\\\hline
$\trpla$ & $[\trpla]$& $s_8$\\
$\trplb$ & $[\trplb]$& $-s_9$\\
$\ta$ & $[\trplcurve]-3[\trpla]$& $12 s_4$\\
$\tb$ & $[\trplcurve]-2[\trpla]-[\trplb]$& $4 s_3$\\
$\tc$ & $[\trplcurve]-[\trpla]-2[\trplb] $ & $4 s_2$\\
$\td$ & $[\trplcurve]-3[\trplb] $& $12 s_1$\\
$\phibar$ & $-2K_B -[\trplcurve] + [\trplb] + [\trpla]$ & $1$\\
$\ha$ & $-2 K_B - 2[\trpla]$ & $0$\\
$\hb$ & $-2 K_B - [\trpla]-[\trplb]$ & $0$\\
$\hc$ & $-2 K_B - 2[\trplb]$& $0$\\
$\lambdaa$ & $-4 K_B - [\trplcurve]-[\trpla]$ & $0$\\
$\lambdab$ & $-4 K_B - [\trplcurve] - [\trplb]$ & $0$\\
$f_2$ & $-4 K_B -2[\trplcurve]$ & $0$\\
$g_2$ & $-6 K_B -2[\trplcurve]$ & $0$\\\hline
\end{tabular}
\end{center}
\caption{Homology classes for the $SU(2)$ model tuned on a generic cubic $\trplcurve \equiv \ta \trpla^3 + 3 \tb \trpla^2 \trplb + 3 \tc \trplb^2 \trpla + \td \trplb^3$. Homology classes are given in terms of the canonical class $K_B$ of the base and the homology classes of $\trpla$, $\trplb$, and $\trplcurve$. The third column gives the map between the parameters used here and those for the $\gsu(2)$ models in \cite{KleversTaylor}.}
\label{tab:su2parameters}
\end{table}

\subsection{The matter spectrum}
\label{eq:MatterTriplePoints}

Equipped with the general non-Tate Weierstrass model described by
\eqref{eq:WSSU2} through \eqref{eq:SU2ExpF0F1G0G1end}, we proceed with
the determination of the singular locus and corresponding matter spectrum of the corresponding F-theory
model. As before, we will focus on two-dimensional base manifolds $B$
of the elliptic fibration yielding a 6D supergravity theory, although
the following results also carry over to non-chiral F-theory
compactifications to 4D.  The discussion will be very similar to the
one in \S\ref{eq:MatterDoublePoints}.
 
We recall that the matter content of F-theory (except for the adjoint matter) is encoded in the codimension two singularities of the elliptic fibration specified by the Weierstrass 
model \eqref{eq:WSSU2}-\eqref{eq:SU2ExpF0F1G0G1end}. 
In the case at hand, we have two types 
of codimension two singularities. First, there are the common zeros at codimension two in $B$ of 
$\Delta_2=\trplcurve=0$ with $\Delta_2$ being defined via $\Delta=4f^3+27g^2= \trplcurve^2 \Delta_2$. 
These contain, as we demonstrate below, the conventional matter representations of SU(2), \textit{i.e.}~the $\mathbf{2}$ representation.  Second, there are codimension two singularities
from the singularities of $\trplcurve=0$ at $\trpla=\trplb=0$, which support the triple symmetric
matter representations $\mathbf{4}$ of SU(2). 
As before in \S\ref{eq:MatterDoublePoints}, the determination of the irreducible components  of the codimension two 
loci in $\Delta_2=\trplcurve=0$ is the most challenging, and in general involves performing a  primary decomposition  
following the procedure outlined in \cite{CveticGrassiKleversPiragua} to which we refer for further 
details.

Before going into the details of this computation, we summarize the found matter content based on the analysis of codimension two singularities of the general 
non-Tate Weierstrass form  \eqref{eq:WSSU2}-\eqref{eq:SU2ExpF0F1G0G1end} along with the corresponding 6D matter content of F-theory in Table \ref{tab:SU2matter}. 
\begin{table}[ht!]
\begin{center}
\footnotesize
\renewcommand{\arraystretch}{1.2}
\begin{tabular}{|c|c|c|@{}c@{}|}
\hline
 SU(2)-rep & Multiplicity & Fiber & Locus  \\ \hline
$\mathbf{4}$ & $x_{\mathbf{4}}=\frac12[\trpla]\cdot [\trplb]$ & $I_0^{*ns}$  & $V(I_{\text{Sing}})=\{\trpla = \trplb = 0\}$ \rule{0cm}{.4cm}  \\[.1cm] \hline

$\mathbf{3}$ &  $x_{\mathbf{3}}=\frac12[\trplcurve]\cdot([\trplcurve]+K_B) + 1 - 6  x_{\bf 4}$ & $I_2$
 &  $V_{\text{SU}(2)}=\{\trplcurve=0\}$ \rule{0cm}{.4cm}
 \\[0.1cm] \hline
$\mathbf{2}$  &\rule{0cm}{0.2cm}
$  x_{\mathbf{2}}=[\trplcurve]\cdot(-8K_B-2[\trplcurve])+6[\trpla]\cdot [\trplb]+2 x_{\mathbf{4}} $ & $I_3$ & $ V(\mathfrak{p}_1)\cup V(I_{\text{Sing}})$\\[0.1cm] \hline
\end{tabular}
\caption{\label{tab:SU2matter} Matter spectrum of the elliptic
  fibration \eqref{eq:WSSU2}-\eqref{eq:SU2ExpF0F1G0G1end} with a
  singularity of type $I_2$ over a divisor $\trplcurve=0$ with
  ordinary triple point singularities. Shown are the SU(2)
  representations, the multiplicity of full hypermultiplets in a 6D
  theory, corresponding fiber type and locus in the base. We denote
  the variety described by the vanishing set of an ideal $I$ by
  $V(I)$.}
\end{center}
\end{table}

We begin with the discussion of the non-localized matter, i.e.~the adjoint matter, as well as the matter localized at the singularities of $\trplcurve=0$. The geometric genus $p_g$ of $\trplcurve=0$, which counts adjoints, is given by the arithmetic genus $g$ corrected by the contribution from the triple point singularities:
\beq
	p_g=g- 3 [\trpla]\cdot[\trplb]=1+\frac12 [\trplcurve]\cdot([\trplcurve]+K_B)-3 [\trpla]\cdot[\trplb] \,.
\eeq
Here we employ that every triple point contributes $3$ to the
arithmetic genus $g$, which we compute via adjunction in the second
equality.
We assume in analogy to the discussion of matter with double point singularities that we are working with a construction in which all the
triple points contribute 3-symmetric matter representations.  Identifying
$\tfrac12 [\trpla]\cdot[\trplb]$ as the multiplicity $x_{\mathbf{4}}$
of matter fields in the representation $\mathbf{4}$, we thus arrive at the matter multiplicities in the
first and third lines
in Table \ref{tab:SU2matter}. We note that each
triple point contributes only one half-hypermultiplet as the
representation $\mathbf{4}$ is pseudo-real.

Next we turn to the conventional matter localized at the intersection loci $\Delta_2=\trplcurve=0$.
We first gain some intuition about the possible matter loci by solving $\trplcurve=0$ locally away from its triple point singularities and inserting the solution into 
$\Delta_2=0$.  We immediately observe a factorization of $\Delta_2$ into two components, which indicates the existence of two irreducible varieties 
inside $\trplcurve=\Delta_2=0$.

Indeed, we can perform a rigorous primary decomposition of $\trplcurve=\Delta_2=0$ 
in the auxiliary ring 
$\mathbb{C}[\trpla,\trplb,\td,\tc,\tb,\ta]$ 
using Singular \cite{Singular} to obtain two prime ideals denoted by $\mathfrak{p}_1$ and $\mathfrak{p}_2$.\footnote{As before, these ideals may
factor further once specific elements of the ring $R$
are chosen to represent the variables in the auxiliary ring.}
 Explicitly, we find
\bea \label{eq:SU2p2}
	\mathfrak{p}_2=&&\left\{\trpla( \trpla\phibar\tb-\trpla^2\ha - 2\trplb\trpla\hb - \trplb^2\hc) + \phibar \trplb(\trplb\td + 
   2\trpla\tc) ,\right. \\ && \phantom{.}\left. \trplb( \trplb\phibar\tc+\trpla^2\ha + 2\trplb\trpla\hb + \trplb^2\hc) +\phibar \trpla ( 
   \trpla\ta+ 2\trplb\tb)\right \} / (I_{\text{sing}}\cup I')\,,
\nn
\eea
where we have to quotient by the ideals 
\beq \label{eq:SU2redundant}
	I_{\text{sing}}=\{\trpla,\trplb\}\,,\qquad I'=\{\ha \trpla^2 + 2 \hb \trpla \trplb + \hc \trplb^2,\phibar\}
\eeq	
in order to obtain a prime ideal. The prime ideal $\mathfrak{p}_1$ is too lengthy to be reproduced here; it is generated by several large polynomials
in the parameters in $f$ and $g$. It can be obtained by computing  the saturation ideal of the ideal $\trplcurve=\Delta_2=0$ w.r.t.~the ideal $\mathfrak{p}_2$:
\footnote{Due to the complexity of the involved algebra and the limited available computing power, we were only able to determine the ideal 
$\mathfrak{p}_1$ in the case where $\tb$ is a random rational number between $-1000$ and $1000$.}
\beq
	\mathfrak{p}_1=\{\Delta_2,\trplcurve\} / \mathfrak{p}_2\,.
\eeq

Next we analyze the singularity type of the elliptic fibration along the  varieties $V(\mathfrak{p}_1)$, $V(\mathfrak{p}_2)$ defined by the vanishing loci 
of $\mathfrak{p}_1$, $\mathfrak{p}_2$ inside the variety $\Delta_2=\trplcurve=0$. By investigation of the orders of vanishing of $(f,g,\Delta)$, we find that
\beq
		V(\mathfrak{p}_1):\quad (f,g,\Delta)\sim (0,0,3)\,, \qquad V(\mathfrak{p}_2):\quad (f,g,\Delta)\sim (1,2,3)\,,
\eeq
which indicates a singularity of type $I_3$ and $III$ respectively \cite{Kodaira,Tate}. This means that the variety $V(\mathfrak{p}_1)$
supports matter in the fundamental $\mathbf{2}$ of SU(2) \cite{Bershadsky-all}, whereas $V(\mathfrak{p}_2)$ is the locus of a degeneration of 
an $I_2$ singularity to $III$,. This does not correspond to the emergence of additional physical degrees of freedom due to the lack of new 
holomorphic curves to be wrapped by M2-branes. 

Finally, for the computation of the matter multiplicity $x_{\mathbf{2}}$ of doublets, we need to know the multiplicities of   $V(\mathfrak{p}_1)$, $V(\mathfrak{p}_2)$ inside $\Delta_2=\trplcurve=0$. Using the resultant technique discussed in \cite{CveticGrassiKleversPiragua}, we find the multiplicities to be $1$ and $2$, respectively, i.e.~we obtain the following relation in homology
\beq \label{}
	[\Delta_2]\cdot[\trplcurve]=[V(\mathfrak{p}_1)]+2[V(\mathfrak{p}_2)]\,.
\eeq
The homology class of the left hand side of this equation is readily computed using the explicit expression for $\trplcurve$ in \eqref{eq:Tgen} and $\Delta_2$ 
as it follows from \eqref{eq:SU2ExpF0F1G0G1}. We then compute $[V(\mathfrak{p}_2)]$ using its definition \eqref{eq:SU2p2} as being contained in 
a complete intersection among the additional components specified by the ideals in \eqref{eq:SU2redundant}. The homology classes of the latter are 
easily computed noting their definition as irreducible complete intersections. Their multiplicities inside the complete intersection in \eqref{eq:SU2p2} is 
computed using the resultant as $4$ and $1$, respectively. Thus, we obtain 
\bea \label{eq:SU2p2Homology}
	[V(\mathfrak{p}_2)]&=&(-2K_B+[\trpla])\cdot (-2K_B+[\trplb])-4[\trpla]\cdot[\trplb]-(-2K_B-[\trplcurve]+[\trpla]+[\trplb])\cdot (-2 K_B)\nn\\
	&=& -2K_B\cdot[\trplcurve]-3[\trpla]\cdot [\trplb]\,,
\eea
where we used the homology classes of all relevant sections given in Table \ref{tab:su2parameters}. The first term in the first equality is the homology class of the complete intersection in \eqref{eq:SU2p2} and the second and third terms are the homology classes of the varieties corresponding to the ideals in \eqref{eq:SU2redundant}.
Putting everything together, we obtain the homology class of $[V(\mathfrak{p}_1)]$ as
\beq
	[V(\mathfrak{p}_1)]=[\trplcurve]\cdot(-12K_B-2[\trplcurve])-2(-2K_B\cdot[\trplcurve]-3[\trpla]\cdot [\trplb])=[\trplcurve]\cdot(-8K_B-2[\trplcurve])+6[\trpla]\cdot [\trplb]\,.
\eeq
The first term in the first equality is the homology class of $\Delta_2=\trplcurve=0$ and the second term is \eqref{eq:SU2p2Homology}, which has to be 
subtracted with the correct multiplicity $2$. We also double check the result  for $[V(\mathfrak{p}_1)]$ by directly working with the lengthy ideal 
$\mathfrak{p}_1$. 

We thus obtain the contribution from $I_3$  loci inside $\Delta_2=\trplcurve=0$ to the multiplicity $x_{\mathbf{2}}$ of $\mathbf{2}$ matter fields as shown in 
the first term of the fourth line of Table \ref{tab:SU2matter}.
We note that there are additional matter fields in the $\mathbf{2}$  representation from the ordinary triple point singularities as noted in 
\cite{KleversTaylor}: group-theoretically, the $\mathbf{4}$ representation arises in the decomposition $\mathbf{2}\otimes\mathbf{2}\otimes\mathbf{2}=\mathbf{4}\oplus \mathbf{2}\oplus\mathbf{2}$ at each ordinary triple point $\trpla=\trplb=0$ of $\trplcurve=0$. This shows that each triple point also supports one full hypermultiplet in the representation $\mathbf{2}$, which leads to the second contribution in the last line of Table \ref{tab:SU2matter}.

We conclude by double-checking  the derived SU(2) matter spectrum by testing anomaly freedom of the 6D theory. Following the discussion of
\S\ref{sec:anomalies}, we identify $b=[\trplcurve]$ and $a=K_B$. We then see that anomaly cancellation follows immediately for the spectrum in Table \ref{tab:SU2matter} upon the identification $r=[\trpla]\cdot[\trplb]$,  $g=1+[\trplcurve]\cdot([\trplcurve]+K_B)$.

%---------------------------------------------
\section{Matter transitions}
\label{sec:mattertransitions}

In many situations, the 6D anomaly conditions specify a unique charged
matter content given a few parameters. But for the models considered
here, these same parameters are not enough to fully determine the
matter spectrum, as mentioned in \S\ref{sec:anomalies}. Even if
$a\cdot b$, $b\cdot b$, the gauge group, and the representations are
fixed, the 6D anomaly cancellation conditions still admit multiple
solutions for the matter multiplicities. Given a particular $\gsu(2)$
spectrum, one can find another consistent spectrum through the
exchange 
(\ref{eq:3-transition})
\begin{equation}
3\times \textbf{Adj} + 7\times \mathbf{1} \leftrightarrow \frac{1}{2}{\tiny \yng(3)}+7\times{\tiny \yng(1)}.
\end{equation}
The $\gsu(N)$ models with $N\geq 3$ meanwhile admit multiple spectra
related by the exchanges
(\ref{eq:n-equivalence})
\begin{equation}
\textbf{Adj}+\mathbf{1}\leftrightarrow {\tiny \yng(2)} + {\tiny \yng(1,1)}.
\end{equation}
The ``anomaly equivalent'' theories related by the exchanges have the same number of tensor and vector multiplets. From the perspective of the Weierstrass tunings, the homology classes of the various parameters are not fully fixed by the gauge curve homology class and $-K_B$. In turn, there may be multiple charged matter spectra for a given base and gauge curve homology class.

A natural next step is to ask how these anomaly equivalent models fit into the space of vacua. Specifically, is there a process to move between the anomaly equivalent models? 
As discussed in \cite{transitions}, there are a number of possible ways
to connect 6D models.  The simplest of these, which can be seen as
a purely field-theoretic phenomena, is the Higgs mechanism:  there can
be a theory with a larger gauge group whose Higgs branch contains both
of the models in question.  A more exotic type of transition, known 
since
\cite{Seiberg-Witten, Morrison-Vafa-II}, is the {\em tensionless string}\/
or {\em small instanton}\/ transition in which the number of tensor
multiplets in the theory changes.  The theory is superconformal at the
transition point.
A third possibility was raised in \cite{transitions}, where it was shown that anomaly equivalent models for a variety of gauge groups, such as $\gsu(6)$ and $\gsu(3)$, are connected by 
``matter transitions'' that occur within the Higgs branch of a superconformal
theory.
As in the case of a Higgs transition, 
the original gauge theory must be enhanced, but to a superconformal theory
rather than to a theory with larger gauge group.  In the context of
F-theory, this occurs
through a series of tunings and deformations of the Weierstrass model.
In field theoretic terms, such transitions relate two theories
with the same gauge group and the same number of tensor multiplets
via an intermediate superconformal theory which contains both of them
in its Higgs branch.  
The models derived here have similar matter transitions. 
In addition to being interesting phenomena in their own right, the matter transitions clarify the relationships between exotic higher-genus matter, curve singularities, and non-Tate tuning structures.

\begin{figure}[h]
\begin{center}
\includegraphics[scale=0.5]{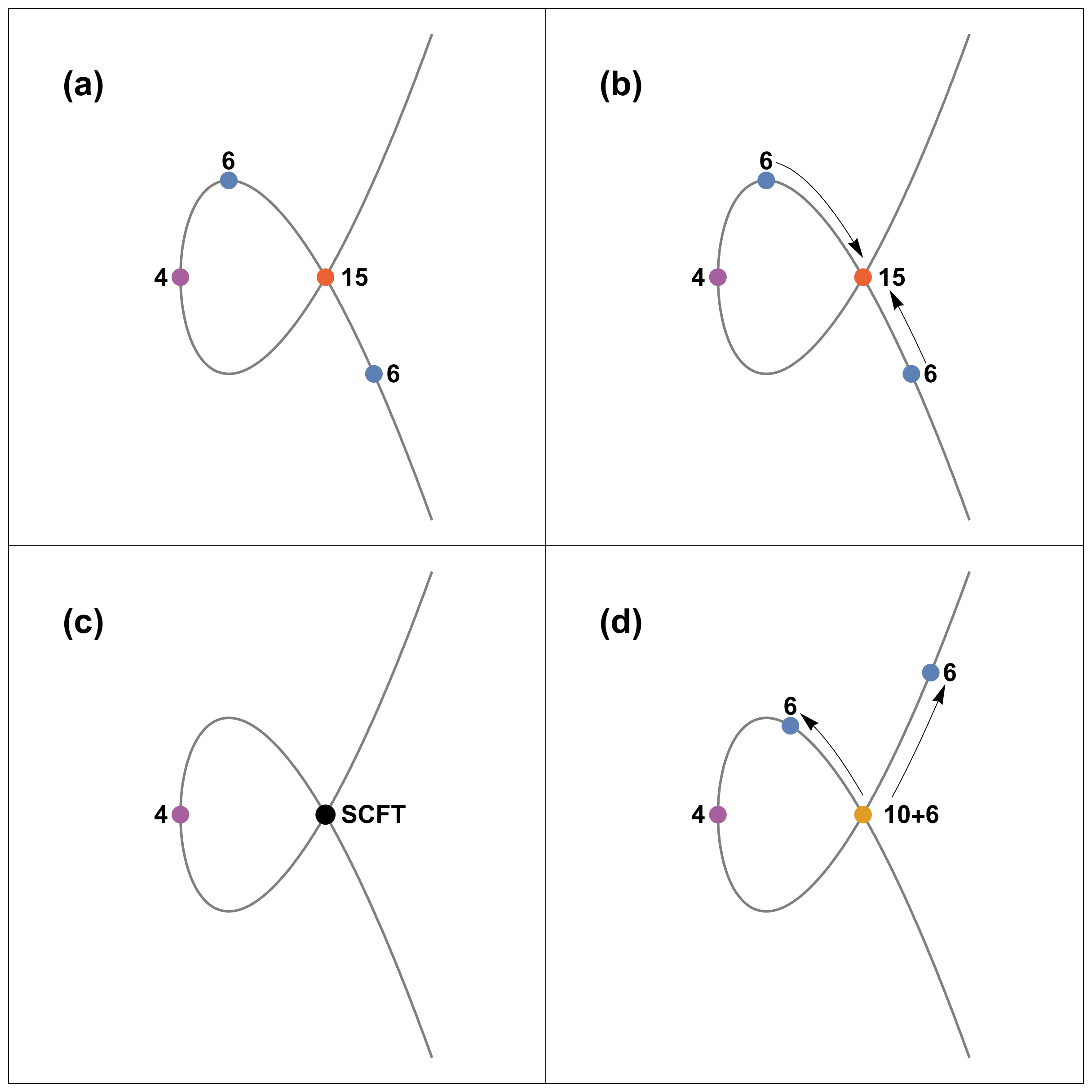}
\caption{Schematic illustration of the transition for $\gsu(4)$. The curve represents (a slice of) $\dblcurve$, the curve along which the $\gsu(4)$ singularity is tuned. Points represent codimension-two loci contributing charged matter; the labels give the $\gsu(4)$ representations associated with particular codimension-two loci. (a) Initially, there are several codimension-two loci supporting charged matter. We assume that a double point has already been tuned on the curve, with $\mathbf{15}$ matter localized at the point. (b) The Weierstrass model is tuned so that two $\mathbf{6}$ loci move to the double point. (c) When the $\mathbf{6}$ loci reach the double point, the singularity type at the double point worsens, giving an SCFT. (d) Deformations in the Weierstrass model remove the SCFT by pulling away two $\mathbf{6}$ loci. The charged matter at the double point is now  $\mathbf{10}+\mathbf{6}$, rather than the initial $\mathbf{15}$. }
\label{fig:transdiag}
\end{center}
\end{figure}

Before turning to the matter transitions in specific models, let us
first describe how these transitions work in general. The process is
illustrated in Figure \ref{fig:transdiag}  for $\gsu(4)$. We assume we
are working with a 6D F-theory model, although the transitions may
occur in 4D models as well. Initially, the F-theory model will have
matter located at several codimension two loci. The first step of the
transition is to tune the Weierstrass model so that a collection of
these loci are moved to a single point. $f$ and $g$ now vanish to at
least orders $4$ and $6$ at this point, signaling the appearance of a
superconformal sector
\cite{Morrison-Vafa-II,Heckman-Morrison-Vafa,DelZotto-Heckman-Tomasiello-Vafa}. Next,
the Weierstrass model is deformed, separating the superconformal locus
again into multiple codimension-two matter loci. The representations supported at these new points are different from those at the beginning of the transition.
The SCFT in this example is the ``single E-string theory''
\cite{Witten:1995gx,Ganor:1996mu, Seiberg-Witten}
corresponding to 
a $-1$ curve in F-theory.
It has a Higgs branch of dimension 29
with a variety of gauge groups and matter representations realized at
different points of that Higgs branch.  

Overall, the setup can be described in terms of the Higgs branch
of a fixed SCFT, within which  one finds
several distinct loci with a previously selected gauge group, although the
matter spectra may differ among the loci.  (There are two loci
for the $\gsu(4)$ example illustrated in Figure \ref{fig:transdiag} below.)
The initial tuning moves us from the first locus to the SCFT. One could then, in principle, move onto the tensor branch through a blow-up on the base. Instead, we follow Weierstrass deformations that move us to a new locus within the Higgs branch with 
the same gauge group but  a different matter spectrum. In the simplest cases, the SCFT at the transition point is an E-string theory consisting of a single $-1$ curve on the tensor branch. The SCFTs for some of the transitions considered here may have extra gauge groups tuned on the $-1$ curve. However, the simplest transitions do not have additional tensor branch gauge symmetries, suggesting that these gauge groups are not a general requirement for transitions. 

Finally, we note a few general observations about the transitions. First, progress through the transition can be described using a single parameter. This may not be immediately obvious from the general discussion or from the detailed analysis of models below. But the transition can be thought of as moving through a single parameter family of models. If desired, one can even write the deformations and tunings explicitly in terms of a single parameter, as done in \cite{transitions}. Second, the description of the transition process above did not mention introducing or deforming singularities along the gauge curve. This omission is a reflection of an important point regarding singular matter: gauge curve singularities do not automatically contribute exotic matter. In particular, introducing a curve singularity only localizes adjoints at the singular point and does not change the matter content. To make the singular points support exotic matter, the theory must additionally undergo a more dramatic change, such as obtaining a superconformal sector at the transition point. In the examples considered here, the tunings and deformations of the transition that actually change the matter content do not affect the number of singular points along the gauge curve. The elliptic curve singularity type at a double point or triple point may change during the course of the transition. However, the gauge curve singularity is present through the core part of the transition.

%--------------------------------

\subsection{$\gsu(3)$ Model with Symmetric Matter}
In order to make the discussion more concrete, we first focus on the $\gsu(3)$ model with symmetric matter. Recall  that if the $\gsu(3)$ singularity is tuned on the curve 
\begin{equation}
\dblcurve = \etaa \dbla^2 + 2 \etab \dbla \dblb + \etac \dblb^2=0,
\label{eq:h-curve}
\end{equation}
the $\gsu(3)$ Weierstrass model is described by
\begin{align}
f&= -\frac{1}{48}\phidbl^2 + f_1 \dblcurve + f_2 \dblcurve^2 & g&= \frac{1}{864}\phidbl^3 -\frac{1}{12}\phidbl f_1 \dblcurve + g_2 \dblcurve^2 + g_3 \dblcurve^3,
\end{align}
with
\begin{align}
\phidbl =& \left(\nua \dbla + \nub \dblb\right)^2 - 2 \nua \nubar \left(\etab \dbla + \etac \dblb\right) \notag\\
& + 2 \nub \nubar\left(\etaa\dbla+\etab \dblb\right) + \nubar^2 \left(\etab^2-\etaa\etac\right),\\
f_1 =& \left(\nua\dbla+\nub\dblb\right)\left(\psia\dbla+\psib\dblb\right)-\left(\psia\nubar + \nua\psibar\right)\left(\etab\dbla+\etac\dblb\right) \notag\\
&+ \left(\psib\nubar + \nub\psibar\right)\left(\etaa \dbla + \etab \dblb\right) + \nubar\psibar\left(\etab^2-\etaa\etac\right),\\
g_2 =& \left(\psia\dbla+\psib\dblb\right)^2 - 2 \psia\psibar\left(\etab\dbla+\etac\dblb\right)\notag\\ 
&+ 2\psib\psibar\left(\etab\dblb+\etaa\dbla\right) +
\psibar^2\left(\etab^2-\etaa\etac\right) -\frac{1}{12}\phidbl
f_2.\label{eq:3-g2}
\end{align}
This model admits transitions that cause a net exchange of matter of
\begin{equation}
\textbf{Adj}+\mathbf{1}\leftrightarrow {\tiny \yng(2)} + {\tiny \yng(1)}.
\end{equation}
For this discussion, we focus on the transitions that exchange an adjoint and a singlet for ${\tiny \yng(2)} + {\tiny \yng(1)}$. These transitions involve three basic steps described in more detail below:
\begin{enumerate}
\item Introduce a new double point along the gauge curve $\dblcurve=0$, localizing adjoint matter at the double point. As mentioned previously, this step does not change the matter content of the theory and thus should not be considered as a core part of the transition. 
\item Further tune the Weierstrass model to obtain an SCFT at the double point.
\item Deform the Weierstrass model in a different way to remove the SCFT, leading to a model with a different matter spectrum. 
\end{enumerate}
The transitions that cause the opposite exchange of matter (${\tiny \yng(2)} + {\tiny \yng(1)} \rightarrow \textbf{Adj}+\mathbf{1}$) can be obtained by inverting the above three steps. 

\paragraph{Step 1: Introduce a new double point.}

Symmetric matter can occur only at double point singularities along
the gauge curve. To convert adjoint matter into symmetric matter, we
therefore must
first introduce a new double point singularity along the curve. This can be done through the tunings
\begin{align}
\etaa &\rightarrow \etaa^\prime & \etab &\rightarrow a\etab^\prime & \etac &\rightarrow a^2\etac^\prime. \label{eq:transtuning1}
\end{align}
$\dblcurve$, now given by
\begin{equation}
  \dblcurve = \etaa^\prime \dbla^2 + \etab^\prime a\dbla\dblb +
  \etac^\prime a^2\dbla^2, \label{eq:transtuning1a}   
\end{equation}
has double point singularities at $\dbla = a = 0$. The tuning
introduces a total of $[a]\cdot[\dbla]$ double points; in the minimal
case, $[a]\cdot[\dbla]=1$, and there is only one new double
point. Note that this tuning does not change the matter spectrum. A
double point singularity along an $\gsu(3)$ curve can correspond to
either $\mathbf{8}+\mathbf{1}$ or $\mathbf{6}+\mathbf{3}$ matter
\cite{mt-singularities}. The $a=\dbla=0$ locus currently supports
localized adjoints, as these double points can be deformed away. Thus,
while the tunings in \eqref{eq:transtuning1} localize adjoints that
previously could propagate throughout the gauge curve, there is no
change in the matter spectrum.  In other words, one cannot introduce
symmetric matter simply by tuning more double points. Changing the
matter spectrum requires a more dramatic change, which here entails
passing through a superconformal point.

Note that in the description here we have assumed the initial curve
$h$ has the form (\ref{eq:h-curve}) and already contains some double
points.  More generally, this need not be the case.  An arbitrary
smooth curve could be put in this form where $\eta_b = 1$ so there are
no initial double point singularities; in this case, for example, we
could tune a single double point singularity from a smooth curve with
none by starting with $[\eta_a] = 1$ and taking $[a] = 1$.

\paragraph{Step 2: Move to the SCFT.}

Currently, symmetric matter (if present) resides at the
$\dbla=\dblb=0$ locus. To convert the localized adjoints at
$\dbla=a=0$ into symmetric matter, one would like to perform a
deformation such as $a\dblb \rightarrow \dblb^\prime$. At this point
in the transition, such a deformation is not possible: there are
factors of $\dblb$ in the Weierstrass model
(\ref{eq:h-curve}-\ref{eq:3-g2}) without corresponding factors of $a$. We therefore
need to perform the additional tunings
\begin{align}
\nub &\rightarrow a \nub^\prime & \psib &\rightarrow  a\psib^\prime. \label{eq:transtuning2}
\end{align}
These tunings make $f$ and $g$ vanish to orders 4 and 6 at $a=\dbla=0$, signaling the presence of an SCFT. In order to produce this superconformal matter, the tuning must have pushed other matter loci to the $a=\dbla=0$ double points. Prior to the tunings in Equations \eqref{eq:transtuning1} and \eqref{eq:transtuning2}, the discriminant takes the form
\begin{equation}
\Delta = \Delta_3 \dblcurve^3,
\end{equation}
where $\Delta_3$ is equivalent to $\PhizeroL^3 \tilde{\Delta}_3$ in
the normalized intrinsic ring. The discriminant locus contributes
$[\tilde{\Delta}_3]\cdot[\dblcurve]$ fundamentals. After the tunings,
$\tilde{\Delta}_3$ vanishes to order 3 on the $a=\dbla=0$ locus, while
$\dblcurve$ vanishes to order 2. Since the $a=\dbla=0$ locus now
supports an SCFT, it should be subtracted out when counting the number
of fundamentals after the transition. For each $a=\dbla=0$ point,
$3\times 2 = 6$ fundamentals therefore disappear as a result of the
tunings in Equations \eqref{eq:transtuning1} and
\eqref{eq:transtuning2}. A careful consideration of the available
degrees of freedom and redundancies in the Weierstrass model indicates
that in the tuning
\eqref{eq:transtuning2}  two singlets are fixed for each
$a=\dbla = 0$ point. This missing matter corresponds exactly to the
loci that were pushed to $a=\dbla=0$ in order to form the SCFT. The
effect of this second step can therefore be summarized as
\begin{equation}
\mathbf{8} + 6\times \mathbf{3} + 3\times\mathbf{1} \rightarrow \textbf{SCFT}.\label{eq:su3step2}
\end{equation}

The SCFT at the transition point is an E-string theory. On the tensor branch, where the $(4,6)$ singularity is resolved by a blowup on the base, there is a single $-1$ curve with no additional gauge groups. The gravitational anomaly condition therefore suggests that exactly 29 hypermultiplets should be removed at the transition point in order to produce the SCFT. Indeed, exactly 29 hypermultiplets participate in this step for each $a=\dbla=0$ point, as seen in the left-hand side of \eqref{eq:su3step2}.  

\paragraph{Step 3: Deform the SCFT.} There are now enough factors of
$a$ in the Weierstrass model to absorb $a$ into $\dblb$. However,
there will be factors of $a$ remaining after such an association. The $\gsu(3)$ model at the end of the transition should have the same basic structure as before, suggesting that a tuning with remaining factors of $a$ is a specialized model that can be further deformed. To remove the extra factors of $a$, we must perform the deformations
\begin{align}
 a\nubar &\rightarrow \nubar^\prime & a\psibar &\rightarrow
 \psibar^\prime \label{eq:transtuning3}
\end{align}
After this deformation, $\nubar^\prime$ and $\psibar^\prime$ are not proportional to $a$. There is no longer an SCFT, as there is no codimension two locus where $f$ and $g$ vanish to orders 4 and 6. The six fundamentals and two singlets that had disappeared are now restored, and the $\dbla=a=0$ double points introduced by Equation \eqref{eq:transtuning1} now contribute $\mathbf{6}+\mathbf{3}$ matter. Every factor of $\dblb$ is accompanied by exactly one factor of $a$, so we can now perform the redefinition
\begin{equation}
a\dblb \rightarrow \dblb^\prime
\end{equation}
and remove all factors of $a$. The transition is now complete, with a total change of matter of the form
\begin{equation}
\mathbf{8} + 6\times\mathbf{3}+3\times\mathbf{1} \rightarrow \textbf{Superconformal Matter} \rightarrow \mathbf{6} + 7\times \mathbf{3} + 2\times \mathbf{1}. \label{eq:su3transmatter}
\end{equation}
The net effect of the transition is therefore to exchange $\mathbf{8}+\mathbf{1}$ for $\mathbf{6}+\mathbf{3}$, as expected from the anomaly cancellation conditions. We can also exchange $\mathbf{6}+\mathbf{3}$ for $\mathbf{8}+\mathbf{1}$ by inverting the steps in the transition.

The Weierstrass model in fact allows for a second type of transition, although this transition will have the same physical effect. This second transition involves transferring factors into $\dbla$ rather than into $\dblb$. To convert adjoints into symmetrics, we initially perform the tunings
\begin{align}
\etac &\rightarrow \etac^\prime & \etab &\rightarrow a \etab^\prime & \etaa &\rightarrow a^2 \etaa^\prime, \label{eq:trans2tuning1}
\end{align}
thereby introducing new double points at the $a=\dblb=0$ loci. The additional tunings
\begin{align}
\nua &\rightarrow a \nua^\prime & \psia &\rightarrow a \psia^\prime\label{eq:trans2tuning1a}
\end{align}
take us to a superconformal point. The transition is completed with the deformations
\begin{align}
a\dbla &\rightarrow \dbla^\prime & a\nubar &\rightarrow \nubar^\prime & a\psibar &\rightarrow \psibar^\prime.\label{eq:trans2tuning2}
\end{align}
In total, the transition exchanges $[a]\cdot[\dblb]$ adjoints for symmetrics through a process identical to that in Equation \eqref{eq:su3transmatter}. Of course, one can perform the transition in the opposite direction by reverting the steps. 

As mentioned in \S\ref{sec:su3doubletuning}, the previously derived
$\gsu(3)$ models with symmetric matter are specializations of the ones
derived here. In particular, these previous models restrict homology
classes by setting certain parameters to constants. These models will
not exhibit all of the possible transitions laid out here, but they
may still allow some subset of the transitions.  For example, the
model in \cite{transitions} essentially forces $\etaa$ to be a
constant, so transitions where $\dbla$ changes are not possible. The
transitions where $\dblb$ changes, however, are still possible. For
the $\gsu(3)$ model in \cite{ckpt}, $\nubar$ is set to one, while both
of the transitions considered here change the homology class of
$\nubar$. These transitions are therefore not possible in this
model. This model has other transitions that connect the different
matter spectra, but the theory undergoes more extreme changes during
the transition. For instance,
when the gauge curve factorizes at the
transition point. As a result, the transition point theory has a new
$\gsu(3)$ gauge group in addition to the SCFT.

Finally, let us examine how non-Tate structure appears as part of the
transition. Consider a situation where $\dblb$ is initially set to
1. There are no double points, so the Weierstrass model has Tate
structure. For instance, $f_1$, which has non-Tate structure in
general, can be written as
\begin{multline}
f_1 = \left(\psib+\psia\dbla+\left(\etab+\etaa\dbla\right)\psibar\right)\left(\nub+\nua\dbla+\left(\etab+\etaa\dbla\right)\nubar\right)\\ 
-\left(\etac+2\etab\dbla+\etaa\dbla^2\right)\left(\psia\nubar+\nua\psibar+\psibar \nubar \etaa\right) \label{eq:F1tateform}
\end{multline}
when $\dblb$ is set to $1$. Since the second term on the right-hand side is proportional to $\dblcurve$, we could move it to $f_2$, leaving the Tate form expression where $f_1$ factors into two components. Now imagine performing the transition using the steps in Equations \eqref{eq:transtuning1}, \eqref{eq:transtuning2}, and \eqref{eq:transtuning3}. In the $\etaa\dbla\psibar$ and $\etaa\dbla\nubar$ parts of the first term, there would not be enough factors of $a$ to absorb into $\nubar$ and $\psibar$, and the term would seem to develop $a^{-1}$ factors. The Tate-form expression for $f_1$ is thus no longer valid after the transition. However, the second term in \eqref{eq:F1tateform} would also have parts with $a^{-1}$ factors. In fact, all the $a^{-1}$ parts cancel between the first and second terms of \eqref{eq:F1tateform}. As long as we keep the second term as part of $f_1$, we can maintain a valid expression for $f_1$. $f_1$ is therefore forced to have non-Tate structure after the transition. The terms proportional to $\dblcurve$ in $f_1$ when $\dblb=1$, which were ``optional'' before the transition, are necessary after the transition. In this way, the transition generates the expected non-Tate structure from a Tate-form model.

\subsection{$\gsu(N)$ Model with Symmetric Matter for $N\geq 4$}
In general, $\gsu(N)$ models with symmetric matter should admit transitions that cause the net exchange
\begin{equation}
\textbf{Adj} + \mathbf{1} \leftrightarrow {\tiny \yng(2)} + {\tiny \yng(1,1)}.
\end{equation}
While the net exchange during the transition should be the same regardless of whether $N$ is even or odd, the Weierstrass models for the two cases are somewhat different. In turn, the details of the transitions will be slightly different for even and odd $N$; in particular, the transition for $\gsu(2k-1)$ is a Higgsed version of the $\gsu(2k)$ transition. We first focus on the case where $N$ is even before turning to the odd $N$ case.

The Weierstrass model for $\gsu(2k)$ is described by
\begin{align}
f &= - \frac{1}{3}\upsilon^2 + \mathcal{O}(\dblcurve^k) & g &= -\frac{1}{27}\upsilon^3 - \frac{1}{3}\upsilon f + \mathcal{O}(\dblcurve^{2k})
\end{align}
with
\begin{align}
\upsilon =& \frac{1}{4}\phidbl + \phi_1 \dblcurve + \ldots \phi_{k-1}\dblcurve^{k-1} \\
\dblcurve =& \etaa\dbla^2 + 2 \etab \dbla \dblb + \etac \dblb^2\\
\phidbl =& \left(\nua\dbla + \nub\dblb\right)^2 - 2 \nua \nubar\left(\etab\dbla+\etac\dblb\right)\notag\\
&+ 2 \nub\nubar\left(\etaa\dbla+\etab\dblb\right) + \nubar^2\left(\etab^2-\etaa\etac\right).
\end{align}
As mentioned in \S\ref{sec:symmonodromy} and \S\ref{sec:highsun-double}, the only non-UFD structure is contained within $\phidbl$, and all discriminant cancellations are exact. The discriminant takes the form
\begin{equation}
\Delta = \phidbl^2 \Delta_{2k}\dblcurve^{2k},
\end{equation}
with the $\Delta_{2k}=\dblcurve=0$ loci contributing fundamental matter. 

Let us consider the analogue of the $\gsu(3)$ transition, where we convert adjoint matter into symmetric matter by transferring a factor $a$ into $\dblb$. As before, we introduce a new double point by tuning
\begin{align}
\etaa & \rightarrow \etaa^\prime & \etab &\rightarrow a \etab^\prime & \etac &\rightarrow a^2 \etac^{\prime}.
\end{align}
Again, this tuning does not actually change the matter content of the theory; instead, it localizes an adjoint at each $a=\dbla=0$ double point and fixes one singlet per double point. We then perform the tuning
\begin{align}
\nub &\rightarrow a \nub^\prime
\end{align}
to take us to the SCFT. After this tuning, $\phidbl$ vanishes to order 2 at the $a=\dbla =0$ loci. Combined with the fact that $\dblcurve$ vanishes to order 2 wherever $a=\dbla=0$, this indicates that two ${\tiny \yng(1,1)}$ loci have been moved to each $a=\dbla=0$ point. Additionally, the tuning removes one singlet for each $a=\dbla=0$ point. Note that $\Delta_{2k}$ does not vanish at loci where $a=\dbla=0$, and no fundamental loci are moved to the $a=\dbla=0$ points. To move to the branch with symmetric matter, we perform the deformations
\begin{align}
a\dblb &\rightarrow \dblb^\prime & a \nubar &\rightarrow \nubar^\prime.
\end{align}
These deformations move away two antisymmetric loci and restore a singlet for each double point, leaving ${\tiny \yng(2)}+{\tiny \yng(1,1)}$ at the double points. 

The full transition can therefore be summarized as
\begin{equation}
\textbf{Adj} + 2\times{\tiny \yng(1,1)} + 2\times\mathbf{1}\rightarrow \textbf{Superconformal Matter} \rightarrow {\tiny \yng(2)} + 3\times{\tiny \yng(1,1)} + \mathbf{1}. \label{eq:sunevenfulltransition}
\end{equation}
As expected, the net change in matter is
\begin{equation}
\textbf{Adj}+\mathbf{1}\rightarrow {\tiny \yng(2)} + {\tiny \yng(1,1)}.
\end{equation}
A total of $8k^2 -2k + 1$ hypermultiplets participate in the
transition per double point. For $\gsu(4)$, this number is 29. Blowing
up the singular point gives a single $-1$ curve with no additional
gauge groups tuned, indicating the appearance of a new tensor
multiplet. Therefore, the gravitational anomaly constraint suggests
that 29 hypermultiplets should disappear at the transition point, as
observed. For general $k$, the number of hypermultiplets will not be a
multiple of 29. However, one should still be able to move to the
tensor branch at the superconformal point and introduce new tensor
multiplets. If there is no change in the number of vector multiplets,
29 hypermultiplets should be lost for each new tensor multiplet. Since
the number of hypermultiplets participating in the transition is not a 
multiple of 29, the tensor branch of the SCFT must also include
additional gauge symmetry. In fact, performing the blow up explicitly
shows that, on the tensor branch, an $\gsp(2k-4)$ gauge group is tuned
on the $-1$ curve.  There are also $4k$ fundamentals charged under
this $\gsp(2k-4)$ group. Accounting for this new gauge group, the
expected change in the number of hypermultiplets is $8k^2-2k +1$, in
agreement with the number of hypermultiplets participating in
\eqref{eq:sunevenfulltransition}.

For $\gsu(2k-1)$, the Weierstrass model is described by
\begin{align}
f &= -\frac{1}{3}\upsilon^2 + \digamma_{k-1}\dblcurve^{k-1} + \mathcal{O}(\dblcurve^k) & g&= -\frac{1}{27}\upsilon^3 -\frac{1}{3} f \phidbl + \gamma_{2k-2}\dblcurve^{2k-2}+\mathcal{O}(\dblcurve^{2k-1}), 
\end{align}
with
\begin{align}
\upsilon =& \frac{1}{4}\phidbl + \phi_1 \dblcurve + \ldots \phi_{k-2}\dblcurve^{k-2}, \\
\dblcurve =& \etaa\dbla^2 + 2 \etab \dbla \dblb + \etac \dblb^2,\\
\phidbl =& \left(\nua\dbla + \nub\dblb\right)^2 - 2 \nua \nubar\left(\etab\dbla+\etac\dblb\right)\notag\\
&+ 2 \nub\nubar\left(\etaa\dbla+\etab\dblb\right) + \nubar^2\left(\etab^2-\etaa\etac\right),\\
\digamma_{k-1} =& \left(\nua \dbla + \nub \dblb\right)\left(\psia\dbla+\psib\dblb\right)-\left(\nua\psibar + \psia \nubar\right)\left(\etab\dbla+\etac\dblb\right)\notag\\
&+\left(\nub\psibar+\psib\nubar\right)\left(\etaa\dbla+\etab\dblb\right) + \nubar\psibar\left(\etab^2-\etaa\etac\right),\\
\gamma_{2k-2}=& \left(\psia\dbla+\psib\dblb\right) - 2\psia\psibar\left(\etab\dbla+\etac\dblb\right)\notag\\
&+ 2\psib\psibar\left(\etaa\dbla+\etab\dblb\right)+\psibar^2\left(\etab^2-\etaa\etac\right).
\end{align}
The transition follows three steps similar to those of the $\gsu(3)$ transition. First, we introduce new double points at $a=\dbla=0$ through the tunings
\begin{align}
\etaa &\rightarrow \etaa^\prime & \etab &\rightarrow a \etab^\prime & \etac &\rightarrow a^2\etac^\prime.
\end{align}
At this point in the transition, each new double point corresponds to an adjoint and a singlet of matter. We then perform the further tunings
\begin{align}
\nub &\rightarrow a \nub^\prime & \psib &\rightarrow a \psib^\prime,
\end{align}
taking us to the SCFT. The tunings move two ${\tiny \yng(1,1)}$ loci and four fundamental loci to each $a=\dbla=0$ double point while fixing two singlets per double point. Finally, the SCFT is removed by the Weierstrass deformations
\begin{align}
a\dblb &\rightarrow \dblb^\prime & a\nubar &\rightarrow \nubar^\prime & a\psibar&\rightarrow \psibar^\prime.
\end{align}
Two ${\tiny \yng(1,1)}$ loci and four fundamental loci are pushed from each double point, and two singlets are reintroduced per double point. This leaves ${\tiny \yng(2)}+{\tiny\yng(1,1)}$ at each $a=\dbla=0$ double point. 

The full $\gsu(2k-1)$ transition can be summarized as
\begin{equation}
\textbf{Adj}+2\times{\tiny \yng(1,1)}+4\times{\tiny \yng(1)}+3\times\mathbf{1}\rightarrow \textbf{SCFT} \rightarrow {\tiny \yng(2)}+3\times{\tiny \yng(1,1)} + 4\times{\tiny\yng(1)} + 2\times\mathbf{1}. \label{eq:sunoddfulltransition}
\end{equation}
The net change in matter is therefore
\begin{equation}
\textbf{Adj}+\mathbf{1}\rightarrow {\tiny \yng(2)} + {\tiny \yng(1,1)},
\end{equation}
as expected. Note that this transition is essentially a Higgsed version of the $\gsu(2k)$ transition in \eqref{eq:sunevenfulltransition}. Specifically, one can break the $\gsu(2k)$ representations in \eqref{eq:sunevenfulltransition} to $\gsu(2k-1)$ representations to obtain \eqref{eq:sunoddfulltransition}. There are still $8k^2-2k+1$ hypermultiplets that participate in the transition. Additionally, the $\gsu(2k-1)$ transition point SCFT is the same as the $\gsu(2k)$ SCFT: the tensor branch consists of a single $-1$ curve with a tuned $\gsp(2k-4)$ gauge group.

\subsection{$\gsu(2)$ with Triple-Index Symmetric Matter} 
From the anomaly cancellation conditions, there are anomaly equivalent $\gsu(2)$ matter spectra related by the exchanges
\begin{equation}
3\times\textbf{Adj} + 7\times \mathbf{1} \leftrightarrow \frac{1}{2}{\tiny \yng(3)} + 7\times{\tiny \yng(1)}.
\end{equation}
The corresponding F-theory tunings admit transitions between models with these different spectra. Transitions in these $\gsu(2)$ models follow a similar set of steps as the $\gsu(N)$ models with symmetric matter. For concreteness, suppose the $\gsu(2)$ singularity occurs on a curve
\begin{equation}
\trplcurve= \ta\trpla^3 + 3 \tb\trpla^2 \trplb + 3 \tc \trpla \trplb^2 + \td \trplb^3 = 0.
\end{equation}
The Weierstrass model for this case is given in Appendix \ref{app:su2summary}.
To exchange adjoint matter for ${\tiny \yng(3)}$ matter, one
\begin{enumerate}
\item introduces a new triple point in the curve, 
\item performs additional tunings to reach an SCFT, and
\item performs deformations to move away from the SCFT.
\end{enumerate}
As before, simply introducing a triple point does not change the matter content. Before the theory passes through the SCFT, the triple point supports three adjoint hypermultiplets, not ${\tiny \yng(3)}$ matter. To exchange ${\tiny \yng(3)}$ matter for adjoint matter, the steps should be performed in reverse.

To see how these transitions work at the level of the $\gsu(2)$ Weierstrass model, consider exchanging adjoint matter for ${\tiny \yng(3)}$ matter. The first step is to introduce new triple points through the tunings
\begin{align}
\ta &\rightarrow \ta^{\prime} & \tb &\rightarrow a \tb^\prime & \tc &\rightarrow a^2 \tc^\prime & \td &\rightarrow a^3 \td^\prime
\end{align}
$\trplcurve$, now given by
\begin{equation}
\trplcurve= \ta^\prime\trpla^3 + 3 \tb^\prime\trpla^2 a \trplb + 3 \tc^\prime \trpla a^2 \trplb^2 + \td^\prime a^3 \trplb^3 ,
\end{equation}
has new triple point singularities at $a=\trpla=0$. Each of the
$[a]\cdot[\trpla]$ newly introduced triple points supports 3 localized
adjoints and 4 singlets.

To convert the adjoints into symmetric matter, we next need to perform tunings to reach the superconformal point. For the $\gsu(2)$ Weierstrass model, the specific tunings are
\begin{align}
\ha &\rightarrow \ha^\prime & \hb &\rightarrow a \hb^\prime & \hc &\rightarrow a^2 \hc^\prime & \lambdaa &\rightarrow \lambdaa^\prime & \lambdab &\rightarrow a \lambdab^\prime
\end{align}
With these tunings, $f$ and $g$ now vanish to orders $(4,6)$ at $a=\trpla=0$, indicating the appearance of a superconformal sector. The additional tunings move six fundamental loci to each $a=\trpla=0$ triple point while fixing four singlets per triple point. Combined with the three adjoints and four singlets from the first step, there are now a total of 29 hypermultiplets associated with each new triple point. Again, this number matches the number of hypermultiplets expected from the gravitational anomaly constraint. The SCFT at transition point is the same as that for $\gsu(3)$ and $\gsu(4)$: the tensor branch consists of a single $-1$ curve with no additional gauge symmetry. 

We now need to deform the Weierstrass model to get to the model with new ${\tiny \yng(3)}$ multiplets. The specific deformations needed are
\begin{align}
a \trplb &\rightarrow \trplb^\prime & a \phibar &\rightarrow \phibar^\prime
\end{align}
The introduced triple points at $a=\trpla$ are now part of the full triple point locus described by $\trpla=\trplb^\prime = 0$. Each of these triple points supports a half-multiplet of ${\tiny \yng(3)}$ matter and two fundamental hypermultiplets. The deformations additionally move away eleven fundamental loci and introduce a new singlet for each $a=\trpla=0$ triple point. Thus, the total matter change in the transition can be written as
\begin{equation}
3\times\textbf{Adj} + 8\times \mathbf{1} + 6\times {\tiny \yng(1)} \rightarrow \textbf{Superconformal Matter} \rightarrow \frac{1}{2}{\tiny \yng(3)} + 13\times{\tiny \yng(1)} + \mathbf{1}.
\end{equation}
The net change of matter in the transition is therefore
\begin{equation}
3\times\textbf{Adj} +7 \times \mathbf{1} \rightarrow \frac{1}{2}{\tiny \yng(3)} + 7\times{\tiny \yng(1)},\end{equation}
in agreement with the expectations from anomaly cancellation.

\section{Allowed and disallowed matter combinations}
\label{sec:allowed}

We have shown that the two- and three-index representations of SU($N$)
and SU(2) can be realized in F-theory when the gauge group lives on a
divisor with a double or triple point singularity.  Even for these
representations that can be realized locally, there is a more general
question, which pertains to the combinations of matter
representations that can appear together in a given model.  We now
focus on this question, first for symmetric matter representations of
SU($N$), and then for triple-symmetric representations of SU(2).  We
consider in particular the case of models with $T = 0$, corresponding
to F-theory models on $\P^2$, where the existence of a global model in
F-theory depends
at least in part on the geometric question of whether a curve exists
with a given combination of singularity types.

\subsection{SU($N$) symmetric matter}
\label{sec:double-p2}
We now investigate the
question of which combinations of adjoint and two-index symmetric
SU($N$) matter fields can be realized geometrically in F-theory.
We investigate this question at two levels.  First,
whether the corresponding
combination of singularity types is geometrically allowed at the level
of curves on $\P^2$.  Second,
whether explicit Weierstrass models can be identified for the
geometrically allowed configurations  of singularity types.

\subsubsection{Geometry of double points on curves}

For an arbitrary number ($n \leq g$) of symmetric representations of
SU($N$) that is allowed in the low-energy theory
to be possible in F-theory, it is necessary that a generic
curve of arithmetic genus $g$ can be tuned by fixing $n$ moduli
so that the
resulting curve has $n$  simple double point singularities and geometric
genus $p_g=g-n$.
For the simplest case, $n =g = 1, T = 0$, this
corresponds to the existence of a plane cubic with a single double
point.  This can easily be arranged, for example through the cubic
\begin{equation}
u^2 -v^2  + u^3 = 0 \,.
% \label{eq:}
\end{equation}
The next case is a plane quartic $(g = 3, T = 0)$ with up to three double
points.  Any of the possibilities $n \leq g$ can be realized simply by
choosing a quartic where all terms constant and linear in $n$ pairs of
homogeneous coordinates are set to vanish; {\it i.e.} for $n = 3$ we
have the general quartic
\begin{equation}
 au^2 + buv+ cv^2 + dvu^2 + ev^2 u + fv^2 u^2 \,.
% \label{eq:}
\end{equation}

More generally, it is
possible for a curve
of degree $d$ on $\P^2$ to realize any number $n \leq (d-1)(d-2)/2$ of
double points \cite{harris-severi}.\footnote{Note that the arithmetic genus of a curve of
degree $d$ is $(d-1)(d-2)/2$.}
A simple argument for this conclusion
in the case $n = (d-1)(d-2)/2$ (i.e., $p_g=0$)
proceeds as follows:
consider the map from $\P^1$ to
$\P^d$ given by
\begin{equation}
[u:v] \rightarrow[v^d, uv^{d-1}, \ldots, u^{d -1} v, u^d] \,.
% \label{eq:}
\end{equation}
We then map $\P^d\rightarrow \P^2$ by taking the projection onto a
generic subplane, {\it e.g.}
\begin{equation}
[u_0:u_1:u_2:\cdots:u_{d -1}:u_d] \rightarrow
[l_0:l_1:l_2] \,,
l_i = a_{i0} u_0 + a_{i1} u_1 + a_{i2} u_2 \,.
% \label{eq:}
\end{equation}
This gives a well-defined map from $\P^1 \rightarrow \P^2$, since for
generic $l_i$, no
point in $\P^1$ maps to a point in $\P^d$ with $l_0 = l_1 =l_2 = 0$.
The image curve is algebraic
since it is explicitly parameterized by algebraic
functions.  And
the curve has degree $d$, since the intersection with the line $l_0 = 0$ 
gives a generic $d$th degree polynomial in $u, v$ with $d$ roots.
Thus, 
the image in
$\P^2$ is a genus 0 curve of degree $d$.  The singularities in this
curve are generically double point singularities, giving the desired
curve with $n = g$ simple double points.

More generally given any F-theory base surface $B_2$, we can ask whether
it is possible to tune a generic curve of arithmetic genus $g$
 to produce an arbitrary number of double point singularities up
to the limit of available moduli.  As in the case of $\mathbb{P}^2$,
the technical challenge is to ensure that the condition for imposing
each double point is independent of the others.  We are unaware of
any general results of this kind in the mathematics literature.

\subsubsection{Explicit Weierstrass models with multiple double points}
\label{subsec:exp-weier-dbl}
Constructing an explicit global Weierstrass model for a theory with an
arbitrary number of two-index SU($N$) representations presents a
nontrivial challenge.  While many cases are
covered by the explicit constructions in \cite{ckpt, transitions}
and in \S\ref{sec:local-double}--\S\ref{sec:tuning-double},
there are also cases that cannot be realized directly in this way.  
In particular, these approaches 
give SU(N) models
realized on a divisor of the form 
\begin{equation}
A \xi^2 + B \xi \eta + C \eta^2 \,.
\label{eq:simple-quadratic}
\end{equation}
This constrains the range of possible combinations of singularities
that may be realized.
We
do not attempt to give a completely general analysis here but consider
various cases for $T = 0$ and small degree $d$. We focus on the $\gsu(3)$ models, although similar phenomena can be seen in higher $\gsu(N)$ models.

For the case of degree $d = 3, g = 1$, 
a single ($n = 1$) double point can be
easily realized through (\ref{eq:simple-quadratic}) through $[\xi] =[\eta] = 1$, which is compatible with a divisor of degree $d = 3$.

For the case of a quartic, $d = 4, g = 3$, we can realize $n = 1$ or
$n = 2$ through (\ref{eq:simple-quadratic}) by taking $[\xi] = 1,[\eta] = n$, but this does not work for $n = 3$.  Thus, the general
classes of explicit constructions do not provide a Weierstrass model
for the quartic with three double points on $\P^2$.  From the analysis
of the previous subsection we know that such a quartic exists, and in
\S\ref{sec:quartic-3} we give an explicit demonstration that a
Weierstrass model can be found where all three double points in this
quartic support symmetric + antisymmetric SU(3) representations.

For $d=5$, $g=6$, the curve \eqref{eq:simple-quadratic} can realize $n=1,2,$ or $4$ double points. These three cases can be constructed from the Weierstrass tunings in \S\ref{sec:double-points}. The $n=3,5,6$ cases, however, would require a more general construction.

For $d>5$, there are consistent spectra that cannot be realized using the Weierstrass models developed here, even if the curve \eqref{eq:simple-quadratic} can support the appropriate number of double points. As the degree increases, parameters in the Weierstrass model are more likely to become ineffective. While one can address this by setting an ineffective parameter to zero, there may be issues with the resulting Weierstrass model. If $\nubar$ and $\psibar$ are ineffective, $(f,g,\Delta)$ vanish to orders $(4,6,12)$ at the double points that should give symmetric matter. This implies, for example, that the $n=1$ models for $d\geq6$ cannot be realized using the constructions here. In other cases, enough parameters may be ineffective to force the discriminant to vanish exactly. An example of this occurs when $d=8$ and $n=6$. While such models cannot be constructed using the tunings presented here, it may be possible that some alternative Weierstrass tuning can realize these models.

Finally, recall from \S\ref{sec:anomalies} that there are cases such
as those on $\P^2$ with $d > 9$ where the choice of $a, b$, associated
  with the topology of the curve $\Sigma$, seem to force the presence of
  symmetric SU($N$) matter representations.  Specifically, the anomaly conditions imply that the number of $\gsu(3)$ fundamentals should be negative unless there is symmetric matter. Thus, in these cases
  there is no ``generic'' Weierstrass model on $\sigma$ without
  non-UFD structure at some double points.  While this may seem surprising, it can be seen directly from the UFD Weierstrass models. The discriminant of the UFD $\gsu(3)$ model takes the form
  \begin{equation}
  \Delta = \dblcurve^3\left(\phi_0^3\Delta_{\text{fund}} + \mathcal{O}(\dblcurve)\right),
  \end{equation}
  where the $\Delta_{\text{fund}}=\dblcurve=0$ loci support fundamental matter. On $\P^2$, $\phi_0$ and $\Delta_{\text{fund}}$ are respectively sections of $\mathcal{O}(3H)$ and $\mathcal{O}(27-3d)H$. For $d>9$, $\Delta_{\text{fund}}$ is ineffective, as are the higher-order coefficients in the discriminant. The discriminant is therefore forced to vanish identically, which is clearly problematic.\footnote{Alternatively, this can be seen in Tate form from the fact that the Tate coefficient $a_3$ \cite{Bershadsky-all} is of degree 9 for SU(3), and would vanish identically as would all the other Tate coefficients for an SU(3) tuning on a curve of degree $d > 9$.}
Thus, there is no ``generic'' Tate model for SU(3) on a degree $d > 9$
curve in $\P^2$, which agrees with the observation that there is no 6D
supergravity model with $T = 0$ and an SU(3) with $b = 10H$ and only
adjoint and fundamental matter.

While the $d>9$ models with only adjoint and fundamental charged
  matter are inconsistent, the anomaly conditions suggest there are
  anomaly-equivalent models with symmetric matter that are
  consistent. From the analysis of \S\ref{sec:anomalies}, there is a
  seemingly consistent $d=10$ model with 30 symmetrics, 6 adjoints,
  and zero fundamentals of SU(3). Additional spectra can be generated by
  exchanging an adjoint for a symmetric and a fundamental. (For
  $d>10$, the number of symmetrics required is greater than the genus
    of the curve, so there are no consistent models with
    $d>10$). However, none of these models seem to have corresponding
      Weierstrass models. Recall that a double point singularity that
      supports a symmetric hypermultiplet also supports a fundamental
      hypermultiplet. Any Weierstrass model with some number of
      symmetric hypermultiplets should therefore have at least as many
      fundamental hypermultiplets. But the supposedly consistent
      $d=10$ models all have fewer fundamentals than symmetric hypermultiplets,
      suggesting that such models cannot be realized using double
      point singularities. This behavior can be seen directly in the
      Weierstrass models in \S\ref{sec:double-points}, as the
      discriminant vanishes just as in the $d=10$ Tate model even when
      there are $\dbla=\dblb=0$ double points. Note that the
      general argument is fairly independent of the specific
      Weierstrass tuning; the key issue is the singularity type,
      specifically that any double point that supports symmetric
      matter should also support a fundamental hypermultiplet.  One might hope that
      some singularity type may contribute symmetric matter without
      the corresponding fundamental hypermultiplet; if this is the
      case, one could potentially construct the $d=10$ spectra from
      some alternative Weierstrass tuning. But without such a
      development, it seems unlikely that any Weierstrass model, not
      just the one described in \S\ref{sec:double-points}, could
      give the supposedly consistent $d=10$ spectra.

The upshot is that the Weierstrass models developed in 
\S\ref{sec:double-points}  do not realize all of the models with SU($N$)
gauge groups and two-index symmetric representations that look
consistent from the low-energy anomaly perspective. Some of these
models may be realizable in F-theory through a different Weierstrass
tuning than the one described here. However, the $d=10$ models
described above seem to be difficult to obtain in F-theory and may be
candidates for the F-theory ``swampland.'' It would be interesting
to investigate these models further in future work.

\subsubsection{Example: quartics with 3 double points}
\label{sec:quartic-3}

A quartic on $\mathbb{P}^2$ has genus 3 and should be able to support three double points. As mentioned previously, the form of the curve $\dblcurve$ used in \S\ref{sec:double-points} does not allow a quartic to have more than 2 double points. It is, however, possible to construct a quartic with three double points if one goes beyond the structure used earlier. Specifically, the curve
\begin{equation}
\qcurve = \qetaa \quarta^2 \quartc^2 + \qetab \quarta \quartb \quartc^2 + \qetac \quartb^2 \quartc^2 +\qetad \quarta^2 \quartb \quartc + \qetae\quarta \quartb^2 \quartc + \qetaf \quarta^2 \quartb^2 = 0
\end{equation}
has double point singularities at $\quarta = \quartb = 0$, $\quartb = \quartc = 0$, and $\quartc=\quarta = 0$.  If the $\qetaa$ through $\qetaf$ coefficients are constants and if $\quarta, \quartb, \quartc \in \mathcal{O}(H)$, $\qcurve$ is a quartic curve with three double points.

Identifying the curve is only the first part of the tuning. To proceed further, we must describe the normalized intrinsic ring $\normring{\qcurve}$. In fact, $\normring{\qcurve}$ resembles the normalized intrinsic ring used in \S\ref{sec:double-points}. We introduce three parameters, $\qLbc$, $\qLca$, and $\qLab$, described by the relations
\begin{align}
\qLbc =& \frac{1}{\quartc}\left[\left(\qetaf \quarta^2 + \qetae \quarta \quartc\right)\quartb + \frac{1}{2}\left(\qetad \quarta^2 +\qetac\quartb \quartc\right)\quartc\right] \label{eq:quartic-3-normbegin}\\
=& -\frac{1}{\quartb}\left[\left(\qetaa \quarta^2 + \qetab \quarta \quartb\right)\quartc + \frac{1}{2}\left(\qetad \quarta^2 + \qetac\quartb \quartc\right)\quartb \right],\\
\qLca =& \frac{1}{\quarta}\left[\left(\qetac \quartb^2 + \qetab \quartb \quarta\right)\quartc + \frac{1}{2}\left(\qetae \quartb^2 + \qetaa \quartc \quarta\right)\quarta\right]\\
=&- \frac{1}{\quartc}\left[\left(\qetaf\quartb^2 + \qetad \quartb \quartc\right)\quarta+\frac{1}{2}\left(\qetae \quartb^2 + \qetaa \quartc \quarta\right)\quartc\right],\\
\qLab=& \frac{1}{\quartb} \left[ \left(\qetaa \quartc^2 + \qetad \quartc \quartb\right)\quarta + \frac{1}{2}\left(\qetab \quartc^2 + \qetaf \quarta \quartb\right)\quartb\right]\\
=& - \frac{1}{\quarta}\left[\left(\qetac \quartc^2 + \qetae \quartc \quarta\right)\quartb + \frac{1}{2}\left(\qetab \quartc^2+\qetaf \quarta \quartb\right)\quarta\right].
\end{align}
These relations are analogous to the expressions
\begin{align}
\dblL = \frac{1}{\dblb}\left[\etaa \dbla+\etab \dblb\right]= -\frac{1}{\dbla}\left[\etab \dbla+\etac\dblb\right]
\end{align}
used in \S\ref{sec:double-points}. Moreover, $\qLbc$, $\qLca$, and $\qLab$ satisfy the relations
\begin{align}
\qLbc^2 &= \frac{1}{4}\left(\qetad \quarta^2 + \qetac \quartb \quartc\right)^2 - \left(\qetaf \quarta^2 + \qetae \quarta \quartc\right)\left(\qetaa \quarta^2 + \qetab \quarta \quartb\right),\\
\qLca^2 &= \frac{1}{4}\left(\qetae \quartb^2 + \qetaa \quartc \quarta\right)^2 - \left(\qetac \quartb^2 + \qetab \quartb\quarta\right)\left(\qetaf \quartb^2 +\qetad \quartb\quartc\right),\\
\qLab^2 &= \frac{1}{4}\left(\qetab \quartc^2 + \qetaf \quarta \quartb\right)^2-\left(\qetaa \quartc^2 + \qetad \quartc\quartb\right)\left(\qetac \quartc^2+\qetae \quartc \quarta\right)\label{eq:quartic-3-normend},
\end{align}
just as $\dblL$ satisfied the relation $\dblL^2 = \etab^2-\etaa\etac$. 

The general tuning process proceeds as before. We expand $f$ and $g$ as
\begin{align}
f &= f_0 + f_1 {\qcurve} + f_2 {\qcurve}^2 + \ldots & g &= g_0 + g_1 {\qcurve} + g_2 {\qcurve}^2 + \ldots
\end{align}
and impose conditions on the $f_i$ and $g_i$ to force the discriminant to vanish to certain orders. To obtain an $\gsu(3)$ gauge group in an UFD model, $f$ and $g$ would take the form
\begin{align}
f &= - \frac{1}{48}\PhizeroL^4 + \PhizeroL \PsiL \qcurve + f_2 {\qcurve}^2 + \ldots \label{eq:quart3-UFD-f}\\ 
g&= \frac{1}{864}\PhizeroL^6 -\frac{1}{12}\PhizeroL^3 \PsiL \qcurve +\left(\PsiL^2 - \frac{1}{12}\PhizeroL^2f_2\right){\qcurve}^2 + \ldots \label{eq:quart3-UFD-g}.
\end{align}
As in \S\ref{sec:double-points}, we use the UFD tunings, but we let $\PhizeroL$ and $\PsiL$ be elements of $\normring{\qcurve}$. Instead of expanding $\PhizeroL$ and $\PsiL$ as in Equations \eqref{eq:phi0exp} and \eqref{eq:lambdaexp}, we use
\begin{align}
\PhizeroL &= \qnua \quartb \quartc + \qnub \quartc \quarta + \qnuc \quarta \quartb + \qnubara \quarta \qLbc + \qnubarb \quartb \qLca + \qnubarc \quartc \qLab \label{eq:phi0exp3}\\
\PsiL &= \qpsia \quartb \quartc + \qpsib \quartc \quarta + \qpsic \quarta \quartb + \qpsibara \quarta \qLbc + \qpsibarb \quartb \qLca + \qpsibarc \quartc \qLab.\label{eq:lambdaexp3}
\end{align}
The products $\PhizeroL^2$, $\PhizeroL \PsiL$, and $\PsiL^2$ now lie
in $\quotring{\qcurve}$. The explicit expressions are lengthy, so we
do not write them here; however, they can be found by expanding out
the products and using relations \eqref{eq:quartic-3-normbegin}
through \eqref{eq:quartic-3-normend} to remove all occurrences of
$\qLbc$, $\qLca$, and $\qLab$. We can now plug in the expressions for
$\PhizeroL^2$, $\PhizeroL\PsiL$ and $\PsiL^2$ into
\eqref{eq:quart3-UFD-f} and \eqref{eq:quart3-UFD-g}, giving valid
expressions for $f$ and $g$. 
Note that a product of two distinct $\tilde{Q}$'s in these expressions
is always accompanied by the appropriate $\eta$ factors to immediately
put the term in $\quotring{\qcurve}$.
The zeroth and first order terms of the
discriminant vanish exactly, while the second order term is
proportional to an additional factor of $\qcurve$. Therefore, we have a non-Tate model in
which the discriminant is proportional to ${\qcurve}^3$. The $\quarta
= \quartb = 0$, $\quartb = \quartc = 0$, and $\quartc=\quarta = 0$
double points cannot be deformed away, and they support ${\tiny
  \yng(2)}+{\tiny \yng(1)}$ matter.

Note that if we let either $\quarta$, $\quartb$, or $\quartc$ be a constant, we recover expressions nearly identical to those in the Weierstrass model of \S\ref{sec:double-points}. This behavior should be expected. If, say, $\quartc$ is set to a constant, the only remaining double points occur at the $\quarta=\quartb$ loci. This is exactly the situation encountered in \S\ref{sec:double-points}, and we expect that the models should be identical up to trivial shifts and scalings of the parameters, as observed. 

%Notice that if we let $\quarta \rightarrow 1$,$\quartb \rightarrow 1$, or $\quartc \rightarrow 1$, we recover expressions similar to those of Equations \eqref{eq:phi0exp} and \eqref{eq:lambdaexp}. 
%Apart from this change, the tuning process is nearly identical to that given above. The $f_i$ and $g_i$ are defined in terms of $\PhizeroL$ and $\PsiL$ in the same way as before, and we use the above relations to convert local expressions for the $f_i$ and $g_i$ into global expressions. Order 0 and 1 parts of the discriminant vanish exactly, as before. For the order 2 part, the local ring expansions of Equations \eqref{eq:phi0exp3} and \eqref{eq:lambdaexp3} lead to
%\begin{equation}
%\Delta_2 \equiv 12 f_0 f_1^2 + 12 f_0^2 f_2 + 27 g_1^2 + 54 g_0 g_2 \propto \qcurve,
%\end{equation}
%indicating we have used the correct local ring construction for the curve with three double points. In total, the tunings
%\begin{align}
%f_0 &\sim -\frac{1}{48} \PhizeroL^4 & f_1 &\sim \PhizeroL \PsiL 
%\end{align}
%and
%\begin{align}
%g_0 &\sim \frac{1}{864}\PhizeroL^6 & g_1 &\sim -\frac{1}{12}(\PhizeroL^2)(\PhizeroL\PsiL) & g_2 &\sim \PsiL^2 - \PhizeroL^2 f_2
%\end{align}
%lead to an $\gsu(3)$ model.

\subsection{Triple points on $\P^2$}
\label{sec:triple-p2}

We now consider the general question of how many triple-symmetric
representations of SU(2) can be realized in a given class of models.
In this case, there are nontrivial constraints from geometry,
associated with the inability to tune curves with certain combinations
of singularities.  These correspond to theories in the apparent
swampland \cite{swamp, universality}, which look acceptable from the
low-energy point of view but cannot be realized in F-theory.

\subsubsection{Triple points on curves}

We investigate whether arbitrary numbers of triple points can be
realized on curves, specializing for simplicity to the case of the
base $\P^2$.  

The simplest example is a quartic, with genus $g = 3$.
It is straightforward to choose a quartic with a simple triple point
singularity, taking
\begin{equation}
\sigma = A u^3 + B u^2
v + C u v^2 + D v^3 \,,
% \label{eq:}
\end{equation}
with $A, B, C, D$ linear functions of the homogeneous coordinates
$[u:v:w]$.  

Next consider a quintic, with genus $g = 6$.  We can ask whether a
quintic can be tuned with two triple point singularities.  This is not
possible.  Were this possible,
without loss of generality we could put both triple points on
the line $v= 0$.  Then a quintic restricted to $v= 0$ would have to
vanish to third order at two points, {\it e.g.} $u = 0, u = 1$, but
this cannot happen since a quintic only has five roots.  So there cannot
be a quintic with two triple points.
This represents an interesting contribution to the apparent F-theory
swampland; there is an apparent low-energy model with $T = 0$ and an
SU(2) gauge group with anomaly coefficient $b = 5$ and two
triple-symmetric matter fields, but this cannot be realized in
F-theory.  It would be nice to understand if there is some nontrivial
low-energy explanation for the inconsistency of this theory.

Continuing, for a degree $b = 6$ curve we have $g = 10$.  We can check
the existence of such a curve with three triple points by performing a
{\it Cremona transformation}. In a Cremona transformation on the plane
we blow up three points $a, b, c$ and then blow down the three -1
curves associated with the lines $ab, bc, ac$.  Assuming the existence
of a degree 6 curve $C$ with three triple points at $a, b, c$ we perform
the Cremona transformation.  Each blow-up removes a triple point
singularity and drops the self-intersection of the curve by $-9$, so
the resulting curve has self-intersection $36-27 = 9$.  
 The lines $ab$ {\it etc.} do not intersect $C$ anywhere except at $a,  b, \ldots$, so blowing down these lines does not affect the
 self-intersection and the final curve after the Cremona
 transformation is a cubic with no self-intersections, which is
 certainly allowed.  We simply invert this process to create the
 desired curve $C$, {\it i.e.} we perform a Cremona transformation on
 the plane $\P^2$ carrying a smooth cubic $C'$ where all three points
 $a', b', c'$ are disjoint from $C'$.

In a similar way we can check the possibilities for $b$ up to $b = 9$.
For $b = 7, g = 15$.  Starting with the hypothetical maximum configuration of five
triple points on a curve $C$, the Cremona transformation on three of
the triple points gives a curve of self-intersection $49-27 + 3 = 25$,
{\it i.e.}  a quintic with two triple points, which is not allowed as
discussed above, so the maximal $b = 7$ configuration with five triple
points is also not allowed.  
Note that here the extra 3 in the self-intersection comes from the
fact that each of the lines blown down intersects the original septic
once, so blowing each down raises the self-intersection by 1.
For $b = 8, g = 21$, we can do a Cremona
transformation on three triple points, giving three double points, so
in the maximum case of $C$ degree 8 curve with 7 triple points this
gives a degree 7 curve with 4 triple points and three double points.
Iterating this process shows that this is allowed.

Further arguments are needed to go to $b = 9$ and beyond, but this
analysis shows that there are cases, like the situations of a degree 5
curve with two triple points and a degree 7 curve with five triple
points, which appear acceptable from anomaly considerations but are
not allowed simply from the geometry of singularities on the F-theory side.

\subsubsection{Explicit Weierstrass models with triple points}

From the general cubic construction with $\sigma = A \xi^3 + B \xi^2 \eta + C \xi \eta^2 + D \eta^3$ as described in \cite{KleversTaylor}
and in \S\ref{sec:details-3},
we can only have certain combinations
of triple points for each degree of $\sigma$.  Denoting $(x, y) =({\rm deg}\ \xi, {\rm deg}\ \eta) \rightarrow xy$ as the number of triple points,
we have:

For a quartic with
$b = 4$: $g = 3,$ $ (x, y) =(1, 1) \rightarrow xy =1$, so this
possibility has an explicit F-theory realization.

For a quintic with
$b = 5$:  $g = 6,$ $(x, y) =(1, 1) \rightarrow xy =1$.  We cannot
realize two triple points, which matches with the analysis above.

For $b = 6$: $g = 10$, $x = 1, y = 1, 2$, we can have $xy = 1, 2$ but
not 3 triple-symmetric representations.  Thus, while as discussed
above there is a sextic with 3 triple points, a more sophisticated
analysis, likely along the lines of \S\ref{sec:quartic-3}, is needed to
determine if a Weierstrass model can be realized with triple-symmetric
representations at each of these singular points.  

For $b = 7$, $g = 15$ we can realize $xy = 1, 2, 4$.  As discussed
above the case $r = 5$ is not allowed by geometry, so the only open
case is $r = 3$.

For $T = 0$, 
the largest allowed value of $b$
for a model with no 3-index symmetric representations of SU(2) is $b = 12$.  In this case, there are
55 adjoints and  111 neutral scalar fields, which can be read off from
Table~\ref{tab:mattermult}.
This is only enough neutral scalar fields to convert 45 of the
adjoints to triple-symmetric representations.
This matches with the observation that the Tate form for SU(2) on
$\P^2$ is only possible up to $b = 12$ (since $a_4$ is of degree 12).\footnote{Alternatively, the discriminant for the Tate model takes the form $\Delta = \phi^2 \Delta_{\text{fund}} \sigma^2 + \mathcal{O}(\sigma^3)$, and $\Delta_{\text{fund}}$ is ineffective for $b>12$.}
As discussed in \S\ref{sec:anomalies}, however, there are allowed
low-energy models with $b > 12$ that must have triple-symmetric
  matter.  Some of these can be realized through the general cubic
  Weierstrass construction.  For example (see Table (3.22) in
  \cite{KleversTaylor}), there is a model with $b = 13, x = 3, y = 4$
  which has  as matter content (from 12 triple points)
$6 \times\bs{4} + 30 \times\bs{3} + 58 \times\bs{2}$.  The anomaly
  equivalence would only allow the number of $\bs{4}$'s to be reduced
  by 4 before running out of fundamental representations needed to
  make the transformation, leaving at least 2 ($r \geq 4$
  half-hypermultiplet) 3-index symmetric matter representations in any
  valid model. Another example from \cite{KleversTaylor} has $b = 18,   x = y = 6$, with a total matter content of
$18 \times\bs{4} + 28 \times\bs{3} + 36 \times\bs{2}$.  Of the 18
3-index symmetric representations, only 5/2 can be removed through
anomaly-equivalent transitions. We speculate that the required presence of 3-index symmetric matter may  be related to the tallness constraints that force $q=3$ matter in some $U(1)$ models \cite{Morrison-Park-tallness}; however, we leave a full analysis of any possible connection to future work. 

The upshot of this analysis is that there are some cases that appear
allowed from low-energy anomaly considerations that definitely do not
have F-theory models since the required singularity combinations are
not allowed.  In other cases it remains to be shown whether an
explicit Weierstrass model can be realized even when the geometry
allows the singularity combination.
There is no general obstruction, however to finding Weierstrass models in
those cases where the generic matter content must contain 3-index
symmetric matter of SU(2).

\section{Allowed and disallowed representations}
\label{sec:allowed-representations}

So far in this paper we have focused on two- and three-index
representations of SU($N$), associated with double and triple
intersection points on the divisor carrying the gauge group.  In this
section we consider more generally what other kinds of exotic
representations and associated singularities may be allowed in
F-theory.  We begin with some comments on the generalization of the
algebraic analysis to quadruple point singularities and then consider
the constraints from F-theory geometry more generally.

\subsection{Higher singularities}
\label{sec:higher-singularities}

It is natural to ask whether a similar construction to those described above
could be carried
out for higher singularities, such as a quadruple intersection point.
Following the spirit of the simple examples in
\S\ref{sec:simple-examples}, for example, we can try to identify a
simple tuning of SU(2) on the divisor $\sigma = \xi^4-B \eta^4$.  In a
similar fashion to \S\ref{sec:example-triple}, we
can adjoin the element $\alpha$ satisfying $\alpha^4 = B$ to form the
normalized intrinsic ring and then can try various monomials
such as $\phi = \alpha^2 \eta$.  With this Ansatz, we have 
\begin{equation}
f_0 = -B \eta^2/48, \;\;\;\;\;
g_0 = B\xi^2 \eta/864,
% \label{eq:}
\end{equation}
and
\begin{equation}
\Delta_0 = B^2 \eta^2 \sigma/27648 \,.
% \label{eq:}
\end{equation}
At the next order, however, we have
\begin{equation}
\Delta_1 =  (B\xi^2 \eta)g_1/16 + (B^2 \eta^4) f_1/192
 +B^2 \eta^2/27648 \,.
% \label{eq:}
\end{equation}
This cannot be solved for $f_1, g_1$ as there are not enough powers of
$\xi, \eta$ in the last term
on the RHS.  So this Ansatz for the monomial $\phi$ does
not work.  The possibility $\phi = \alpha^3 \eta$ also leads to
similar problems. While the tuning can be completed for $\phi=\alpha^3 \eta^2$, the resulting $f$ and $g$ vanish to orders (4,6) on $\xi=\eta=0$, and the quadruple points seem to support superconformal matter rather than a standard $\gsu(2)$ representation.

%This suggests that it is not possible, at least in a straightforward
%fashion, to generalize the kind of construction described here for two-
%and three-index symmetric representations to the four-index symmetric
%representation of SU(2).  We have not proven in all generality that
%this is not possible algebraically, but we now argue that F-theory
%cannot realize the four-index symmetric representation of SU(2) or
%other more exotic representations on more general grounds, which
%explains the difficulty presented by this direct attempt at algebraic
%construction.
One might be interested in searching for a geometric realization of
some specific more complicated representation such as the
four-symmetric representation of SU(2) in a fashion analogous to the
realizations discussed here of the two- and three-symmetric
representations.  This immediately presents some issues, however.  To
begin with, the genus that would be needed for the four-symmetric
representation) would seem to be 21.
This does not match at all the pattern of the two-
and three-symmetric representations that are realized at double and
triple points, since a quadruple point has arithmetic genus of only 6.
While we do not have a direct conclusive argument that no
singularity type can be constructed that carries this representation,
we now argue that F-theory cannot realize the four-index symmetric
representation of SU(2) or other more exotic representations on more
general grounds, which explains the difficulties that would seem to be
present in any direct attempt at algebraic construction.

\subsection{Dynkin diagrams and higher representations}

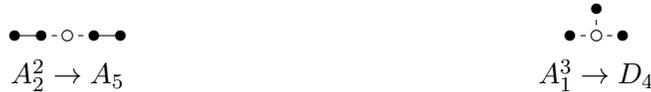
\begin{figure}
\begin{center}
\begin{picture}(200,50)(- 100,- 35)
%\put(0,0){\makebox(0,0){$\longrightarrow$}}
\put(-120,0){\line(1, 0){10}}
\put(-90,0){\line(1, 0){10}}
\multiput(- 120,0)(10,0){2}{\circle*{4}}
\multiput(- 90,0)(10,0){2}{\circle*{4}}
\put(-100, 0){\circle{4}}
\multiput( -106,0)(10,0){2}{\line(1,0){2}}
\multiput( 94,0)(10,0){2}{\line(1,0){2}}
\multiput(90,0)(20,0){2}{\circle*{4}}
\put(100,10){\circle*{4}}
\put(100, 0){\circle{4}}
\put(100,4){\line( 0,1){2}}
\put(- 100,- 15){\makebox(0,0){$A_2^2 \rightarrow A_5$}}
\put( 100,- 15){\makebox(0,0){$A_1^3 \rightarrow D_4$}}
\end{picture}
\end{center}
\caption[x]{\footnotesize Allowed embeddings of Dynkin diagrams
  corresponding to Kodaira singularities
  at codimension two points for (a)
the
two-index symmetric representation of SU(3), (b) the three-index
symmetric representation of SU(2).
Solid circles represent the Dynkin diagrams of the gauge group and the
open circle represents the matter fields.
}
\label{f:allowed-embeddings}
\end{figure}

One way of understanding the representations we have described so far
is in terms of the Dynkin diagrams associated with the Kodaira
singularity types on the divisor carrying the gauge group and at the
singular point.
In the simplest cases where the rank is enhanced by one and the
Katz-Vafa analysis \cite{Katz-Vafa}
applies, an $A_{N -1}$ singularity associated with SU($N$) is enhanced
to $A_N$ to give a fundamental representation, $D_N$ to give the
two-index antisymmetric tensor representation, and $E_N$ to give the
three-index antisymmetric tensor representation.
In the case of the two-index symmetric representation of SU($N$) studied in
\S\ref{sec:double-points}, the $A_{N -1}$ Dynkin diagram
is embedded twice in the $A_{2 N -1}$ Dynkin diagram
\cite{mt-singularities}.
Similarly, in the case of the three-index symmetric representation of
SU(2) studied in \S\ref{sec:details-3}, the $A_1$ Dynkin diagram
is embedded three times in $D_4$ \cite{KleversTaylor}.  Both of these
situations are illustrated in Figure~\ref{f:allowed-embeddings} and
represent embeddings of Dynkin diagrams that can be realized through
singularities in valid F-theory models.

The geometric interpretation of this embedding, following
\cite{Katz-Vafa}, is that the discriminant locus is being sliced by
a 1-parameter family of parallel curves.  Over the central curve,
one finds a singularity represented by the enlarged Dynkin diagram.
When the curve moves to a nearby, parallel curve, the singularities
present correspond to the subdiagram which has been embedded.
For example, over a generic curve passing through
the triple point of the discriminant corresponding to a
three-index symmetric representation of $SU(2)$ we find a singularity
of type $D_4$, but when that curve is moved to a nearby parallel curve,
it meets the discriminant locus in three separate points, each corresponding
to an $A_1$ singularity.

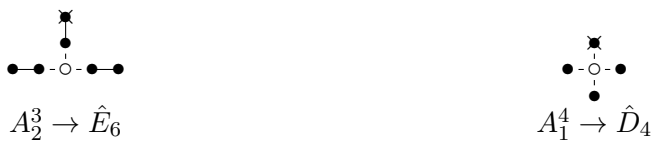
\begin{figure}
\begin{center}
\begin{picture}(200,70)(- 100,- 35)
%\put(0,0){\makebox(0,0){$\longrightarrow$}}
\put(-120,0){\line(1, 0){10}}
\put(-90,0){\line(1, 0){10}}
\put(-100, 10){\line( 0,1){ 10}}
\multiput(- 120,0)(10,0){2}{\circle*{4}}
\multiput(- 90,0)(10,0){2}{\circle*{4}}
\multiput(- 100,10)(0,10){2}{\circle*{4}}
%\put(-104,16){\line(1,1){8}}
%\put(-104,24){\line(1,-1){8}}
\put(-100,20){\makebox(0,0){$\times$}}
\put(-100, 0){\circle{4}}
\multiput( -106,0)(10,0){2}{\line(1,0){2}}
\put(-100, 4){\line(0,1){2}}
\multiput( 94,0)(10,0){2}{\line(1,0){2}}
\multiput(90,0)(20,0){2}{\circle*{4}}
\put(100,10){\circle*{4}}
\put(100,-10){\circle*{4}}
\put(100, 0){\circle{4}}
\put(100,4){\line( 0,1){2}}
\put(100,-6){\line( 0,1){2}}
\put(100,10){\makebox(0,0){$\times$}}
%\put(104,6){\line(1,1){8}}
%\put(104,14){\line(1,-1){8}}
\put(- 100,- 20){\makebox(0,0){$A_2^3 \rightarrow \hat{E}_6$}}
\put( 100,- 20){\makebox(0,0){$A_1^4 \rightarrow \hat{D}_4$}}
\end{picture}
\end{center}
\caption[x]{\footnotesize Disallowed embeddings of Dynkin diagrams
  corresponding to Kodaira singularities
  at codimension two points for (a)
the
three-index symmetric representation of SU(3), (b) the four-index
symmetric representation of SU(2).
Solid circles represent the Dynkin  diagrams of the gauge group and the
open circle represents the matter fields.
Circle with a cross represents the extra node of the extended Dynkin diagram.
}
\label{f:disallowed-embeddings}
\end{figure}

We now consider, however, what a Dynkin diagram embedding would look
like for representations that go beyond those considered here.  A
triple-symmetric representation of SU(3) would correspond to an
embedding of $A_2^3$ into a Dynkin diagram of rank $7$, but
no such Dynkin diagram exists.
It is tempting to instead try to use the extended Dynkin diagram $\hat{E}_6$, and
the embedding of $A_2^3$ illustrated in Figure~\ref{f:disallowed-embeddings}.
(Similarly, one could try to use an embedding of $A_1^4$ into 
the extended Dynkin diagram $\hat{D}_4$, also 
illustrated in Figure~\ref{f:disallowed-embeddings}.)
The flaw here is that the extended diagram can never correspond to
a singularity (over a curve on the base): since an extended Dynkin
diagram does not have a negative-definite intersection matrix, it
is not possible to contract {\em all}\/ of its curves simultaneously to
a single point.  In fact, since the linear combination of curves corresponding
to the maximal root of the associated Lie algebra is linearly equivalent
to the fiber of the elliptic fibration, if we shrink all of the curves in
the extended diagram to zero area, we necessarily shrink the elliptic
fiber itself to zero area (giving the F-theory limit of M-theory).

Thus, we argue that none of these embeddings are allowed as singularities
producing the desired matter representation.
Similar considerations  rule out the
exotic 4-index antisymmetric ({\bf 70}) representation of SU(8) and
the ``box'' ({\bf 20'}) representation of SU(4), as discussed in
\cite{mt-singularities}.  Despite some effort (see
e.g. \cite{transitions}), no F-theory Weierstrass model for either of
these representations has been found.

Note that this analysis suggests a number of nontrivial matter
configurations that are charged under two SU($N$) groups.  For
example, we could embed SU(2) $\times$ SU(3)$^2$ into $E_6$ or SU(2)
$\times$ SU(4)$^2$ into $E_8$ to realize matter charged under the
fundamental of the SU(2) and the two-index symmetric representation of
the other SU($N$) group.

One might ask whether other gauge groups besides SU($N$) can give
analogous exotic matter representations in F-theory associated with
singularities on the divisor supporting the gauge group.  In general,
this does not seem to be possible.  For all gauge groups other than
SU($N$) and Sp($N$) the Kodaira singularity involves a vanishing of
the Weierstrass coefficients $f, g$ to at least degrees $(2, 3)$.
Even a double point or cusp singularity at such a point thus would
involve a vanishing of $f, g$ to at least degrees $(4, 6)$.  This
implies that outside of the context of 6D SCFT's
\cite{Heckman-Morrison-Vafa, DelZotto-Heckman-Tomasiello-Vafa} no
higher gauge group can be supported on a divisor with an intrinsic
singularity.  So in a supergravity model without a superconformal
sector associated with a codimension two (4, 6) locus, we do not
expect exotic matter of the type considered in this paper for any
gauge groups other than SU($N$) and Sp($N$).  For Sp(1) the gauge
group is equivalent to SU(2), and the exotic matter is the 3-index
symmetric matter representation.  For Sp(2) and higher, the symmetric
matter representation is indistinguishable from the adjoint
representation, so this does not represent exotic matter.  This
matches with the discussion in \S\ref{sec:symmonodromy}, where we
identified the role of the non-UFD structure in SU($N$) symmetric
representations in terms of the field appearing as the square root in
the split condition; if this root can be defined near a given
singularity in terms of the ring of functions on the divisor then the
associated model has an adjoint matter representation, while if the
root lives in an extension associated with the normalized intrinsic
ring, then the model has a symmetric representation.

The upshot of this analysis is that we expect that the only exotic
representations associated with singular divisors in supergravity
theories without superconformal sectors are the 3-index symmetric
representation of SU(2) and the 2-index symmetric representation of
SU($N$) that we have studied here.  It may be interesting to try to
understand better to what extent this constraint on matter fields is
special to F-theory or may be more general.  Certainly from the point
of view of the low-energy theory, as discussed in
\cite{KumarParkTaylor} and in \S\ref{sec:background}, there are
anomaly-consistent models that contain higher exotic matter
representations such as the 4-index symmetric representation of SU(2),
and there are also anomaly-consistent models that contain exotic
matter representations of higher Kodaira groups such as
$G_2$.\footnote{Thanks to Andrew Turner for discussions and identifying
  some models of this type.}
It would be nice to understand whether these are actually inconsistent
models or part of the ``swampland''.  For example, heterotic
constructions may be able to give
rise to theories with the 4-index symmetric SU(2) representation and
exotic representations of e.g. SU(5) and SU(6)
with higher-level constructions \cite{Dienes-mr}, although it is not
clear if such constructions can preserve supersymmetry.
It would be nice to
understand whether such constructions are indeed possible in
consistent supersymmetric theories in 6D and/or 4D,
and whether they can be related to or bounded by
the physics of F-theory models.

\section{Conclusions}
\label{sec:conclusions}

In this paper we have developed a general approach to analyzing exotic
matter representations that can appear in F-theory when the gauge
group lives on a singular divisor $D$.  We analyzed symmetric
representations of SU($N$) and 3-index symmetric representations of
SU(2), and argued that these are the only exotic representations of a
single simple gauge factor that
can arise in this context.  These representations are realized through
unusual Weierstrass models in which the cancellation of the
discriminant to guarantee the $I_n$ Kodaira singularity type over $D$
is realized in a nontrivial way that is only possible when the ring of
functions on $D$ is not a unique factorization domain.
These results extend further the known correspondence between the
geometry of elliptic fibrations at codimension two singularities in
the base and the representation theory of matter in the associated
F-theory model.

We have used a variety of approaches to confirm the matter content in
the non-UFD constructions presented here.  In the examples studied in
previous work \cite{ckpt, transitions, KleversTaylor} the symmetric
SU($N$) matter content of the non-UFD Weierstrass models was
determined implicitly by connecting to other
abelian or nonabelian models through Higgsing and unHiggsing
transitions.  Here we have shown more explicitly how the structure of
the non-UFD models relates to the resolved geometry describing these
matter contents, with full details of the resolution worked out for a
concrete example in Appendix C.  It would be interesting to understand these
structures better from the geometric point of view, possibly also
using the string junction motivated approach of deformations explored
in \cite{Grassi-Halverson-Shaneson, Grassi-Halverson-Shaneson-2014}.

While the constructions of non-UFD models we have carried out here are
much more general than those encountered previously, it would be good
to have a more complete picture of the range of possibilities.  In
particular, this could be done by removing redundancies in the
Weierstrass models for these compactifications and explicitly
determining the number of independent degrees of freedom and comparing
to anomaly cancellation.

Among other things,  the analysis presented here gives a clear picture of which
spectra that appear to be allowed from low-energy anomaly constraint
considerations can be realized in F-theory.  This has allowed us to
identify a specific subset of models that are in the apparent ``swampland''
with no string realization and no clear low-energy inconsistency.
Further study of these models could lead to an improved understanding
of quantum consistency conditions for supergravity, and to a better
understanding of whether string theory is in fact universal for 6D
supergravity theories \cite{universality}.

The analysis done here of possible matter content for nonabelian
factors in 6D F-theory models also fits into the general program of
systematically classifying all elliptic Calabi-Yau threefolds by
identifying all allowed bases $B$ \cite{mt-clusters, mt-toric,
  Martini-WT, Wang-WT} and then carrying out all possible tunings of
Weierstrass models over each base \cite{Johnson-WT, Johnson-WT-2}.  In
particular, the analysis here seems to complete the picture of what
possible codimension two singularities may arise in principle from nonabelian charged matter,
associated with distinct Calabi-Yau threefold tunings over a given
base.  The detailed questions, however, of which anomaly-allowed
combinations of two- and three-index symmetric matter can be
explicitly realized in F-theory, however, must also be addressed to
complete the classification of associated Calabi-Yau threefold
geometries.

\vspace{.5in}

{\bf Acknowledgements}: We would like to thank 
Lara Anderson, Keith Dienes, I\~naki Etxebarria,
James Gray, 
and
Andrew Turner
for helpful
discussions.   
DRM and WT thank the Aspen Center for Physics,
which is supported by National Science Foundation grant PHY-1607611,
and the Institut Henri
Poincar\'e for hospitality during various stages of this project.
The work of DRM was supported in part by National Science Foundation grants 
PHY-1307513 and PHY-1620842 (USA) and by the Centre National de la Recherche Scientifique 
(France). 
The work of NR and WT was supported by
the DOE under contract
\#DE-SC00012567, and was also supported in
part by the National Science Foundation under Grant
No. PHY-1066293.

\appendix

\section{Weierstrass models with symmetric matter}
\label{app:summaryOfModels}
For the Weierstrass models that admit symmetric matter summarized below, we assume the gauge group is tuned on a curve of form
\begin{equation}
\dblcurve \equiv \etaa \dbla^2 + 2 \etab \dbla \dblb +\etac \dblb^2 = 0.
\end{equation}
The double points at $\dbla=\dblb=0 $ support symmetric matter.

Below we give the tunings for $\gsu(3)$, $\gsu(4)$, $\gsu(2k)$ and $\gsu(2k+1)$. The $\gsu(3)$ and $\gsu(4)$ tunings can be generated from the expressions for $\gsu(2k+1)$ and $\gsu(2k)$. However, since $\gsu(3)$ and $\gsu(4)$ are referenced frequently, we include the explicit expressions for convenience.
\subsection{$\gsu(3)$ with symmetric matter}
$f$ and $g$ are given by
\begin{align}
f &= -\frac{1}{48}\phidbl^2 + f_1 \dblcurve + f_2 \dblcurve^2+f_3 \dblcurve^3\\
g&= \frac{1}{864}\phidbl^3 -\frac{1}{12}\phidbl f_1 \dblcurve + g_2 \dblcurve^2 + g_3 \dblcurve^3.
\end{align}
$\phidbl$, $f_1$ and $g_2$ are given by 
\begin{multline}
\phidbl = \left(\nua \dbla + \nub\dblb\right)^2 - 2\nua
\nubar\left(\etab \dbla + \etac \dblb\right)\\ + 2 \nub
\nubar\left(\etaa \dbla +
\etab\dblb\right)+\nubar^2\left(\etab^2-\etaa\etac\right),
\label{eq:appendix-phi}
\end{multline}
\begin{multline}
f_1 = \left(\nua\dbla+\nub\dblb\right)\left(\psia\dbla+\psib\dblb\right) - \left(\nua\psibar+\psia\nubar\right)\left(\etab\dbla+\etac\dblb\right)\\+\left(\nub\psibar+\psib\nubar\right)\left(\etaa\dbla+\etab\dblb\right)+\nubar\psibar\left(\etab^2-\etaa\etac\right),
\end{multline}
and
\begin{multline}
g_2 = \left(\psia \dbla+\psib \dblb\right)^2 -2 \psia \psibar\left(\etab\dbla+\etac\dblb\right) \\
+ 2 \psib \psibar \left(\etaa\dbla+\etab\dblb\right)+\psibar^2\left(\etab^2-\etaa\etac\right)-\frac{1}{12}\phidbl f_2.
\end{multline}
Other parameters are untuned. The leading order term in the
discriminant is given by
\begin{multline}
\Delta = \Big[\frac{\phidbl^2}{16}\big[2\left(\nub \psia - \nua \psib\right)\left(\nubar\left(\psia \dbla + \psib \dblb\right)-\psibar\left(\nua \dbla + \nub \dblb\right)\right)\\
- \nubar^2\left(\etac\psia^2-2 \etab\psia\psib+\etaa\psib^2\right) - \psibar^2\left(\etac \nua^2 - 2 \etab \nua \nub + \etaa \nub^2\right) \\
+ 2\nubar\psibar\left(\etac\psia\nua - \etab \psia \nub - \etab \psib \nua + \etaa \psib \nub\right)\big]\\ 
  +4 f_1^3 -\frac{\phidbl^2}{2}f_1 f_2-\frac{9}{2}\phidbl f_1 g_2+
\frac{1}{192}\phidbl^4 f_3
+  \frac{1}{16}\phidbl^3 g_3\Big]\dblcurve^3 + \mathcal{O}(\dblcurve^4) \label{eq:SU3discapp}
\end{multline}

\subsection{$\gsu(4)$ with symmetric matter}
$f$ and $g$ are given by
\begin{align}
f=& -\frac{1}{48}\phidbl^2 - \frac{1}{6}\phidbl \phi_1 \dblcurve + f_2 \dblcurve^2+f_3\dblcurve^3\\
g =& \frac{1}{864}\phidbl^3+\frac{1}{72}\phi_1 \phidbl^2 \dblcurve + \frac{1}{36}\phidbl\left(\phi_1^2 - 3 f_2\right)\dblcurve^2\notag\\
&-\frac{1}{27}\left(9\phi_1 f_2+
\frac{9}{4} \phidbl f_3+
\phi_1^3\right)\dblcurve^3 + g_4 \dblcurve^4
\end{align}
Here, $\phidbl$ is
as given in (\ref{eq:appendix-phi})
%\begin{multline}
%\phidbl = \left(\nua \dbla + \nub\dblb\right)^2 - 2\nua \nubar\left(\etab \dbla + \etac \dblb\right)\\ + 2 \nub \nubar\left(\etaa \dbla + \etab\dblb\right)+\nubar^2\left(\etab^2-\etaa\etac\right),
%\end{multline}
and all other parameters are untuned. The leading order term in the discriminant is given by
\begin{equation}
\Delta = \frac{1}{576}\phidbl^2\left(36\phidbl g_4+12 \phidbl \phi_1 f_3 -4\left(3 f_2+\phi_1^2\right)^2\right)\dblcurve^4 + \mathcal{O}(\dblcurve^5) \label{eq:Delta4}
\end{equation}
\subsection{$\gsu(2k)$ with symmetric matter} 
$f$ and $g$ are given by
\begin{align}
f &= -\frac{1}{3}\upsilon^2 + \mathcal{O}(\dblcurve^k) \\ g&= -\frac{1}{27}\upsilon^3 - \frac{1}{3}\upsilon f + \mathcal{O}(\dblcurve^{2k}),
\end{align}
with
\begin{equation}
\upsilon = \frac{1}{4}\phidbl + \phi_1 \dblcurve + \phi_2 \dblcurve^2 + \ldots + \phi_{k-1}\dblcurve^{k-1}.
\end{equation}
$\phidbl$ is again
as given in (\ref{eq:appendix-phi})
and all other parameters are untuned.

\subsection{$\gsu(2k+1)$ with symmetric matter}
$f$ and $g$ are given by
\begin{align}
f&= -\frac{1}{3}\upsilon^2 + \digamma_k \dblcurve^k + \mathcal{O}(\dblcurve^{k+1}) \\
g&= -\frac{1}{27}\upsilon^3 - \frac{1}{3}\upsilon f + \gamma_{2k} \dblcurve^{2k} + \mathcal{O}(\dblcurve^{2k+1}),
\end{align}
with
\begin{equation}
\upsilon = \frac{1}{4}\phidbl + \phi_1 \dblcurve + \phi_2 \dblcurve^2 + \ldots + \phi_{k-1}\dblcurve^{k-1}.
\end{equation}
$\phidbl$
is again
as given in (\ref{eq:appendix-phi}),
and $\digamma_k$ and $\gamma_{2k}$ are given by
\begin{multline}
\digamma_k = \left(\nua\dbla+\nub\dblb\right)\left(\psia\dbla+\psib\dblb\right) - \left(\nua\psibar+\psia\nubar\right)\left(\etab\dbla+\etac\dblb\right)\\+\left(\nub\psibar+\psib\nubar\right)\left(\etaa\dbla+\etab\dblb\right)+\nubar\psibar\left(\etab^2-\etaa\etac\right),
\end{multline}
and
\begin{equation}
\gamma_{2k} = \left(\psia \dbla+\psib \dblb\right)^2 -2 \psia \psibar\left(\etab\dbla+\etac\dblb\right) + 2 \psib \psibar \left(\etaa\dbla+\etab\dblb\right)+\psibar^2\left(\etab^2-\etaa\etac\right).
\end{equation}
Other parameters are untuned. 

\section{$\gsu(2)$ Weierstrass model with three-index symmetric matter}
\label{app:su2summary}

The $\gsu(2)$ gauge group is tuned on a curve of form
\begin{equation}
\trplcurve \equiv \ta \trpla^3 + 3 \tb \trpla^2 \trplb + 3 \tc \trpla \trplb^2 + \td \trplb^3 = 0.
\end{equation}
The $\trpla=\trplb=0$ triple points support three-index symmetric matter. 
$f$ and $g$ are given by
\begin{align}
f&= f_0 + f_1 \trplcurve + \mathcal{O}(\trplcurve^2) & g &= g_0 + g_1 \trplcurve + \mathcal{O}(\trplcurve^2),
\end{align}
with
\begin{multline}
f_0 = -\frac{1}{48}\Bigg[\left(\ha \trpla^2 + 2 \hb \trpla \trplb + \hc \trplb^2\right)^2 + 2 \phibar\left(\ha \trpla + \hb\trplb\right)\trplLzero\\
+ 2 \phibar\left(\hb \trpla + \hc\trplb\right)\trplLone+\phibar^2 \trplLsq\Bigg],
\end{multline}
\begin{multline}
g_0 = \frac{1}{864}\Bigg[\left(\ha \trpla^2 + 2 \hb \trpla \trplb + \hc \trplb^2\right)^3 \\
+ 3 \phibar\left(\ha \trpla^2+2 \hb \trpla \trplb + \hc \trplb^2\right)\left[\left(\ha\trpla+\hb\trplb\right)\trplLzero+\left(\hb\trpla+\hc\trplb\right)\trplLone\right]\\
+3\phibar^2\left(\ha\trpla^2+2 \hb\trpla\trplb+\hc\trplb^2\right)\trplLsq + \phibar^3 \trplLcu\Bigg],
\end{multline}
\begin{align}
f_1 &= \lambdaa \trpla + \lambdab \trplb,\\
g_1 =&\frac{\phibar^2}{576}\left[\trpla\left(\hc \ta - 2 \hb \tb + \ha \tc\right)+\trplb \left(\hc \tb - 2 \hb \tc + \ha \td\right)-\frac{\phibar}{3}\left(\ta \td-\tb\tc\right)\right]\notag\\
& -\frac{1}{12}\left(\ha\trpla^2 + 2 \hb \trpla\trplb + \hc\trplb^2\right)\left(\lambdaa\trpla+\lambdab\trplb\right) - \frac{1}{12}\phibar\left(\lambdaa \trplLzero+\lambdab \trplLone\right)
.
\end{align}
$\trplLzero$, $\trplLone$, $\trplLsq$ and $\trplLcu$ are defined as
\begin{align}
\trplLzero=& - \tb \trpla^2 - 2 \tc \trpla \trplb - \td \trplb^2 \\
\trplLone =& \ta \trpla^2 + 2 \tb \trpla \trplb + \tc\trplb^2 \\
\trplLsq =& \left(\tb^2-\ta \tc\right)\trpla^2 + \left(\tb \tc - \ta \td\right)\trpla \trplb + \left(\tc^2-\tb \td\right)\trplb^2\\
\trplLcu =& \left( \ta \tb \tc - \tb^3 \right)\trpla^3
+3\left(\ta \tc^2 -\tc \tb^2\right)\trpla^2 \trplb\\
&+3\left(\ta \tc \td-\tb^2\td\right)\trpla \trplb^2 
+ \left( \tc^3 - 2 \tb \tc \td + \ta \td^2\right)\trplb^3.
\end{align}
All other parameters are untuned.

\section{Symmetric matter and resolutions}
\label{sec:resolution}

The non-UFD nature of the split condition plays a key role in the singularity structure for symmetric matter. To see this, we must analyze the resolution of elliptic fibration singularities at double points, which is described in \cite{mt-singularities}. Consider an $\gsu(4)$ model with symmetric matter located at double points, where the $A_3$ gauge singularity enhances to a codimension-two $A_7$ singularity. For simplicity, we assume we are working in six dimensions. Near a double point, the $\gsu(4)$ gauge curve will appear to consist of two components. The global structure of the gauge curve connects these two components, but if one focuses on a sufficiently small region near the double point, the two components look disconnected except for their intersection at the double point. For the purposes of the resolution, it is therefore sufficient to consider the limit in which the $\gsu(4)$ gauge curve factorizes. Suppose the curve takes the form
\begin{equation}
\sdcurve = \sda^2 - \sdcoeff \sdb^2,
\end{equation}
as in \S\ref{sec:simple-examples}. To analyze the double points at $\sda=\sdb=0$, we can consider the case where $\sdcoeff$ becomes a perfect square:
\begin{equation}
\sdcoeff \rightarrow \sdcosqrt^2.
\end{equation}
Then, the $\gsu(4)$ gauge curve factorizes into two components, given by
\begin{equation}
\sda \pm \sdcosqrt \sdb. \label{eq:factcomp}
\end{equation}
Of course, we eventually have to account for the fact that these two components are in fact connected. The details of this connection determine whether the double point contributes symmetric or adjoint matter.

Before turning to the specific way in which the split condition affects the resolution, let us outline the basic resolution procedure. There are $A_3$ singularities along the two components. These singularities are resolved via blow-ups that introduce three exceptional $\mathbb{P}^1$'s per component, giving six exceptional curves in total. The intersection pattern of the three exceptional curves forms an $A_3$ Dynkin diagram, and the three curves correspond to the positive simple roots of $A_3$. Other $-2$ curves, given by linear combinations of the three exceptional curves, correspond to the other $A_3$ roots. At the double point, the singularity type enhances from $A_3\times A_3$ to $A_7$. There are now seven exceptional curves forming an $A_7$ Dynkin diagram, and appropriate linear combinations of these curves fill out the $A_7$ roots. Some of these $A_7$ curves correspond to the $A_3\times A_3$ roots for the two components. The remaining $A_7$ curves correspond to the weights of charged matter localized at the double point. The intersection numbers of these remaining curves with the $A_3\times A_3$ exceptional curves give (the negative of) the Dynkin indices for the weights. From this information, one can read off the representations supported at the double point. As described so far, this process would seem to give $A_3\times A_3$ representations, such as $(\mathbf{4},\mathbf{4})$ or $(\mathbf{4},\mathbf{\bar{4}})$. Based on the global structure of the curve, an $A_3$ exceptional curve for one component can be identified with an exceptional curve in the other component. The identification then allows one to convert the $A_3\times A_3$ representations into $A_3$ representations such as the symmetric or adjoint representations.

To proceed further, we consider the standard $I_4$ Weierstrass model
\begin{equation}
y^2 = x^3 + f x + g,
\end{equation}
with
\begin{align}
f&= -\frac{1}{48}\phi^2 -\frac{1}{6}\phi \phi_1 \sdcurve +f_2 \sdcurve^2 \\ 
g&= \frac{1}{864}\phi^3+\frac{1}{72}\phi_1 \phi^2 \sdcurve + \frac{1}{36}\phi\left(\phi_1^2 - 3 f_2\right)\sdcurve^2-\frac{1}{27}\left(9\phi_1 f_2+\phi_1^3\right)\sdcurve^3 + g_4 \sdcurve^4.
\end{align}
If we define
\begin{equation}
x^\prime =  x - \frac{1}{12}\phi -\frac{1}{3} \phi_1 \sigma,
\end{equation}
the Weierstrass model can be written as
\begin{equation}
y^2 = {x^\prime}^3 +  \left(\frac{\phi}{4}+ \phi_1 \sigma\right){x^\prime}^2+ \left(f_2 + \frac{1}{3}\phi_1^2\right)\sigma^2x^{\prime} +  g_4 \sigma^4.
\end{equation}
We let $\sdcurve$ factorize into the two components in Equation \eqref{eq:factcomp}. In addition, we assume the split condition is satisfied with $\phi=\phi_0^2$. For now, we do not specify the form of $\phi_0$, although we will return to this issue shortly. Up to the inclusion of higher order terms in $x^\prime$ and $\sigma$, this model is similar to that in \cite{mt-singularities}.  However, $\phi_0$ and $f_2+12\phi_1^2$ were set to constants in \cite{mt-singularities}. Thus, the expressions for the exceptional curves given there have a hidden dependence on $\phi_0$. This dependence on $\phi_0$ must be considered to obtain a full understanding of the double points. Nevertheless, the steps of the resolution are identical, so we do not describe the full resolution procedure. Below, we discuss the end result of the resolution, focusing in particular on how parameters such as $\phi_0$ and $\beta$ affect the exceptional curves.

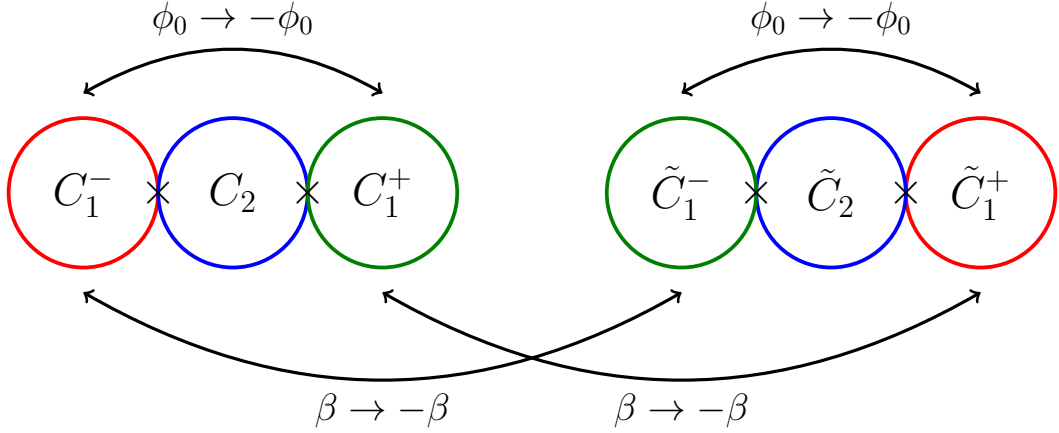
\begin{figure}
\begin{center}
\begin{tikzpicture}
\draw[color=red, line width=0.5mm] (-36ex,0) circle (6ex);
\draw (-36ex,0) node{\LARGE$C_1^-$};
\draw[color=blue, line width=0.5mm] (-24ex,0) circle (6ex);
\draw (-24ex,0) node{\LARGE$C_2$};
\draw[color=green!50!black, line width=0.5mm] (-12ex,0) circle (6ex);
\draw (-12ex,0) node{\LARGE$C_1^+$};

\draw (36ex,0)[color=red, line width=0.5mm] circle (6ex);
\draw (36ex,0) node{\LARGE$\tilde{C}_1^+$};
\draw (24ex,0)[color=blue, line width=0.5mm] circle (6ex);
\draw (24ex,0) node{\LARGE$\tilde{C}_2$};
\draw (12ex,0)[color=green!50!black, line width=0.5mm] circle (6ex);
\draw (12ex,0) node{\LARGE$\tilde{C}_1^-$};

\path [<->, very thick] (-36 ex,8ex) edge[bend left] (-12ex, 8 ex);
\draw (-24ex,14ex) node{\Large $\phi_0\rightarrow -\phi_0$};
\draw (24ex,14ex) node{\Large $\phi_0\rightarrow -\phi_0$};
\path [<->, very thick] (36 ex,8ex) edge[bend right] (12ex, 8 ex);

\path [<->, very thick] (36 ex,-8ex) edge[bend left] (-12ex, -8 ex);
\draw (12ex,-17.5ex) node{\Large$\beta\rightarrow-\beta$};
\draw (-12ex,-17.5ex) node{\Large$\beta\rightarrow-\beta$};
\path [<->, very thick] (-36 ex,-8ex) edge[bend right] (12ex, -8 ex);
\draw (-30ex,0) node{\LARGE$\mathbf{\times}$};
\draw (-18ex,0) node{\LARGE$\mathbf{\times}$};
\draw (18ex,0) node{\LARGE$\mathbf{\times}$};
\draw (30ex,0) node{\LARGE$\mathbf{\times}$};
\end{tikzpicture}
\end{center}
\caption{Exceptional curves for the $A_3\times A_3$ resolution. Circles denote the exceptional curves, with $x$'s marking the intersections. Arrows indicate how exceptional curves are exchanged under $\phi_0\rightarrow-\phi_0$ and $\beta\rightarrow-\beta$. Colors indicate which $C$ and $\tilde{C}$ curves are identified for the case with symmetric matter.}
\label{fig:a3xa3res}
\end{figure}

Along either of the two components, the blow-ups introduce three exceptional curves that form an $A_3$ Dynkin diagram, as illustrated in Figure \ref{fig:a3xa3res}.  For one of the components, we label the exceptional curves  $C_1^-$, $C_2$, and $C_1^+$. The plus and minus subscripts describe the dependence of the exceptional curves on $\phi_0$. The explicit expressions for $C_1^+$ and $C_1^-$ are nearly identical, except for the replacement of $\phi_0$ with $-\phi_0$. As a result, sending $\phi_0$ to $-\phi_0$ while leaving the other parameters unchanged exchanges $C_1^+$ and $C_1^-$. In $C_2$, $\phi_0$ only appears in even powers, so there is no exchange involving $C_2$ when $\phi_0 \rightarrow -\phi_0$. For the second component, the resolution produces three different exceptional curves, $\tilde{C}_1^{-}$, $\tilde{C}_2$, and $\tilde{C}_1^+$, that form a distinct $A_3$ Dynkin diagram. The $\tilde{C}_1^\pm$ have similar expressions related by $\phi_0\rightarrow -\phi_0$ and are thus exchanged when the sign of $\phi_0$ is flipped. Moreover, the $C_1^+$ and $\tilde{C}_1^+$ expressions are nearly identical, except for the fact that they are associated with different components. If one were to exchange the two components by sending $\beta \rightarrow -\beta$ (while keeping other parameters fixed), $C_1^+$ and $\tilde{C}_1^+$ would then be exchanged. The same is true for $C_1^-$ and $\tilde{C}_1^-$.

\begin{figure}
\begin{center}
\begin{tikzpicture}
\draw[fill=black, thick] (-36ex,0) node[circle, fill=black,label=below:{\large$\gamma_1^-$},draw]{} -- (-24ex,0) node[circle, fill=black,label=below:{\large$\gamma_2^-$},draw]{} -- (-12ex,0) node[circle, fill=black,label=below:{\large$\gamma_3^-$},draw]{} -- (0ex,0) node[circle, fill=black,label=below:{\large$\gamma_4^{\phantom{+}}$},draw]{} -- (12ex,0) node[circle, fill=black,label=below:{\large$\gamma_3^+$},draw]{}-- (24ex,0) node[circle, fill=black,label=below:{\large$\gamma_2^+$},draw]{} -- (36ex,0) node[circle, fill=black,label=below:{\large$\gamma_1^+$},draw]{};
\draw[line width=0.5mm, dashed,color=red] (-36ex, 0) circle(6ex);
\draw[color=red] (-36ex, 8ex) node{\large$C_1^-$};
\draw[line width=0.5mm, dashed,color=green!50!black] (-12ex, 0) circle(6ex);
\draw[color=green!50!black] (-12ex,8ex) node{\large$\tilde{C}_1^-$};
\draw[line width=0.5mm, dashed,color=green!50!black] (36ex, 0) circle(6ex);
\draw[color=green!50!black] (36ex,8ex) node{\large$C_1^+$};
\draw[line width=0.5mm, dashed,color=red] (12ex, 0) circle(6ex);
\draw[color=red] (12ex,8ex) node{\large$\tilde{C}_1^+$};
\draw[line width=0.5mm, dashed,color=blue] (0ex, 0) circle(6ex);
\draw[color=blue] (0ex, 8ex) node{\large$\tilde{C}_2$};
\draw[line width=0.5mm, dashed,color=blue] (-18ex, 12ex) arc(90:270:12ex) -- (18ex, -12ex) arc(-90:90:12ex) -- (-18ex, 12ex) ;
\draw[color=blue] (0ex, -14ex) node{\large$C_2$};

\draw (-30ex,0) node{\LARGE $\mathbf{\times}$};
\draw (-6ex,0) node{\LARGE $\mathbf{\times}$};
\draw (6ex,0) node{\LARGE $\mathbf{\times}$};
\draw (30ex,0) node{\LARGE $\mathbf{\times}$};
\end{tikzpicture}
\end{center}
\caption{Embedding of $A_3 \times A_3 \rightarrow A_7$ at a double point. Black dots represent exceptional curves for the $A_7$ singularity, with the lines between them denoting intersections between the exceptional curves. Colored lines indicate the combinations of $\gamma$ curves corresponding to the $A_3\times A_3$ exceptional curves. Colors indicate which $C$ and $\tilde{C}$ curves are identified for the case with symmetric matter.}
\label{fig:a7res}
\end{figure}
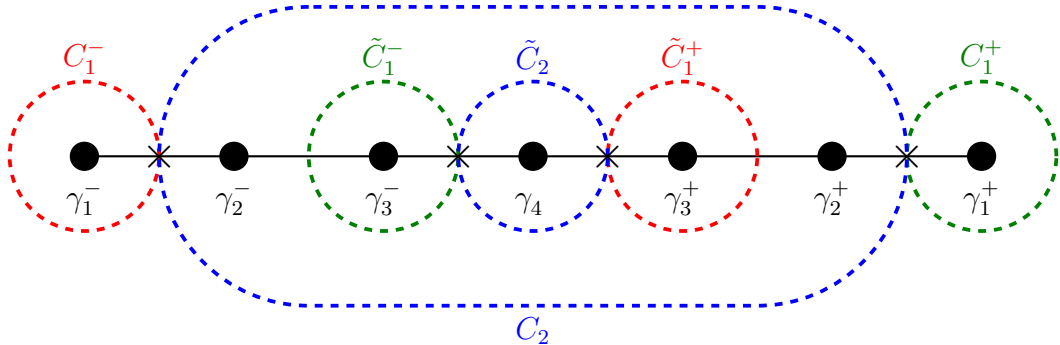

At the $\sda=\sdb=0$ points, the singularity type enhances to $A_7$. We now have seven exceptional curves, denoted by the symbol $\gamma$, whose intersections are summarized by the Dynkin diagram in Figure \ref{fig:a7res}. These curves, together with other $-2$ curves given by linear combinations of the $\gamma$'s, fill out the 28 positive roots of $A_7$. Some of these positive roots correspond to the $A_3$ exceptional curves from before. In particular, the $C$ and $\tilde{C}$ exceptional curves become linear combinations of the $\gamma$ curves at the double point:
\begin{align}
C_1^{\pm} &\rightarrow \gamma_{1}^{\pm}& 
C_2 &\rightarrow \gamma_2^- + \gamma_3^- + \gamma_4 + \gamma_3^{+} + \gamma_2^+& 
\tilde{C}_1^{\pm} &\rightarrow \gamma_3^{\pm} & 
\tilde{C}_2 &\rightarrow \gamma_4 \label{eq:excurvesa3toa7}
\end{align}
Likewise, the other $A_3$ roots, formed by linear combinations of the $C$'s and $\tilde{C}$'s, become linear combinations of the $\gamma$'s at the double point. Thus, 12 of the 28 positive $A_7$ roots represent the positive $A_3$ roots from before. The remaining positive $A_7$ roots correspond to the weights of the charged matter localized at the double point. One can calculate the intersection numbers of these curves with those in Equation \eqref{eq:excurvesa3toa7} to obtain the (negative of) the Dynkin indices of the weights. An explicit analysis of the weights shows that the charged matter consists of bifundamentals; since we are essentially dealing with an $\gsu(4)\times\gsu(4)$ gauge group, this is the expected result. In particular, the curve $\gamma_* = \gamma_3^-+\gamma_4+\gamma_3^++\gamma_2^++\gamma_1^+$ has the intersection numbers
\begin{align}
\gamma_*\cdot C_1^- &= 0 & \gamma_*\cdot C_2 &= 0 &  \gamma_*\cdot C_1^+ &= -1 \\
\gamma_*\cdot \tilde{C}_1^- &= -1 & \gamma_*\cdot \tilde{C}_2 &= 0 &  \gamma_*\cdot \tilde{C}_1^+ &= 0
\end{align}
The corresponding root therefore has Dynkin indices $[0,0,1]$ and $[1,0,0]$, which are those for the highest weight of the bifundamental. 

Now, we can return to the situation where the gauge curve does not factorize. Near the double point, the curve still appears to consist of two distinct components, but the two components are connected by the global structure of the curve. The two $A_3$'s from the two components should therefore be identified with each other. In particular, an exceptional curve for one component should be identified with a specific exceptional curve for the other branch. In the setup described above, the two components are essentially identical except for the sign of $\beta$. We therefore need to examine how the forms of the exceptional curves change when $\beta$ is sent to $-\beta$ while other parameters are unchanged. Suppose we have a standard, UFD tuning, where $\phi_0$ does not depend on $\beta$. Then, when $\beta \rightarrow -\beta$, the curve $C_1^+$ becomes $\tilde{C}_1^+$, indicating that $C_1^+$ and $\tilde{C}_1^+$ should be identified. Importantly, $\phi_0$ was unaffected by letting $\beta \rightarrow -\beta$, implying that $C_1^+$ should be identified with $\tilde{C}_1^+$ and not $\tilde{C}_1^-$. The $A_7$ curve $\gamma_*$ corresponding to the highest weight intersects two curves that are not identified, $C_{1}^+$ and $\tilde{C}_{1}^-$. Once the global structure of the gauge curve is accounted for, the Dynkin index is $[1,0,1]$, that for the highest weight for adjoint matter. This implies that in the UFD situation, the double point contributes adjoint matter.

For the non-UFD tuning from \S\ref{sec:simple-examples}, $\phi_0$ is proportional to $\sdL$. Note that $\beta$ in some sense plays the same role as $\sdL$, so $\phi_0$ is essentially proportional to $\beta$. Taking $\beta \rightarrow -\beta$ therefore changes the sign of $\phi_0$ as well. $C_{1}^+$ is now identified with $\tilde{C}_{1}^-$, not with $\tilde{C}_1^+$. Since $\gamma_*$ intersects both $C_{1}^+$ and $\tilde{C}_{1}^-$, the highest weight now has Dynkin indices $[2,0,0]$, signaling the appearance of symmetric matter. This alternative identification relies crucially on the fact that $\phi_0$ has a particular structure based on the form of the gauge curve. An arbitrary $\phi_0$, such as that in the UFD tuning, leads to an identification corresponding to adjoint matter. The non-UFD implementation of the split condition is thus a vital feature of the models with symmetric matter.

\end{document}